\documentclass[universe,review,accept,pdftex,moreauthors]{Definitions/mdpi}

\usepackage{amsmath}    
\usepackage{amsfonts}   
\usepackage{mathtools}  
\usepackage{tensor}     
\usepackage{graphicx}   
\usepackage{caption}     
\usepackage{subcaption}  
\usepackage{accents}     

\renewcommand\arraystretch{1.1} 
\newcommand{\be}{\begin{equation}}
\newcommand{\ee}{\end{equation}}
\firstpage{1} 
\pubvolume{1}
\issuenum{1}
\articlenumber{0}
\pubyear{2024}
\copyrightyear{2024}
\externaleditor{Firstname Lastname}
\datereceived{29 April 2025} 
\daterevised{14 June 2025} 
\dateaccepted{15 June 2025} 
\datepublished{ } 
\hreflink{https://doi.org/}

\Title{From Barthel--Randers--Kropina Geometries to the \linebreak  Accelerating Universe: A Brief Review of Recent \linebreak  Advances in Finslerian Cosmology}

\TitleCitation{From Barthel--Randers--Kropina Geometries to the    Accelerating Universe: A Brief Review of Recent    Advances in Finslerian Cosmology}

\Author{Amine Bouali 
$^{1,2}$, Himanshu Chaudhary   $^{3,4}$, Lehel Csillag    $^{3,5}$, Rattanasak Hama   $^{6}$, Tiberiu Harko   $^{3,7,}$*, \linebreak  Sorin V. Sabau   $^{8,9}$, Shahab Shahidi   $^{10}$}

\AuthorNames{Amine Bouali, Himanshu Chaudhary, Lehel Csillag, Rattanasak Hama, Tiberiu Harko, Sorin V. Sabau, and Shahab Shahidi}

\AuthorCitation{Bouali, A.; 
 Chaudhary, H.; Csillag, L.; Harko, T.; Hama, R.; Sabau, V.S.; Shahidi, S..}

\address{%
$^{1}$ \quad Laboratory of Physics of Matter and Radiation, Mohammed I University,  Oujda BP 717, Morocco; a1.bouali@ump.ac.ma \\ 

$^{2}$ \quad Higher School of Education and Training, Mohammed I University,  Oujda BP 717, Morocco\\

$^{3}$ \quad Department of Physics, Babeș-Bolyai University, Kogălniceanu Street 1, 400084 Cluj-Napoca,   Romania; himanshu.chaudhary@ubbcluj.ro; (H.C.); lehel.csillag@ubbcluj.ro   (L.C.)    \\

$^{4}$ \quad Research Center of Astrophysics and
Cosmology, Khazar University, 41 Mehseti Street,  \linebreak   Baku     AZ1096, 
Azerbaijan\\ 

$^{5}$ \quad Department of Mathematics and Computer Science, Transilvania University,  Iuliu Maniu Street 50, \linebreak    500091  Brașov, Romania    \\ 

$^{6}$ \quad Faculty of Science and Industrial Technology, Prince of Songkla University, Surat Thani Campus, \linebreak   Surat Thani     84000, Thailand; rattanasak.h@psu.ac.th \\

$^{7}$ \quad Astronomical Observatory, 19 Ciresilor Street, 400487  Cluj-Napoca, Romania\\

$^{8}$ \quad School of Biological Sciences, Department of Biology, Tokai University, Sapporo 005-8600, Japan; sorin@tokai.ac.jp \\

$^{9}$ \quad Graduate School of Science and Technology, Physical and Mathematical Sciences, Tokai University, \linebreak    Sapporo 005-8600, Japan\\

$^{10}$\quad School of Physics, Damghan University, Damghan   41167-36716, Iran; s.shahidi@du.ac.ir}

\corres{Correspondence:   tiberiu.harko@aira.astro.ro}

\abstract{We present a review of recent developments in cosmological models based on Finsler geometry, as well as geometric extensions of general relativity formulated within this framework.  Finsler geometry generalizes Riemannian geometry by allowing the metric tensor to depend not only on position but also on an additional internal degree of freedom, typically represented by a vector field at each point of the spacetime manifold. We examine in detail the possibility that Finsler-type geometries can describe the physical properties of the gravitational interaction, as well as the cosmological dynamics. In particular, we  present and review the implications of a particular implementation of Finsler geometry, based on the Barthel connection, and of the $(\alpha,\beta)$ geometries, where $\alpha$ is a Riemannian metric, and $\beta $ is a  one-form. For a specific construction of the deviation part $\beta$, in these classes of geometries, the Barthel connection  coincides with the Levi--Civita connection of the associated Riemann metric. We review the properties of the gravitational field, and of the cosmological evolution in three types of geometries: the Barthel--Randers geometry, in which the Finsler metric function $F$ is given by $F=\alpha +\beta$, in the Barthel--Kropina geometry, with $F=\alpha ^2/\beta$, and in the conformally transformed Barthel--Kropina geometry, respectively. After a brief presentation of the mathematical foundations of the Finslerian-type modified gravity theories,  the generalized Friedmann equations in these geometries are written down by considering that the background Riemannian metric in the Randers and Kropina line elements is of Friedmann--Lemaitre--Robertson--Walker type. The matter energy balance equations are also presented, and they are interpreted from the point of view of the thermodynamics of irreversible processes in the presence of particle creation. We investigate the cosmological properties of the Barthel--Randers and Barthel--Kropina cosmological models in detail.  In these scenarios, the additional geometric terms arising from the Finslerian structure can be interpreted as an effective geometric dark energy component, capable of generating an effective cosmological constant. Several cosmological solutions—both analytical and numerical—are obtained and compared against observational datasets, including Cosmic Chronometers, Type Ia Supernovae, and Baryon Acoustic Oscillations, using a Markov Chain Monte Carlo (MCMC) analysis.  A direct comparison with the standard $\Lambda$CDM model is also carried out. The results indicate that Finslerian cosmological models provide a satisfactory fit to the observational data, suggesting they represent a viable alternative to the standard cosmological model based on general relativity.}


\begin{document}


\section{Introduction---Early History and Modern~Developments}

In 1918, the~same year in which Herman Weyl proposed his important extension of  Riemannian geometry~\cite{Weyl1, Weyl2}, by~introducing the concept of non-metricity to mathematics and physics, another equally interesting generalization of the Riemannian framework was proposed by Paul Finsler~\cite{Finsler}.  The~work of Weyl was clearly influenced by the progresses in the physics of his time, and~especially by the development of general relativity by Einstein~\cite{Ein0,Ein1} and Hilbert~\cite{Hil}, and~it has been extensively used in physics, especially recently~\cite{Scholtz}. On~the other hand,  Finsler geometry, which represents, at~least from a physicist's perspective, a~much more drastic extension of  Riemannian geometry, was not considered from the point of view of applications in fundamental physics for at least 40~years. 

 Actually, Finslerian geometry was already anticipated by Riemann~\cite{Riemann}, who already assumed the existence of metric structures in a general space based on the  distance element $ds=F\left(x^1,x^2,\dots,x^n;dx^1,dx^2,\dots,dx^n\right) =F\left(x^1,x^2,\dots,x^n;,y^1,y^2,\dots,y^n \right)$, or~simply $F(x,y)$, where the condition that for  $y=0$, $F$ is a positive definite function defined on the tangent bundle $TM$ must also be imposed. In~order to assure the independence of the length of curves on their parameterization, it is further required that $F$ is a homogeneous function of degree one in $y$. 
 Riemann geometry is just a special case of the general distance element $ds$, obtained under the assumption $F^2=g_{IJ}(x)dx^Idx^J$. Thus, Finsler geometry is not a proper generalization of Riemann's geometry, but~it ``...is just Riemannian geometry without the quadratic restriction'' \cite{Chern}. However, it is an established practice especially in the physics literature to describe Finsler geometry as a generalization of Riemann's geometry, and~in the present work, we will also adopt this view, mostly for the simplicity of the~presentation.   

In a simple description, Finsler geometry can be considered as a geometry in which the metric tensor $g_{ij}$ is a function of both the coordinates $x$, defined on the base manifold $M$, and~of the tangent vectors $y$, so that $g_{IJ}=g_{IJ}(x,y)$. In~a physically intuitive sense, Finsler geometry can be considered as a geometry in which the metric tensor, determining the distance between two neighboring points, is an arbitrary function of both coordinates and velocities. For~presentations  of Finsler geometry from a mathematical perspective, see~\cite{F2b,F4b,F5b,F6b}.

The remarkable success of general relativity in the description of gravitational phenomena and~its many observational confirmations have significantly strengthened the position of Riemannian geometry as the ``true'' geometry of nature.  General relativity, in~its Riemannian formulation, successfully passes all the Solar System tests, inside which it can explain  the gravitational phenomenology with high precision. Important confirmations of the validity of general relativity include such diverse phenomena as the perihelion precession  of Mercury, light bending by the Sun, frame dragging,  the~Nordtvedt effect in lunar motion, and~the Shapiro time delay~\cite{Will}. The~experimental detection of  gravitational waves represents a brilliant confirmation of the predictions of general relativity~\cite{Grav1}, which opened a new window in the Universe, and~led  to new perspectives on the black hole properties, their dynamical evolution, and~the mass distribution of  neutron stars~\cite{Grav2}. Inspired by the immense success of general relativity, the~Riemannian geometric framework was also extended, leading to important developments in mathematics like the introduction of the concept of torsion~\cite{Cartan1,Cartan2}, or~of the concept of absolute parallelism~\cite{Weitz}. These new geometries have found important applications in physics, and~they have been used to model the gravitational interaction from new~perspectives. 

From a mathematical point of view, Ehlers Pirani and Schild aimed to develop a constructive axiomatization of gravity. In~their seminal 1972 paper, they built up the kinematical structure of general relativity, purely based on axioms, which have a somewhat empirical content~\cite{Ehlers:2012}. Although~the axiomatization is quite dense and technical, recent pedagogical reviews are present in the literature~\cite{Linnemann:2021, Adlam:2022, Adlam:2022b}. We briefly outline the four steps of the axiomatization in the following~\cite{Pfeifer:2019}:
\begin{enumerate}
    \item In the first step, the~basic physical/empirical entities are postulated, such as the timelike worldlines for freely falling massive particles, the~lightlike worldlines of light rays, and~radar echoes between massive particle worldlines. These empirical elements provide enough structure to define a system of coordinates and allow for the construction of a differentiable structure on the spacetime set $M$, turning it into a smooth manifold.
    \item The conformal structure is established by requiring that, at~each point in spacetime, the~set of all possible directions (i.e., tangent vectors) splits into two components when the directions corresponding to massless (lightlike) trajectories are removed. This splitting reflects the causal distinction between future and past.
Additionally, in~a sufficiently small neighborhood $V$ around the worldline of a massive particle, for~any point $p \in V$ not lying on the particle's path, the~function that maps $p$ to the product of the radar emission time $t_e$ and reception time $t_r$, i.e.,~$p \mapsto t_e t_{r}$, must be at least twice differentiable.
    \item Imposing that through each point in spacetime, and~for each timelike direction, there exists one unique timelike (massive) trajectory passing through that point, which results in a projective structure. Each of these trajectories must admit a parametrization such that, in~local coordinates near the point, the~motion satisfies $\ddot x =0$. This expresses the fact that particles move along straight lines in free fall.
\item In the final step, compatibility between the conformal and projective structures is required. In~particular, light rays must be special cases of particle geodesics in the limit of zero mass. This determines the metric up to a conformal factor, which leads to a Weyl structure. Through some technical steps, 
eliminating the second clock effect leads to a Lorentzian structure, i.e.,~a pseudo-Riemannian manifold.
\end{enumerate}
It turns  
 out that if one weakens the twice differentiability assumption on the map $p \mapsto t_{e} t_{r}$, there are Finsler functions which satisfy all the other EPS axioms. A~well-known example is given by
\begin{equation}
    F(x,y)=e^{2\sigma(x,y)} \sqrt{\left|a_{IJ}(x)y^I y^J \right|},
\end{equation}
where $\sigma(x,y)$ is a non-singular $0$-homogeneous function in $y$ satisfying
\begin{equation}
    \partial_{K} \sigma - y^{ M} \Gamma^{N}_{M K}(x) \frac{ \partial \sigma(x,y)}{\partial y^{N}}=0,
\end{equation}
with $\Gamma^{N}_{MK}$ being the Christoffel symbols associated with the Riemannian metric $a_{IJ}(x)$. This geometry is of Berwald type, i.e.,~it is the closest to a Riemannian geometry, but~it is not~trivial.

Despite the systematic, rigorous and physically appealing nature of Finsler geometry, the~applications in  physics of this geometry appeared relatively late. A~first step in this direction was taken in the work of Randers~\cite{Rand},  who tried to formulate a unified theory of electromagnetism and gravity. Initially, the~theory was formulated in a higher-dimensional Riemannian geometry. The~Finslerian nature of the Randers geometry was recognized by Ingarden~\cite{Ing}.  Essentially, Randers geometry is a specific example of a Finsler geometry, with~the fundamental function $F(x,y)=\left(a_{IJ}(x)dx^Idx^J\right)^{1/2}+b_K(x)y^K$, where $b_K(x)$ is an arbitrary vector field. Presently,  Randers spaces are considered as Finsler spaces $F^n = (M, \alpha +\beta)$, equipped with the Cartan nonlinear connection $N$, i.e.,~they are denoted with $RF^n=(M,\alpha +\beta, N)$ \cite{Mir}. A~comprehensive overview of several Finsler metrics and of some of their physical applications is given in Table~\ref{tab:finsler-metrics}.
\begin{table}[H]
\centering
\caption{Examples of Finsler 
 metrics with their defining functions and characteristic~properties.}
\label{tab:finsler-metrics}
\resizebox{\textwidth}{!}{\begin{tabular}{>{\raggedright\arraybackslash}m{3cm}>{\raggedright\arraybackslash}m{6.3cm}>{\raggedright\arraybackslash}m{6.3cm}}
\toprule
\textbf{Name} & \textbf{Finsler Function \boldmath{\(F(x, y) \)}} & \textbf{Properties} \\
\midrule

Semi-Riemannian & 
\(F(x, y) = \sqrt{|a_{IJ}(x) y^I y^J|} \) & 
Quadratic in $y$, reversible, i.e.,~$F(x,y)=F(x,-y)$. \\

\midrule
Randers & 
\(F(x, y) = \sqrt{a_{IJ}(x) y^I y^J} + b_I(x) y^I \) & 
Non-reversible, i.e.,~$F(x,y) \neq F(x,-y)$; appears in EM analogs and Lorentz-violating physics. \\

\midrule
Kropina & 
\(F(x, y) = \frac{a_{IJ} y^I y^J}{b_I y^I}, \quad b_I y^I \neq 0 \) & 
Singular on hypersurfaces; non-reversible. \\

\midrule
Matsumoto & 
\(F(x, y) = \frac{\sqrt{a_{IJ}y^I y^J}}{{{
\sqrt{a_{IJ}y^I y^J}}} - b_K y^K} \) & 
Non-reversible; used in irreversible mechanics; singular on hypersurfaces. \\

\midrule
Bogoslovsky & 
\(F(x, y) = \left(-\eta_{IJ} y^I y^J \right)^{\frac{1 - b}{2}} \left(n_K y^K\right)^b \) & 
Breaks full Lorentz invariance; \(0 < b < 1 \); used in very special relativity. \\

\midrule
Funk & 
\(F(x, y) = \frac{\sqrt{\delta_{IJ} y^I y^J} - \delta_{IJ} x^I y^J}{1 - \delta_{IJ} x^I x^J} \) & 
Defined on unit ball; forward complete; non-reversible. \\

\midrule
Locally Minkowskian & 
\(F(x, y) = F(y) \) & 
General flat case; depends only on \(y \), independent of \(x\). \\

\midrule
\(\alpha\)-\(\beta\) metrics & 
\(F = \phi(\alpha, \beta),\quad \alpha = \sqrt{a_{IJ} y^I y^J},\; \beta = b_K y^K \) & 
Unifies and generalizes Randers, Kropina, and Matsumoto metrics via the scalar function \(\phi \). \\

\midrule
General Lagrangian & 
\(F(x, y) = \sqrt{L(x, y)} \), \(L \) 2-homogeneous in \(y \) & 
More general case; has some applications in mechanics. \\

\midrule
Optical/media & 
\(F(x, y) = n(x, y) \|y\| \) & 
Includes anisotropic effects in optics through the direction-dependent refractive index $n(x,y)$. \\

\midrule
Non-reversible Finsler & 
General \(F \) such that \(F(x, y) \neq F(x, -y) \) & 
Includes Funk, Randers, and~non-reversible geometries. \\

\bottomrule
\end{tabular}
}
\end{table}

Finsler geometry has found applications in several areas of physics, including the geometric formulation of quantum mechanics~\cite{Tave1,Tave2,Tave3,Tave4}, kinetic theory of gases~\cite{Voicu2, Pfeifer2025}, dispersion relations~\cite{Ratzel:2011} and optics~\cite{Duval:2008}. A~more exotic application is related to an extension of the standard model of particle physics, where a Finsler structure can be given for the nonminimal fermion sector~\cite{Schreck:2016}. It is also interesting to note that Lorentz-violating scalar fields can also be described by Riemann--Finsler geometry~\cite{Edwards:2018}. In~the latter cases, the~Finsler geometry considered is of Berwald type: the closest one to the Riemannan geometry, which is still not trivial and compatible with the weakened Ehlers--Pirani--Schild~axioms.

The first attempts at formulating a Finslerian theory of gravitation, but~still in the framework of unified field theories,  belonged to Horv\'{a}th~\cite{Hor1b}, and~Horv\'{a}th and Mo\'{o}r~\cite{Hor2b}.   Early Finslerian-type gravitational theories were also formulated in~\cite{Hor3b,Hor4b}, where a set of Finslerian-type gravitational field equations were proposed, representing a straightforward extension of the Einstein field equations, and~given by
\begin{equation}\label{h1}
R_{\mu \nu}-\frac{1}{2}g_{\mu \nu}R+\lambda g_{\mu \nu}=\chi T_{\mu \nu},
K_{\mu \nu}-\frac{1}{2}g_{\mu \nu}K+\lambda g_{\mu \nu}=\chi T_{\mu \nu},
\end{equation}
and
\begin{equation}\label{h3}
S_{\mu \nu}-\frac{1}{2}g_{\mu \nu}S-\lambda^{(i)} g_{\mu \nu}=-\chi ^{(i)}T_{\mu \nu}^{i},
\end{equation}
respectively. In~Equations~(\ref{h1}) and (\ref{h3}), $\chi$ denotes the gravitational constant, $\lambda$ is the cosmological constant, and $\lambda ^{(i)}$ and $\chi ^{(i)}$ represent the internal cosmological and gravitational constants. The~quantities  $R_{\mu \nu}$, $R$, $K_{\mu \nu}$, $K$ {{and $S_{\mu \nu}$ and $S$ denote the hh-, hv- and vv-Ricci curvature tensors, and~the hh-, hv- and vv-scalar curvatures of the Finsler space $(M,F)$. Moreover, }}
 $T_{\mu \nu}$ is the energy--momentum tensor of the baryonic matter, and~$T_{\mu \nu}^{i}$ denotes the internal energy--momentum tensor. In~\cite{As0} another Finslerian-type approach to gravity was introduced and developed. The~main focus of this work was the Finslerian description and interpretation of the particle motion in a gravitational field. For~the Finslerian extensions of general relativity along the line of~\cite{As0}, see~\cite{As1}. Finslerian-type generalized Schwarzschild metrics have been investigated  in~\cite{As2,As3}. {For further studies of black holes and wormholes in Finsler geometry, see~\cite{Zohreh1,Zohreh2,Zohreh3,Zohreh4,Zohreh5,Zohreh6,Zohreh7,Zohreh8,Zohreh9,Zohreh10}.}

Miron~\cite{Miron} has proposed the vector bundle point of view to formulate a system of Einstein-type gravitational field equations. The~starting point of this approach is to interpret the field  $y$ as a fiber at the point $x$ of the base  manifold $M$. The~total space of this vector bundle is obtained  as a unification of the fields $x$ and $y$ \cite{Ikeda}. A~nonlinear connection naturally induces an adapted basis {in the fiber $T_{(x,y)}TM$ of the double tangent bundle $TTM$} and a corresponding dual basis {in the fiber $T_{(x,y)}^{*}TM$ of the cotangent bundle $T^*TM$}. These are given by
\begin{equation}\label{eq3}
X_A=\left(\frac{\delta}{\delta x^{\lambda}}=\frac{\partial}{\partial x^{\lambda}}-N_{\lambda}^i\frac{\partial}{\partial y^i}, \frac{\partial}{\partial y^i}\right),
X^A=\left(dx^{\kappa},\delta y^i=dy^i+N_{\lambda}^idx^{\lambda}\right),
\end{equation}
{respectively. }

In Eq.~(\ref{eq3}), the~indices $A,B$ run over $\left(A,B\right)\in \left\{0,1,2,3,...,7\right\}$, while the Greek indices $\lambda, \kappa$ take  values in $\{ 0,1,2,3 \}$.  The~functions $N_{\lambda}^i$ define the nonlinear connection. By~construction, the~basis (\ref{eq3}) is adapted to a Finsler-type metric on $TM$, given by
\begin{equation}
G=g_{\lambda \kappa}(x,y)dx^{\kappa}dx^{\lambda}+g_{ij}(x,y)\delta y^i\delta y^j,
\end{equation}
where $g_{\lambda k}(x,y)$ and $g_{ij}(x,y)$ are the horizontal and vertical metric components, respectively.

In the total space of the vector bundle, the Einstein field equations are assumed to have their standard general relativistic form, $ \mathcal{R}_{AB}-(1/2)\mathcal{R}g_{AB}=\tau _{AB}$,
 where $\tau _{AB}$ is the matter energy--momentum tensor. The~field equations can be decomposed, and~can be written as~\cite{Miron}
\be
R_{\lambda \nu}-\frac{1}{2}(R+S)g_{\lambda \nu}=\tau _{\lambda \nu},
\overset{1}{P}_{i \lambda}=\tau _{i\lambda},  \overset{2}{P}_{ \lambda i}=-\tau _{\lambda i},
\ee
\be
 S_{i j}-\frac{1}{2}(R+S)g_{i j}=\tau _{ij}.
\ee

An interesting Finslerian-type gravitational theory was proposed in~\cite{Rutz}.  The~basic idea of this approach is the assumption that the Einstein vacuum field equations can be obtained from the condition  $H=H_i^i=0$, where $H_k^i$, the~Finslerian deviation tensor, is obtained from the first and second derivatives of the quantity $G^l=\gamma _{jk}^l\dot{x}^j\dot{x}^k/2$ as
\be
H^i_{\;k}=2\frac{\partial G^i}{\partial x^k}-\frac{\partial ^2 G^i}{\partial x^j \partial \dot{x}^k}\dot{x}^j+2\frac{\partial ^2G^i}{\partial \dot{x}^j\partial \dot{x}^k}G^j-\frac{\partial G^i}{\partial \dot{x}^j}\frac{\partial G^j}{\partial \dot{x}^k}.
\ee

If the metric is Riemannian, we obtain the limiting case of the general relativistic gravitational field equations. But~Finslerian solutions of the gravitational field equations proposed in~\cite{Rutz} can also be~obtained.

An important class of Finsler geometries {is} represented by Berwald--Finsler {spaces}.  A~system of gravitational field equations defined in a Berwald geometry 
was proposed and investigated in~\cite{Lixin}. The~derivation of the field equations in the Berwald--Finsler geometry is essentially based on the Bianchi identities satisfied by the Chern curvature tensor. In~general, the geometric part of the gravitational field equation is not symmetric, and~this indicates that the principle of the local Lorentz invariance is not satisfied in this theory. The~field equations as proposed in~\cite{Lixin} are given by
\begin{eqnarray}
\left[Ric_{\mu\nu}-\frac{1}{2}g_{\mu\nu}S\right]+\left\{\frac{1}{2}
B^{~\alpha}_{\alpha~\mu\nu}+B^{~\alpha}_{\mu~\nu\alpha}\right\}=8\pi
G T_{\mu\nu},
\end{eqnarray}
where 
\be
A_{\lambda\mu\nu}\equiv\frac{F}{4}\frac{\partial}{\partial y^\lambda}\frac{\partial}{\partial y^\mu}\frac{\partial}{\partial y^\nu}(F^2),
\ee
is the Cartan tensor, and~$B_{\mu\nu\alpha\beta}=-A_{\mu\nu\lambda}R^{~\lambda}_{\theta~\alpha\beta}y^\theta/F$. The~Cartan tensor measures the deviation of the Finsler geometry from the Riemannian one on a given~manifold.

With the use of a variational principle, Finsler-type gravitational field equations have been derived from a Finsler--Lagrange function $L$ in~\cite{Voicu1}. The~action considered in~\cite{Voicu1} is given by 
\be\label{act0} 
S[L] = \int_{\Sigma\subset TM} \mathrm{vol}(\Sigma) R_{|\Sigma}, 
\ee
where $\Sigma = \{(x,\dot x)\in TM|F(x,\dot x) = 1\}$ denotes the unit tangent bundle, and~$\mathrm{vol}(\Sigma)$, the~volume form on $\Sigma$, is defined by using the Finsler metric. The~field equations obtained from the action (\ref{act0}) are given by
\be
2R - \frac{L}{3}g^{Lij}R_{\cdot i \cdot j} + \frac{2L}{3}g^{Lij}\left[ (\nabla P_{i})_{\cdot j} + P_{i|j} - P_{i}P_{j}\right]= 0,
\ee
where $P$ is the Landsberg tensor, and~
\be
g^L_{ij}=\frac{1}{2}\frac{\partial ^{2}L}{\partial \dot{x}^{i}\partial \dot{x}^{j}}=\frac{1}{2} L_{\cdot i\cdot j}.
\ee
Moreover, $R_{.i.j}$ is the geodesic deviation operator, and~$R$ denotes its~trace.

When extended to gravitational systems much bigger than the Solar System, which involve the presence of galactic and cosmological scales, Einstein's theory of general relativity faces a number of very serious challenges, whose possible solutions could be obtained only if we introduce a fundamental modification in our understanding of the gravitational~force. 

One of the most intriguing discoveries of the recent times is the observational proof that our Universe is in fact in a phase of accelerating expansion~\cite{acc1,acc2,acc3,acc4,acc5}. The~transition from deceleration to acceleration occurred at a small redshift $z$, given by $z\approx 0.5$. Explaining this observation requires a profound modification of the theoretical foundations  of Einstein's general~relativity.

The simplest explanation of the recent exponential de Sitter-type expansion  can be obtained by reintroducing  the cosmological constant $\Lambda$ in the Einstein field equations, introduced by Einstein in 1917~\cite{Ein}, and~later rejected by him as the biggest blunder of his life. Einstein's  main goal was to use $\Lambda$ to construct a static, general relativistic cosmological model of the~Universe. 

The cosmological constant has a complicated history~\cite{Wein}, but~presently, it is adopted as one of the basic physical (geometrical?) parameters to build up the standard  cosmological paradigm, the~$\Lambda$CDM model, which is currently used for the interpretation of the observational data. The~$\Lambda$CDM model also includes in its theoretical structure another basic, equally mysterious component, dark matter~\cite{dark}. Intensive searches for the dark matter particle have yielded no results, and~thus, the only evidence for the  existence of dark matter is gravitational. However, despite these theoretical shortcomings, the~$\Lambda$CDM model fits the observational data very well~\cite{fit1,fit2,fit3,fit4}.

But on the theoretical level, the $\Lambda$CDM model is confronted with the uncertainty of its foundations: no (convincing) physical theory exists that could provide it with a solid basis. The~first major problem is related to the geometrical or physical interpretation of the cosmological constant~\cite{Wein,Wein1a,Wein2a}. If~we  interpret the cosmological constant as the vacuum energy density $\rho_{vac}$ at the Planck scale, we are led to the  “worst prediction in physics” \cite{Lake}. The~vacuum energy density can be computed as
\be
\rho _{vac}\approx \frac{\hbar}{c}\int _{k_{dS}}^{k_{Pl}}{\sqrt{k^2+\left(\frac{mc}{\hbar}\right)^2}d^3k}
\approx  \rho _{Pl}
=\frac{c^5}{\hbar G^2}=10^{93}\; {\rm \frac{g}{cm^3}},
\ee
and this result differs by a factor of  $10^{-120}$  from the observed value of the energy density associated with $\Lambda$, $\rho _{\Lambda}=\Lambda c^2/8\pi G\approx 10^{-30}\;{\rm g/cm^3}$ \cite{fit3}.

Presently, the~$\Lambda $CDM standard paradigm is faced with several important problems.  An~important challenge to the $\Lambda$CDM model is the ``Hubble tension'', which has its origins in the differences in the estimations of the values of the Hubble constant $H_0$ (representing the present-day value of the Hubble function $H$), obtained on one hand from the CMB measurements~\cite{fit4}, and~on the other hand from the local observations of the Type Ia Supernovae~\cite{M1,M2,M3}. The~value of $H_0$ obtained by the SHOES collaboration  for $H_0$ is
$H_0 = 74.03 \pm 1.42$ km/s/Mpc~\cite{M1}. The~early Universe determinations, using the Planck satellite data, give for  $H_0$ the value  $H_0 = 67.4 \pm 0.5$ km/s/Mpc~\cite{fit4}, which differs by $\sim  5\sigma$ from the SH0ES~estimations.

Several other theoretical problems whose solution cannot be found in the framework of the  $\Lambda$CDM paradigm  still exist. Some of these problems are represented by the question of the smallness of $\Lambda$ and its fine tuning.  The~question of why the transition from the decelerating to the accelerating phases took place recently is still waiting for a response. And~perhaps the most important question, if~the cosmological constant is really required to construct successful cosmological models, does not have an answer~yet.
 
Hence, considering alternative approaches for the description of the gravitational interaction may allow us to solve the
observational problems of cosmology, and reconsider the cosmological
constant from a new perspective. There are three major possibilities  that have been proposed for the extension of general relativity: the~dark component approach, the~dark gravity approach, and~the dark coupling approach~\cite{Harko}. In~the dark gravity approach, it is assumed that the geometry of the Universe, as~well as the description of the gravitational interaction, requires a significant departure from the formalism of the Riemann geometry. Geometries that go beyond the Riemannian one, for~example, Weyl geometry, geometries with torsion, or~teleparallel geometries, have been investigated in~\cite{W1,W2,W3,W4,W5,W6,W7,W8,W9,W10,weylcartan}. For~reviews of dark gravity-type theories, and~their applications, see~\cite{R1,R2,R3,R4, R5}.

As a dark gravity candidate, Finsler geometry has an important scientific potential yet to be explored, despite the fact that it has already been considered in various contexts as an important alternative to Riemann geometry, and~the standard $\Lambda$CDM~paradigm. 

 Randers geometry was extensively applied in the study of the gravitational phenomena in~\cite{R1a, R2a,R3a,R4a,R5a,R6a,R7a,R8a, R9a, R10a, R11a,R12a,R13a,R14a,R15a,R16a,R17a,R18a,R19a,R20a,R21a}. 
In~\cite{R5a}, generalized Friedmann equations in a Randers--Finsler geometry of the form
\be\label{1a}
\dot{H}+H^2+\frac{3}{4}HZ_t=-\frac{4\pi G}{3}\left(\rho +3p\right),
\ee
\be\label{1b}
\dot{H}+3H^2+\frac{11}{4}HZ_t=4\pi G\left(\rho -p\right),
\ee
were obtained, where $H$, $\rho$, and $p$ are the Hubble functions, 
and~the baryonic matter energy density and pressure, respectively. The~quantity  $Z_t=\dot{u}_0$, with~$u_0$ denoting the time component of the four-velocity $u_{\mu}$. The~generalized  Friedmann equations above give the relation $3H^2+3HZ_t=8\pi G \rho$. The~term $HZ_t$, induced by the Finsler--Randers geometry, leads to the existence of new phases in the cosmic evolution of the Universe. The~system of Friedmann Equations~(\ref{1a}) and (\ref{1b}) was used in~\cite{R6a} for the investigation of particle creation processes in Finsler-type~geometries.

A scalar--tensor theory that arises effectively from the Lorentz fiber bundle of a Finsler-type geometry was proposed in~\cite{Ikeda2}, where its cosmological implications were also investigated.  The~action in the presence of matter considered in this work is 
\begin{equation}
\mathcal S = \frac{1}{16\pi G} \int \sqrt{|\det\mathbf G|}\,\mathcal L_G dx^{(N)}+ \int
\sqrt{|\det\mathbf G|}\,\mathcal L_M dx^{(N)},
\end{equation}
where $dx^{(N)} = d^4x\wedge \phi^{(1)}\wedge \phi^{(2)}$. Several Lagrangian densities were adopted,   given by
\be
\mathcal L_G = \mathcal{\tilde R} - \frac{1}{\phi}V(\phi),
\ee
where $V(\phi)$ is the potential for the scalar $\phi$, and~\be
\mathcal{\tilde R} = R - \frac{2}{\phi}\square \phi +
\frac{1}{2\phi^2}\partial_\mu\phi\partial^\mu\phi, 
\ee
where $\square$ is the d'Alembert operator, and~ $\mathcal{\tilde R}$ denotes the curvature for the particular situation of a holonomic basis $ [X_M,X_N] = 0 $. In~the case of a non-holonomic basis, the adopted Lagrangian density is given by
$\overline{\mathcal L}_G = \tilde R$. The~generalized Friedmann equations are given by
\begin{align}
    & 3H^2 = 8\pi G\rho_m - \frac{\dot\phi^2}{4\phi^2} +\frac{1}{\phi}\left(\frac{V(\phi)}{2}
-
3H \dot\phi\right)  \label{hol fried 1}, \\
    & \dot H = -4\pi G(\rho_m+P_m)  + \frac{\dot\phi^2}{4\phi^2} + \frac{1}{2\phi}\left(H\dot\phi- \ddot\phi\right)\label{hol fried 2},
\end{align}
\begin{equation}
     \ddot\phi + 3H\dot\phi = - 16\pi G \phi \mathcal \rho_m - 6\phi\left(\dot H +2H^2\right)
+ \phi V'(\phi) + \frac{\dot\phi^2}{2\phi},
\end{equation}
and
\begin{eqnarray}
  &&
  \!\!\!\!  \!\!\!\!\!\!\!\!
  3H^2 = 8\pi G\rho_m - (1+A)\frac{\dot\phi^2}{4\phi^2} - 3H
\frac{\dot\phi}{\phi}
\label{nonhol fried 1},\\
&&\!\!\!\!  \!\!\!\!\!\!\!\!
    \dot H = \! -4\pi G(\rho_m\!+\!P_m) + (1\!+\!A)\frac{\dot\phi^2}{4\phi^2} +
\frac{1}{2\phi}
\left(H\dot\phi\! - \!\ddot\phi\right)\!,
\\
  && \!\!\!\! \!\! \!\! \!\!\!\!(1+A)\left(\ddot\phi + 3H\dot\phi \right) =  - 16\pi G
\phi \mathcal \rho_m -
6\phi\left(\dot H +2H^2 \right) + \frac{\dot\phi^2}{2\phi}\left(1+A+\phi A' \right) - \dot\phi \dot A
     \label{nonhol fried 3},
\end{eqnarray}
respectively.  Here, $A(\phi)$ denotes a real function of $\phi$.  From~the above equations, one can reconstruct the
thermal history of the Universe, and~obtain a sequence of matter and dark-energy-dominated phases. The~effective dark energy equation of state has a parameter that can be either phantom or quintessence type. A~phantom-divide crossing during the cosmological evolution also~appears.

Berwald--Finsler geometries have been investigated as potential candidates for the description of the gravitational fields~\cite{B1,B2}. Spatially
homogeneous and isotropic Berwald spacetimes,  obtained from a Finsler Lagrangian constructed  from a
zero-homogeneous function defined on the tangent bundle, and~which include the velocity dependence of the Finsler
Lagrangian, were discussed  in~\cite{B2}. Cosmological Berwald geometries can also be used for the description of the dynamics of the~Universe.

In a series of recent papers~\cite{Hama, Hama2,Hama3,Hama4}, a systematic investigation of the applications of Finsler geometry in cosmology was considered. The~basic geometric ingredients used for the construction of gravitational theories were the $(\alpha,\beta)$ metrics, and~the osculating Barthel connection. One of the basic properties of the Finsler geometry is that an arbitrary point vector field $y$ is associated with each point of the spacetime manifold, with~the metric $g$ becoming a function of both $x$, the~coordinates defined on the base manifold, and~$y$, $g=g(x,y)$. In~the osculating approach, one assumes the existence of a vector field $Y=Y(x)$, which allows the construction of a Riemannian metric $g(x)=g(x,Y(x))$. In~the case of the $(\alpha,\beta)$ metrics, the~connection associated with these metrics, the~Barthel connection, is nothing but the Levi--Civita connection associated with $g(x)$.

The introduction of the osculating Barthel connection in the $(\alpha,\beta)$ spaces of the Finsler geometry  leads to a significant simplification of the mathematical formalism, and~opens the possibility of constructing unique and well-defined physical models that could be successfully used for the description of the gravitational interaction. In~the present review, we briefly introduce, from~the perspective of the cosmological applications, the~Barthel--Randers, Barthel--Kropina, and~conformal Barthel--Kropina models, and~we also present their mathematical and theoretical~foundations.

From the generalized Friedmann equations of the osculating Barthel–Randers--Kropina cosmological models, obtained by assuming that the background Riemannian metric is of the Friedmann–Lemaitre–Robertson–Walker (FLRW) type, an~effective geometric dark energy component can be always generated, which results from the presence of extra terms in the cosmological evolution~equations. 
 
 A central problem for the physical acceptability of the Finslerian-type models is how successfully they could describe the observational data. The~cosmological tests, and~comparisons with observational data of these dark energy models, considered in the above-mentioned works  are investigated in detail. In~the current review, we perform a detailed analysis of the three cosmological models of Barthel--Randers type, introduced in~\cite{Hama},  constraining the model parameters. In~our investigation, we use 15 Hubble data points (Cosmic Chronometers), the~Pantheon Supernovae Type Ia data sample, and~the most recent
Baryon Acoustic Oscillation (BAO) measurements from the Dark Energy Spectroscopic Instrument (DESI) Data Release 2. The~statistical analysis is performed by using Markov Chain Monte Carlo simulations. The~results of the statistical analysis of the Barthel--Randers cosmological models are compared with a similar analysis of the Barthel--Kropina cosmological models performed in~\cite{Hama3}, and~with the standard $\Lambda$CDM model.  The~Akaike information criterion (AIC) and~the Bayesian information criterion (BIC) are used as the model selection tools. The~statefinder diagnostics, consisting of the study of the jerk and snap parameters, and~the $Om(z)$ diagnostics are also considered for the comparative study of the Barthel--Randers, Barthel–Kropina and $\Lambda$CDM cosmologies. Our statistical results indicate that the osculating Barthel-type $(\alpha,\beta)$ Finslerian  dark energy models give a good description of the observational data, and~thus, they can be considered a viable alternative to the $\Lambda$CDM model, even if not all of them are favorable compared to $\Lambda$CDM.

The present paper is organized as follows. We review some basic concepts and definitions of Finsler geometry in Section~\ref{sect1}. In~Section~\ref{sect2}, we introduce the mathematical foundations of the Barthel--Randers--Kropina gravitational theories, by~reviewing the basic concepts of $(\alpha,\beta)$ metrics, osculating spaces, and~Barthel connections. The~basic principles of constructing $(\alpha,\beta)$ cosmological models are also presented. The~generalized Friedmann equations of the three basic models considered in the present work, Barthel--Randers, Barthel--Kropina, and conformal Barthel--Kropina, are also written, together with the corresponding energy balance equations. The~thermodynamic interpretation of the cosmological models with a non-vanishing matter energy--momentum tensor is also presented and~briefly discussed.  The~cosmological implications of the Barthel--Randers and Barthel--Kropina models are discussed in Section~\ref{sect3}, where constraints on the free parameters of these models are obtained by using a combination of observational datasets, including Type Ia Supernovae, Baryon Acoustic Oscillations, and~Hubble parameter measurements. A~detailed comparison of the Barthel--Randers and Barthel--Kropina cosmological models is performed in Section~\ref{sect4}.  A~discussion of the main results and the relevance of the statistical analysis is presented in Section~\ref{sect7}. Finally, we discuss and conclude our results in Section~\ref{sect8}.

\section{Fundamentals of Finsler~Geometry}\label{sect1}

The assumption that 
spacetime can be described mathematically as a four-dimensional
differentiable manifold $M$, endowed with a pseudo-Riemannian tensor $g_{I
J} $, where $I,J,...=0,1,2,3$, is one of the fundamental assumptions of modern theoretical physics. The~next fundamental concept, the~interval between two
events located at  the points $x^{I}$ and $x^{I} + dx^{I}$ on the worldline of a standard clock, is  
defined by the chronological hypothesis as $ds=\left(g_{I
J}dx^{I}dx^{J}\right)^{1/2}$  \cite{Tava,Tava1}. A~very important metrical generalization of the Riemannian
geometry  is the geometry anticipated by Riemann~\cite{Riemann}, but~which was later on systematically developed 
by Finsler~\cite{Finsler}.

In a simple interpretation, Finsler spaces are metric
spaces with the interval $ds$ between two neighboring points $%
x=(x^{I})$ and $x+dx=(x^{I} + dx^{I})$ given by
\begin{equation}  \label{dsF}
ds=F\left(x,dx\right),
\end{equation}
where $F$, the~Finsler metric function, must be positively homogeneous of degree
one in $dx$, and~thus satisfy the condition
\begin{equation}
F\left(x,\lambda dx\right)=\lambda F\left(x,dx\right) \; \; \text{for} \; \; \lambda>0.
\end{equation}

The Finsler metric function $F$ can be written in terms of the
canonical coordinates of the tangent bundle $(x,y)=(x^I,y^I)$, where $y=y^I\left(\partial/
\partial x^I\right)$ is any  tangent vector $y$ at $x$. Then we can introduce the Finsler metric tensor $%
g_{I J}$ defined according to
\begin{equation}  \label{Hessian mat}
g_{I J}\left(x,y\right)=\frac{1}{2}\frac{\partial ^2F^2\left(x,y\right)}{%
\partial y^{I}\partial y^{J}}.
\end{equation}
Hence Equation~(\ref{dsF}) can be written as 
\be
ds^2=g_{I J}\left(x,y\right)y^{I}y^{J}.
\ee
Riemann spaces are some particular cases of the
Finsler spaces, and~they correspond to $g_{IJ}\left(x,y\right)=g_{IJ
}\left(x\right)$ and $y^{I}=dx^{I}$, respectively.

Given a Finsler function $F$, one can obtain the geodesic equations in the form~\cite{Tava,Tava1}
\begin{equation}
\frac{d^2x^{I}}{d\lambda ^2}+2G^{I}\left(x,y \right)=0,
\end{equation}
where the functions $G^{I}(x,y)$ denote the spray coefficients
\begin{equation}
G^{I}\left(x,y \right)=\frac{1}{4%
}g^{IJ}\left(\frac{\partial ^2F^2}{\partial x^K \partial y^J}y^K-\frac{%
\partial F^2}{\partial x^J}\right).
\end{equation}
Equivalently, more reminiscent of Riemannian geometry, they can be rewritten as
\begin{equation}\label{Finsler geod}
\frac{d^2x^{I}}{d\lambda ^2}+\Gamma^{I}_{J K}\left(x,y \right)\frac{%
dx^{J}}{d\lambda}\frac{dx^{K}}{d\lambda}=0,
\end{equation}
where
\begin{equation}
    \Gamma^{I}_{JK}(x,y) y^{J} y^{K}=2 G^{I}(x,y)=\frac{1}{2%
}g^{IJ}\left(\frac{\partial ^2F^2}{\partial x^K \partial y^J}y^K-\frac{%
\partial F^2}{\partial x^J}\right).
\end{equation}

{Here, the~quantities $\Gamma^{I}_{JK}(x,y)$ are the local coefficients of the Finsler connection used. We point out that the theory of Finsler connections is more complex than in the Riemannian case since the only metrical, torsion-free connection is the Levi--Civita one; hence, the existence of such a connection implies that the Finsler metric is in fact Riemannian. The~obvious conditions imposed for the existence of Finsler connections are then non-metricity and torsion-free (this is the case of the Chern connection) or metrical connection with surviving torsion (this is the case of the Cartan connection). The~precise form of the Finsler connections depends on the concrete form of the non-metricity and surviving torsion, but~these details are beyond the aim of the present paper (one can consult~\cite{F5b,F6b,MHSS} or other textbooks on Finsler geometry). The~local coefficients of a Finsler connection $D$ are defined as
\begin{equation}
		D_{\frac{\delta}{\delta x^K}}\frac{\delta}{\delta x^J}=\Gamma^I(x,y)_{JK}\frac{\delta}{\delta x^I},\quad D_{\frac{\partial}{\partial x^K}}\frac{\partial}{\partial x^J}=V^I(x,y)_{JK}\frac{\partial}{\partial x^I},
		\end{equation}
		where $\frac{\delta}{\delta x^I}$ is the adapted basis  induced by the Cartan nonlinear connection. We prefer to keep the notation $C^I_{JK}=\frac{1}{2}\frac{\partial^3 F^2}{\partial y^I\partial y^J\partial y^K}$ for the Cartan tensor of $(M,F)$. 	
	For instance, the~Cartan connection of a Finsler space $(M,F)$ is the only Finsler connection on $TTM$ that is metrical with $h(hh)$ and $v(vv)$-torsion vanishing (the rest of the torsion survives). The~local coefficients of the Cartan connection are given by
\begin{equation}\label{Cartan conn coeff}
		\begin{split}
		\Gamma ^{I}_{J K}\left(x, y \right)&=\frac{1}{2}g^{I
			L}\left(x,y \right)\Bigg[\frac{\delta g_{L J}\left(x, y \right)}{%
			\delta x^{K}}  
		+\frac{\delta g_{L K}\left(x,y\right)}{\delta
			x^{J}}-\frac{\delta g_{J K}\left(x,y\right)}{\delta x^{L}}\Bigg]{,}\\
			V ^{I}_{J K}\left(x, y \right)&=\frac{1}{2}g^{I
				L}\left(x,y \right)\Bigg[\frac{\partial g_{L J}\left(x, y \right)}{%
				\partial y^{K}}  
			+\frac{\partial g_{L K}\left(x,y\right)}{\partial y^{J}}-\frac{\partial g_{J K}\left(x,y\right)}{\partial x^{L}}\Bigg]{.}
			\end{split}
		\end{equation}	 
		{Likewise, 
 the~Chern connection is defined to be the only Finsler connection on $TTM$ which is almost compatible and torsion-free.  		The~local coefficients of the Chern connection are given by $(\Gamma^{I}_{JK}(x,y),0)$. 
	 However, despite the rather complicated form of the connection coefficients $\Gamma^{I}_{JK}(x,y)$, due to the homogeneity of the geometrical objects involved, \linebreak  Equation~\eqref{Finsler geod} takes the simple form
\begin{equation}
\frac{d^2x^{I}}{d\lambda ^2}+\gamma^{I}_{J K}\left(x,y \right)\frac{%
dx^{J}}{d\lambda}\frac{dx^{K}}{d\lambda}=0,
\end{equation}
where}
\begin{eqnarray}\label{35}
\gamma ^{I}_{J K}\left(x, y \right)=\frac{1}{2}g^{I
L}\left(x,y \right)\Bigg[\frac{\partial g_{L J}\left(x, y \right)}{%
\partial x^{K}}  
+\frac{\partial g_{L K}\left(x,y\right)}{\partial
x^{J}}-\frac{\partial g_{J K}\left(x,y\right)}{\partial x^{L}}\Bigg],
\end{eqnarray}
are the formal Christoffel coefficients of the Finsler space $(M,F)$.

 Berwald spaces are a special class of Finsler geometry, and~they can be obtained
by assuming that the Berwald connection coefficients are independent of the fiber coordinate $y$
\cite{F2b}.

With the help of the spray coefficients $G^{I}$, one can define the vector field $S$ on $%
TM\setminus{0}$ according to
\begin{equation*}
S=y^I\frac{\partial}{\partial x^I}-2G^I\frac{\partial}{\partial y^I}.
\end{equation*}

$S$ is the spray induced by $F$. A~curve $\gamma$ defined on $M$ is a geodesic of $F$ if and only if its canonical lift $\hat{%
\gamma}(t)=(\gamma(t),\dot{\gamma}(t))$ to $TM$ is an integral curve of $S$.

The notion of Finsler metrics that we have recalled here belong to the class
of \textit{classic Finsler metrics}; in~other words, at~each point $x\in M$,
the function $F_x:T_xM\to \mathbb{R} $ is a function defined on the tangent
space $T_xM$ of a differentiable manifold $M$ satisfying the following~conditions:

\begin{enumerate}
\item[(i)] $F_x$ is $C^\infty$ on $\widetilde{T_xM}=T_xM\setminus\{0\}$.

\item[(ii)] $F_x$ is positively 1-homogeneous: $F_x(\lambda y)=\lambda F_x(y)$, for~all $\lambda>0$ and $y\in T_xM$.

\item[(iii)] For each $x\in M$, the~Hessian matrix \eqref{Hessian mat} is
positive definite on $\widetilde{T_xM}$.
\end{enumerate}

At each point $x\in M$, the~indicatrix $\{y\in T_xM:F_x(y)=1\}$ is a closed,
strictly convex, smooth hypersurface around the origin of $T_xM$.

A more general notion is the notion of conic Finsler metrics, that
is, Finsler norms defined only on a conic domain of $T_xM$. Let us recall
that $A_x\subset T_xM$ is called a conic domain of $T_xM$ if $A_x$
is an open, non-empty subset of $T_xM$ such that if $v\in A_x$, then $%
\lambda v\in A_x$, for~all $\lambda>0$. We remark that the origin of $T_xM$
does not belong to $A_x$ except for the case $A_x=T_xM$. 

We can now define a
Finsler norm defined only on a conic domain $A_x\subset T_xM$ with the
properties (i)--(iii) given above for all $y\in A_x$. At~each point $x\in M$,
the indicatrix $S_x:=\{y\in A_x\subset T_xM:F_x(y)=1\}$ is a hypersurface
embedded in $A_x$ as a closed~subset.

Let $A\subset TM$ be an open subset of the tangent bundle $\pi:TM\to M$ such
that $\pi(A)=M$, and~$A$ is conic in $TM$; that is, for each $x\in M$, the~set $A_x:=A\cap T_xM$ is a conic domain in $T_xM$. A~function $F:A\to
\mathbb{R} $ is a conic Finsler metric if its restriction $F_x:A_x\to
\mathbb{R} $ satisfies the conditions (i)--(iii) above, for~each $x\in M$.
The local and global geometry of conic Finsler spaces can be now developed
in a similar way with the case of classical Finsler metrics (see~\cite{JS,YS1}
and references therein).

\section{Osculating \boldmath{\texorpdfstring{$(\alpha, \beta)$}{ab}}-Type Cosmological Models}\label{sect2}

In the present section, we review the mathematical and theoretical foundations of the Barthel--Randers--Kropina-type cosmological models, which we will discuss in the framework of the general $(\alpha, \beta)$ geometries.  We also write down the generalized Friedmann equations obtained for the case of the Friedmann--Lemaitre--Robertson--Walker Riemannian metric. The~thermodynamic interpretation of the models is also~discussed.

\subsection{Mathematical Foundations of the \texorpdfstring{$(\alpha, \beta)$}{ab} Finslerian Cosmologies}

We begin our discussion of the cosmology of the Finslerian-type geometries with a brief discussion of the mathematical properties of the $(\alpha, \beta)$ metrics,~the Barthel connection, and~the osculating~geometries.   

\subsubsection{Kropina and \texorpdfstring{$(\alpha, \beta)$}{ab} Geometries} 

As we have already mentioned, a~special type of Finsler space is the Randers space~\cite{Rand}, with~the metric function given by
\begin{equation}
F=\left[ a_{I J }(x)y^{I }y^{J }\right] ^{1/2}+b_{I }(x)y^{I },
\end{equation}%
where $a_{I J }$ is the metric tensor of a Riemannian space,
and $b_{I}(x)y^{I }$ is a linear $1$-form defined on the tangent bundle $TM$, and we use the parametrization $dx^I=y^I$. 

Another remarkable class of Finsler spaces are the  Kropina spaces~\cite{Krop1,Krop2}, which are Finsler spaces with
metrics of the form
\begin{equation}
F\left(x,y\right) =\frac{a_{IJ }(x)y^{I }y^{J }}{b_{I }(x)y^{I }}.
\end{equation}

By generalizing these results, Matsumoto~\cite{Mat1,Mat2} defined the notion
of the $(\alpha,\beta)$ metrics in the following way: given a Finsler metric function $%
F(x,y)$, it  is called an $(\alpha,\beta)$ metric if $F$ is a positively
homogeneous function $F(\alpha,\beta)$ of first degree in two variables $%
\alpha \left(x,y\right) =\left[ a_{I J }(x)y^{I }y^{J }\right] ^{1/2}$
and $\beta \left(x,y\right) =b_{I }(x)y^{I }$, respectively. 

In our analysis, we suppose that $\alpha $ is a {pseudo-Riemannian} or Riemannian metric, with~the properties that it
is non-degenerate (regular), or~positive definite.  {We point out that the Kropina-type metrics are conic Finsler metrics defined only on some conic domain of the tangent bundle. By~extension, we can regard any $(\alpha,\beta)$ as conic Finsler metrics, allowing the Riemannian part to be a pseudo-Riemannian metric.}

In the special cases of the
Randers and Kropina metrics, the Finsler metric functions are given by $F=\alpha +\beta $, and~$F=\dfrac{\alpha^2}{\beta}$, respectively (see also Table~\ref{tab:finsler-metrics}). Therefore the Randers and Kropina metrics
belong to the general class of the $(\alpha,\beta)$ metrics. 

We can define the general $(\alpha,\beta)$ metrics as metrics having the metric function given by  
\be
F(\alpha,\beta)=\alpha \phi \left(\beta /\alpha \right) =\alpha \phi \left(s\right), 
\ee
where $ s=\beta /\alpha $, and~$\phi =\phi (s)$ is a $%
C^{\infty }$ positive function on an open interval $(-b_{o},b_{o})$.

Many examples of $(\alpha,\beta)$ metrics have been considered in the literature, the~most studied one being the Randers metric $F=\alpha+\beta$ (see
~\cite{BCS} and references within). The~Kropina  metrics  are also classic examples of $(\alpha, \beta)$ metrics (see \cite%
{AIM} for physical motivations of introducing this metric). More recently,
the local and global aspects of Kropina metrics have been extensively
studied in~\cite{YS1,YS2,SSY, Heefer1, Heefer2}. 

\subsubsection{The Barthel~Connection}

Let $(M^n,F)$ be a Finsler space, defined on a base manifold $M^n$. We can also define on $M^n$ a vector field $Y(x)\neq 0$.
We introduce now a specific mathematical object $(M^n,F(x,y),Y(x))$, which represents a Finsler space %
$(M^n,F(x,y))$ with a tangent vector field $Y(x)$ also defined.  If~the vector $Y$ does not vanish {at} any point on $M$, then from the Finslerian metric $\widehat{g}(x,y)$, one obtains the $Y$-Riemann metric $\widehat{g}_{Y}(x)=\widehat{g}(x,Y)$.

{{
Obviously, one can evaluate any geometrical object in a Finsler space $(M,F)$ in the specific direction given by a fixed vector field $Y$.  This remark leads to the natural idea of evaluating the Chern connection coefficients $\Gamma^I_{JK}(x,y)$ given in \eqref{Cartan conn coeff} at $y=Y(x)$. This is the so-called Barthel connection introduced by Barthel himself in~\cite{Ba1,Ba2} and developed later by Ingarden and Matsumoto~\cite{AIM}.

The Barthel connection is an affine connection on $\left(M, \widehat{g}_{Y}\right)$, metrical with torsion
\begin{equation}
	T^I_{JK}(x)=V^I_{LJ}(x,Y(x))Y^L_K-V^I_{LK}(x,Y(x))Y^L_J,
	\end{equation}
	where $Y^L_J(x):=\frac{\partial Y^L}{\partial x^J}-N^L_J(x,Y(x))$ and 
	$N^L_J(x,y)$ is the Cartan nonlinear connection of the Finsler space $(M,F)$. 
	
	The local coefficients of the Barthel connection can be written as
\begin{equation}
	b^I_{JK}(x)=F^I_{JK}(x,Y(x))+V^I_{JL}(x,Y(x))Y^L_J(x).
	\end{equation}
	
	 The absolute differential of the vector $Y$ is defined according to~\cite{In1}
\be\label{DY}
DY^I=dY^I+Y^K b^I_{KH}(x, Y) dx^H,
\ee
where by $b^I_{KH}(x, Y)$ we have denoted  the coefficients of the Barthel connection. By~using the homogeneity of the geometrical objects involved, we obtain 
\be\label{DY2}
DY^I=dY^I+Y^K\left(\widehat{\gamma}^I_{KH}-\widehat{\gamma}^R_{KS}Y^S\widehat{C}^I_{RH}\right)dx^H{.}
\ee

The Barthel connection has a number of interesting properties. It depends on the vector field on which it acts, a~property that does not exist in Riemann geometry. Therefore, the~Barthel connection has very different properties, and~so it significantly differs from the connections in Riemann geometry. Another important property is related to the direction and magnitude dependence of the Barthel connection.  For~anisotropic metrics, like most of the Finsler metrics, all geometric properties depend on the direction. For~the Barthel connection, there is no dependence on the  magnitude of the vector field, but~only on its direction. The~Barthel connection, keeping the metric function unchanged by the parallel transport, is the simplest connection with this property.  For~Finsler vector fields, depending on both $x$ and $y$, the~Barthel connection allows the transition to the Cartan  geometry of the Finsler spaces. Moreover,  the~Barthel connection can be considered as the connection of a point in Finsler~space. 

Finally, we would like to mention that unlike the Levi--Civita connection of the Riemannian geometry, the~Finsler connections (Cartan or Chern connections) are not defined on the base manifold $M$, but~they live on the total space of the tangent bundle~\cite{F5b,F6b}. On~the other hand, even though they are completely different, the~Barthel and Levi--Civita connections are affine connections on the base manifold $M$.  This is a very important characteristic of Barthel-type geometries that leads to significant differences between the geometrical theories of gravity as formulated on Riemann and Finsler~manifolds.

}}

\subsubsection{The \texorpdfstring{$Y$}{Y}-Osculating Riemann~Geometry}

In 1936, Nazim~\cite{Naz} introduced and developed the concept of osculating Riemann spaces of Finsler geometries. This concept was later studied by Varga~\cite{Varga}. In~the osculating approach, one associates to a complex geometric object, like a Finsler geometry and connection, a~simpler mathematical structure, for~example, an~affine or a linear connection, or~a Riemann metric. Hence, in this approach, it is assumed that the osculating geometry approximates  the more complicated one at a certain level. Thus, with~the use of the osculating formalism, one can derive mathematical results that significantly simplify the description of the properties of the mathematically complex~geometries.

Let a  {nowhere-vanishing} local section $Y$ of $\pi_M:TM\to M$ be given. Geometric objects existing on $TM$ can be pulled back to $M$. Since ${g}_{IJ} \circ Y$ is a function defined on $U$,  a~new metric, defined according to
\begin{equation}\label{Y-riem}
 \widehat{g}_{IJ}(x):={g}_{IJ}(x,y)|_{y=Y(x)},\quad x\in U,
\end{equation}
can be introduced. The~pair $\left(U,\widehat{g}_{IJ}\right)$ represents a Riemannian manifold, while  $\widehat{g}_{IJ}(x)$ is the $Y$- osculating Riemannian metric, defined on $(M,F)$.

For the osculating Riemannian metric given by equation~\eqref{Y-riem}, we define the Christoffel symbols of the first kind as
\begin{eqnarray}
  \widehat{\gamma}_{IJK}(x):=\frac{1}{2}\left\{\frac{\partial}{\partial x^J}\left[{g}_{IK}(x,Y(x))\right]
+\frac{\partial}{\partial x^K}\left[{g}_{IJ}(x,Y(x))\right] {- \frac{\partial}{\partial x^I} [{g}_{JK}(x,Y(X))]}\right\}.
\end{eqnarray}
With
 the use of the rule of the derivative of the composed functions, we explicitly find
\begin{equation}\label{Christ for g_Y}
\widehat{\gamma}_{IJK}(x)=\left.{\gamma}_{IJK}(x,y)\right|_{y=Y(x)}
 +2 \left.\left({C}_{IJL}\frac{\partial Y^L}{\partial x^K}+{C}_{IKL}\frac{\partial Y^L}{\partial x^J}-{C}_{JKL}\frac{\partial Y^L}{\partial x^I} \right)\right|_{y=Y(x)},
\end{equation}
where ${C}_{IJL}$ denotes the Cartan tensor. Hence, if~a global section $Y$ of $TM$ exists, so that $Y(x)\neq 0$,  $\forall x\in M$, one can always define the osculating Riemannian manifold $(M,\widehat{g}_{ij})$.

\paragraph{{\it The case of the} 
 $(\alpha, \beta)$ {\it metrics.}} Let us focus now on the case of the $(\alpha,\beta)$ metrics. The~Hessian matrix of an $(\alpha,\beta)$-metric $F=F(\alpha,\beta)$ is given by
\begin{equation}
{g}_{IJ}(x,y)=\frac{L_\alpha}{\alpha}h_{IJ}+\frac{\partial L_{\alpha\alpha}}{\alpha^2}y_Iy_J+\frac{\partial L_{\alpha\beta}}{\alpha}\left(y_Ib_J+y_Jb_i\right)+L_{\beta\beta}b_Ib_J,
\end{equation} 
and the Cartan tensor by
\begin{equation}
2{C}_{IJK}=\frac{L_{\alpha\beta}}{\alpha}\left(h_{IJ}p_K+h_{JK}p_I+h_{KI}p_J\right)+L_{\beta\beta\beta}p_Ip_Jp_K,
\end{equation}
where the notation $L:=\frac{F^2}{2}$ is customary. Here,
$p_I=b_I-\frac{\beta}{\alpha}y_I$, and $h_{IJ}=g_{IJ}-\frac{y_Iy_J}{F^2}$ is the angular tensor of $(M,F)$. 

We choose the vector field as $Y=b$, with~$b^{I}=a^{IJ}b_{J}$. Since the vector field {$Y$} is globally non-vanishing on $M$, we obtain the result that $\beta $ has no zero points on $M$. Hence, we can define for the $(\alpha, \beta)$ metrics the $b$- osculating Riemannian manifold $(M,\widehat{g}_{IJ})$, with~the Riemannian metric given by $\widehat{g}_{IJ}(x):={g}_{IJ}(x,b)$, where $b^{I}=a^{IJ}b_{J}$.

With respect to $\alpha $, the~length $\widetilde{b}$ of $b$ is obtained as  $\widetilde{b}^{2}=b_{I}b^{I}=\alpha ^{2}\left(x,b\right)$. Moreover, we have  $Y_{I}\left(x,b\right)
=b_{I}$, respectively.

The $b$-osculating Riemannian metric can be explicitly written as
\begin{equation}
\begin{split}
   \widehat{g}_{IJ}\left(x\right)  =\left.\frac{L_{\alpha }}{\widetilde{b}}\right|_{y=b(x)}g_{IJ}
    +\left.\left(\frac{L_{\alpha \alpha }}{\widetilde{b}^{2}}+2\frac{L_{\alpha \beta }}{\widetilde{b}}%
+L_{\beta \beta }-\frac{L_{\alpha }}{\widetilde{b}^{3}}\right)\right|_{y=b(x)} b_{I}b_{J}.
\end{split}
\end{equation} 

 Moreover, 
  we obtain $\beta \left(x,b\right) =\widetilde{b}^{2}$, and~ $p_{I}\left(x,b\right) =0$. Hence, by~using the expression of the Cartan tensor for an $(\alpha,\beta)$ metric,  we obtain the important result that ${C}_{IJK}\left(x,b\right) =0$.
For  $Y=b$, we find
\be
\widehat{\gamma}_{IJK}(x)=\left.{\gamma}_{IJK}(x,y)\right|_{y=b(x)}.
\ee

Thus, we have obtained the fundamental result that in the case of an $(\alpha,\beta)$-metric, the~Barthel connection, representing the linear $b$-connection, where  $b^I=\left(a^{IJ}b_{J}\right)$, is the Levi--Civita connection of the $b$-Riemannian metric. Hence, after~the evaluation of the fundamental Finsler tensor $g_{ij}(x,y)$ of $(M,F)$ at the point $(x,Y(x))$, we obtain a Riemannian metric $\widehat{g}_Y$ on $M$, with~its Levi--Civita~connection.

\paragraph{{\it The curvature tensor. }} The Barthel connection with local coefficients $\left(b_{BC}^A(x)\right)$ is an affine connection. We define the curvature tensor of an affine connection, having local coefficients $\left(\Gamma _{BC}^A(x)\right)$, as~\begin{equation}\label{curv}
R^A_{BCD}=\dfrac{\partial \Gamma^A_{BD}}{\partial x^C}-
    \dfrac{\partial \Gamma^A_{BC}}{\partial x^D}+\Gamma^E_{BD}\Gamma^A_{EC}
    -\Gamma^E_{BC}\Gamma^A_{ED}.
    \end{equation}

Thus, it follows that the curvature of the Barthel connection can be obtained from Equation~(\ref{curv})  by taking $\left(\Gamma _{BC}^A(x)\right)=\left(b_{BC}^A(x)\right)$. For~the Kropina metric with  $F=\alpha^2/\beta$, the~Barthel connection is equal to the Levi--Civita connection of the osculating metric $\widehat{g}_{AB}(x)=g_{AB}\left(x,b(x)\right)$, with~$b_I(x)$ denoting the components of the one-form $\beta$. Moreover, $g_{AB}$ is the fundamental tensor of $F$. 

Since $b_{BC}^A=\widehat{\gamma}_{BC}^A$, where $\widehat{\gamma}_{BC}^A$ are the coefficients of the Levi--Civita connection, for~the curvature tensors of the Kropina metric, we find the expressions
\begin{equation}
\widehat{R}^A_{BCD}=\dfrac{\partial \widehat{\gamma}^A_{BD}}{\partial x^C}-
\dfrac{\partial \widehat{\gamma}^A_{BC}}{\partial x^D}+\widehat{\gamma}^E_{BD}\widehat{\gamma}^A_{EC}
-\widehat{\gamma}^E_{BC}\widehat{\gamma}^A_{ED},
\end{equation}
    and
\begin{equation}
\widehat{R}_{BD}=
    \displaystyle\sum_A\left[\dfrac{\partial \widehat{\gamma}^A_{BD}}{\partial x^A}-\dfrac{\partial \widehat{\gamma}^A_{BA}}{\partial x^D}
    +\sum _E\left(\widehat{\gamma}^E_{BD}\widehat{\gamma}^A_{EA}-\widehat{\gamma}^E_{BA}\widehat{\gamma}^A_{ED}\right)\right],
 \end{equation}
respectively, where the indices $A,B,C,D,E$ take the values $\{0,1,2,3\}$, and~we have defined $\widehat{R}_{BD}=\widehat{R}_{BAD}^{A}$, and~$\widehat{R}_{D}^{B}=\hat{g}^{BC}\widehat{R}_{CD}$, respectively~\cite{Hama, Hama2,Hama3,Hama4}. The~generalized Ricci scalar is defined according to $\widehat{R}=\widehat{R}_{B}^{B}$.

\subsection{Building Cosmological Models in \texorpdfstring{$(\alpha, \beta)$}{ab} Geometries}

We proceed now to the investigation of the possibility of using the Finslerian geometric framework, and~the general $(\alpha, \beta)$ metrics, to~build specific cosmological models. Our basic quantities in this respect are the Riemannian metric $a_{IJ}$, which determines $\alpha=a_{IJ}dx^Idx^J=a_{IJ}y^I y^J$, and~the coefficient $b_I$ of the one-form $\beta =b_I dx^I=b_I y^I$. To~construct a cosmological model in the $(\alpha, \beta)$-type Finsler geometries, we introduce several assumptions, \linebreak  detailed below.
\paragraph{{\it The Universe is homogeneous and isotropic.  
}}
As a first approximation of the structure and matter distribution in the Universe, we assume the validity of the cosmological principle. The~cosmological principle implies the homogeneity of the Universe.  The~homogeneity of the Universe imposes the fundamental constraint that on large cosmological scales, all the physical and geometrical properties of the Universe depend globally on the cosmological time only.
\paragraph{{\it The Riemannian metric 
} $a_{IJ}$ {\it is the FLRW metric}.}
In the following, we will restrict our study to the case in which the metric $a$ in the definition of $\alpha =a_{IJ}dx^Idx^J$ is the flat, homogeneous and isotropic Friedmann--Lemaitre--Robertson--Walker (FLRW)  metric. In~a system of coordinates $\left(x^0=ct,x^1=x,x^2=y,x^3=z\right)$ defined on the base manifold $M$, the~Riemannian FLRW metric is given by
\begin{equation}\label{metr}
ds^2_R=g_{IJ}dx^Idx^J=c^2dt^2-a^2\left(x^0\right)\left(dx^2+dy^2+dz^2\right),
\end{equation}
where $t$ is the universal cosmological time, $c$ is the speed of light, and~$a\left(x^0\right)$ is the scale factor, describing the expansionary properties of the Universe. From an~observational point of view, the Hubble function, defined as $H=\left(1/a\left(x^0\right)\right)\left(da\left(x^0\right)/dx^0\right)$, plays an important role, since it allows the in depth comparison of the astrophysical data with the theoretical~predictions.
\paragraph{{\it The Finsler metric depends on  
} $x^0$ {\it only.}}
The cosmological principle, together with the homogeneity postulates, requires that together with $a_{IJ}=a_{IJ}\left(x^0\right)$, the~components of the 1-form $\beta$ are also functions of the cosmological time only, $ b_I=b_I\left(x^0\right)$.
\paragraph{{\it The 1-form
} $b$ {\it has vanishing space-like components.}}
The cosmological principle, requiring the homogeneity of the Universe, as~well as the isotropy condition that follows from the choice of the Riemann metric as the FLRW metric, with only diagonal components, leads to  a strong mathematical constraint on the components of the coefficients of the 1-form  $b$. The~isotropy of the metric imposes the condition 
 that all the space-like components of $A$ identically vanish, so that $b_1=b_2=b_3=0$. If~this restriction is not satisfied, then after performing a spatial rotation in the $(x,y,z)$ three-dimensional space,  we can construct a preferred direction, oriented, for~example, in~the direction of the $x$ coordinate. 
But the possibility of such a transformation and~the existence of a preferred direction would contradict the isotropic condition implemented via the FLRW metric, as~well as the observed large-scale spatial isotropy of the
Universe. Hence, in~the present approach to Finslerian cosmology, we assume that the vector $b$
has only one independent component, so that  $b_0\left(x^0\right)$. Therefore,  in~a homogeneous and isotropic cosmology, the 1-form field $b$ takes the simple form
\begin{align}\label{special A_i}
(b_{I})=\left(a\left(x^0\right)\eta\left(x^0\right),0,0,0\right)=\left(b^{I}\right),
\end{align}
where we have introduced the auxiliary function $\eta \left(x^0\right)$ to obtain a representation of $b_0$ in terms of the scale factor of the Universe. 
\paragraph{{\it Matter moves along the Hubble flow. 
}}
We assume that similarly to the standard general relativistic cosmology, defined in the Riemannian geometric setting, in~Finslerian cosmology, we can also introduce a comoving frame in which the cosmological observers, as~well as ordinary matter,  move along with the Hubble flow, defined by the metric $a_{IJ}(x)$. If~we introduce the Riemannian four-velocities $u^I$ of the matter particles, defined as $u^I=dx^I/ds_R$, then the existence of a moving frame implies that the space-like components of the matter four-velocity vanish identically, and~the four-velocity has only a non-zero temporal component $u^I=(1,0,0,0)$, which can be normalized to one. 
\paragraph{{\it The matter content of the Universe is a perfect fluid.  
}}
We postulate that cosmological matter in the Universe consists of a perfect fluid, whose thermodynamic properties  can be described by two basic thermodynamic quantities only, given by the energy density $\rho c^2$, and~the thermodynamic pressure $p$, respectively. We also assume that the thermodynamic quantities can be defined in the usual way, by~using the standard definitions of statistical physics and thermodynamics. From~assumptions {\it c} and {\it d}  it follows that the matter energy--momentum tensor has only two non-zero components,  $\widehat{T}_0^0=\rho c^2$, and~$\widehat{T}_I^I=-p$, $I=1,2,3$, and~thus, it can be represented in the form
\begin{equation}
\widehat{T}_{I}^{J}=\begin{pmatrix}
{\rho} c^2 & 0 & 0 & 0\\
0 & -{p} & 0 & 0\\
0 & 0 & -{p} & 0\\
0 & 0 & 0 & -{p}
\end{pmatrix} .
\end{equation}
\paragraph{{\it Geometric quantities. 
}} 
Once the above conditions and assumption have been adopted, we obtain the expressions of the Finsler metric, and~of $\alpha$ and $\beta$ in the FLRW cosmological background of the osculating Barthel--$(\alpha,\beta)$ geometry as~follows: 
\begin{enumerate}
    \item[(iii)] $(a_{IJ})=\begin{pmatrix}
    1 & 0 & 0& 0\\
    0 & -a^2\left(x^0\right) & 0 & 0 \\
    0 & 0 &-a^2\left(x^0\right) & 0\\
    0 & 0 & 0 &-a^2\left(x^0\right)
    \end{pmatrix} ;
    $
    \item[(iv)] $\left.\alpha \right|_{y=b(x)}=a\left(x^0\right)\eta\left(x^0\right)$;
    \item[(v)] $\left.\beta \right|_{y=b(x)}=\left[a\left(x^0\right)\eta\left(x^0\right)\right]^2$;
    \item[(vi)] $\left.(h_{IJ}\right|_{y=b(x)})=\begin{pmatrix}
    0 & 0 & 0 & 0\\
    0 & -a^2\left(x^0\right) & 0 & 0 \\
    0 & 0 &-a^2\left(x^0\right) & 0\\
    0 & 0 & 0 &-a^2\left(x^0\right)
    \end{pmatrix}.
    $
\end{enumerate}
where $I,J\in \{0,1,2,3\}$, and~$h_{IJ}(x,y):=a_{IJ}-\frac{y_I}{\alpha}\frac{y_J}{\alpha}$ is the angular metric of $(M,\alpha)$. From~the above expressions, it turns out that the Finsler metric $g$ is also diagonal.  This result follows in a natural way from the definition of the $(\alpha,\beta)$ metric. Thus, in~this  Finslerian modification of general relativity, we maintain one of the essential properties of the FLRW metric (\ref{metr}), namely homogeneity and isotropy of the Riemannian geometry.
\paragraph{{\it Gravitational field equations.
}}
We postulate that the Einstein gravitational field equations, which geometrically describe the properties of the gravitational interaction, can be formulated in a general $(\alpha, \beta)$ Finslerian geometry as
\begin{equation}\label{Ein}
\widehat{G}_{IJ}=\widehat{R}_{IJ}-\frac{1}{2}\widehat{g}_{IJ}\widehat{R}=\frac{8\pi G}{c^4}\widehat{T}_{IJ},
\end{equation}
where $G$ is the gravitational constant. These equations are the natural extension of the Riemannian Einstein equations in the Finslerian framework. They reduce to the standard general relativistic Einstein equations in the limiting case of the Riemann~geometry.

\paragraph{{\it Flowchart of the algorithmic construction of  osculating Barthel-type Finslerian gravitational theories.
}} 

The steps necessary to construct a specific  Barthel--Kropina-type cosmological model, and~the underlying gravitational theory, are presented algorithmically in the form of a flowchart in Figure~\ref{flow}.

\begin{figure}[H]
\includegraphics[width=0.95\linewidth]{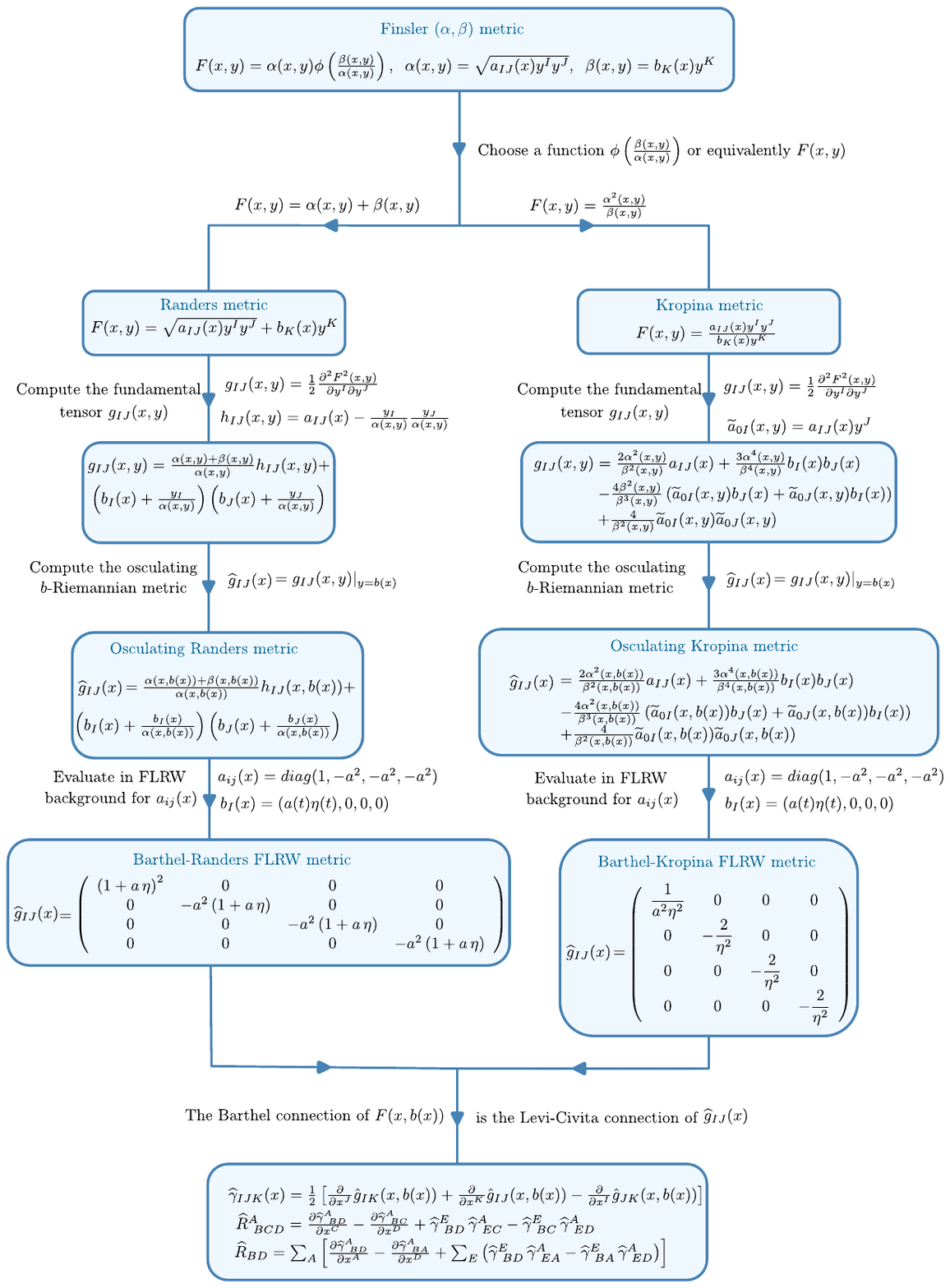}
\caption{Flowchart of the algorithmic approach for the construction of the $(\alpha, \beta)$ gravitational models with the Barthel~connection.}\label{flow}
\end{figure}
 
Although specific examples are restricted to the cases of Barthel--Randers and Barthel--Kropina geometries, the~formalism can be easily extended to any other choices of the Finslerian function $F=\alpha \phi (s)$, and~the implementation of the geometrical model into a gravitational theoretical framework can be performed easily. In~all these cases, some extensions of standard general relativity can be obtained, leading to gravitational field  equations that contain extra Finslerian terms, which in a cosmological context can be interpreted as describing an effective geometric dark energy, generated by the nonlocal geometric structure of the spacetime, with~the metric tensor nonlocalized due to the presence of the vector $y$, representing an internal degree of~freedom.

There are a large number of $(\alpha, \beta)$ metrics, whose physical and gravitational properties are worth investigating, and~which could offer new and important insights into the gravitational and cosmological phenomena. The~flowchart of the algorithmic approach presented in Figure~\ref{flow} may simplify and clarify the necessary steps taken for the investigation of these classes of~theories.

\subsection{Barthel--Randers~Cosmology}

For the case of the Randers geometry, we have $F=\alpha +\beta$. By~using the above assumptions, it turns out that the generalized Friedmann equations in this geometry take the form~\cite{Hama}
\begin{equation}\label{feq1R}
3H^{2}=8\pi G\phi ^2 \rho -\frac{3}{4}\frac{\dot{\phi} ^{ 2}}{\phi ^{2}}-3H\frac{%
\dot{\phi}}{\phi },
\end{equation}
and
\begin{equation}\label{feq2R}
2\dot{H}+3H^{2}=-\frac{8\pi G}{c^{2}}\phi ^{2}p-\frac{\ddot{\phi}}{\phi }+%
\frac{5}{4}\frac{\dot{\phi}^{2}}{\phi ^{2}}-H\frac{\dot{\phi}}{\phi },
\end{equation}
where we have denoted 
\be
\phi \left(x^0\right)=1+a\left(x^0\right)\eta \left(x^0\right).
\ee
Equations~(\ref{feq1R}) and (\ref{feq2R}) give the dynamical evolution of $H$ as 
\be
\dot{H}=-4\pi G \phi^2\left(\rho +\frac{p}{c^2}\right)-\frac{\ddot{\phi}}{2\phi}+\frac{\dot{\phi}^2}{\phi^2}+H\frac{\dot{\phi}}{\phi}.
\ee
Equation~(\ref{feq1R}) can be rewritten as
\be
3\left(H+\frac{\dot{\phi}}{2\phi}\right)^2=8\pi G \phi ^2 \rho.
\ee
By introducing 
 the new Hubble function defined as
\be
\widetilde{H}=H+\frac{\dot{\phi}}{2\phi},
\ee
we obtain the final form of the generalized Friedmann equations in the Barthel--Randers cosmology as
\be
3\widetilde{H}^2=8\pi G \phi ^2 \rho,
\ee
and
\be
2\dot{\widetilde{H}}+3\widetilde{H}^2=-\frac{8\pi G}{c^2}\phi ^2p+\frac{\dot{\phi}^2}{\phi ^2} +2H\frac{\dot{\phi}}{\phi},
\ee
respectively.

\paragraph{{\it The energy conservation equation.  
}} The energy conservation equation of the matter in the present model of the Barthel--Randers cosmology can be obtained by assuming that similarly to the standard Riemannian general relativistic case, the~covariant divergence of the energy--momentum tensor vanishes,  $\widehat\nabla_\mu T^{\mu\nu}=0$, with~the covariant derivative calculated with the help of the Barthel--Randers  connection $\widehat{\gamma}\tensor{}{^\mu_{\nu\alpha}}$. Hence, the~conservation equation in Barthel--Randers cosmology can be written as~\cite{Hama}
\begin{equation}\label{consR}
\dot{\rho}+3\left(H+\frac{\dot\phi}{2\phi}\right)\left(\rho +\frac{p}{c^2}\right) =0,
\end{equation}
or, in~an alternative form, as~\be
\frac{d}{dt}\left(\rho a^3\right)+\frac{p}{c^2}\frac{d}{dt}a^3+\frac{3}{2}a^3\left(\rho +\frac{p}{c^2}\right)=0.
\ee
The conservation 
 equation is not independent, and~can be also derived directly with the use of the  Friedman equations. By~taking the time derivative of Equation~\eqref{feq1R}, and~after substituting $\dot H$ from Equation~\eqref{feq2R}, we find
\begin{align}\label{consRR}
&8\pi G\phi^2\left(\dot\rho+\frac{3H p}{c^2}\right)-\frac{3}{8\phi^3}(\dot\phi^3-2H\phi\dot\phi^2-20H^2\phi^2\dot\phi^2)\nonumber\\&+4\pi G\phi\dot\phi(3p+4\rho)+9H^3=0.
\end{align}
By substituting
 the expression of $H^2$ from the Friedmann equation~\eqref{feq1R}, we recover equation~\eqref{consRR}. 

\subsection{Barthel--Kropina~Cosmology}

In the Kropina geometry, the Finslerian metric function is given by $F=\alpha ^2/\beta$. The~generalized Friedmann equations can be obtained directly from the Einstein equations, and~are given by~\cite{Hama2}
\begin{equation}\label{Fr1B}
\frac{3(\eta')^2}{\eta^2}=\frac{8 \pi G}{c^2}\frac{1}{a^2\eta ^2}\rho,
\end{equation}
and
\begin{equation}\label{Fr2B}
2\frac{\eta''}{\eta}+2\mathcal{H}\frac{\eta'}{\eta}-3\frac{(\eta')^2}{\eta ^2}=\frac{8\pi G}{c^4}\frac{p}{a^2\eta ^2},
\end{equation}
respectively, where $\mathcal{H}$ denotes the generalized Hubble function of the Barthel--Kropina cosmology, defined according to $\mathcal{H}=\left(1/a\left(x^0\right)\right)\left(da\left(x^0\right)/dx^0\right)$. 

In the above equations, and~in the following,  a~prime denotes the derivative with respect to $x^0$, a~dot denotes the derivative with respect to the cosmological time $t$, and~the standard Hubble function is given by $H=c\mathcal{H}$. After~eliminating the term $-3\left(\eta '\right)^2$ with the help of Equations~(\ref{Fr1B}) and  (\ref{Fr2B}), it becomes~\cite{Hama2} 
\begin{equation}\label{Fr3}
a\eta \frac{d}{dx^0}\left(\eta 'a\right)=\frac{4\pi G}{c^4}\left(\rho c^2+p\right).
\end{equation}
In the  
Barthel--Kropina geometry, the full system of the generalized Friedmann equations is represented by two ordinary differential equations with four unknowns $\left(a,\eta, \rho, p\right)$. By~considering an equation of state for the baryonic matter, $p=p(\rho)$, the~number of unknowns in the system of generalized Friedmann equations becomes three, and~the system is still underdetermined. Therefore, to~obtain solvable cosmological models, and~to close the system, we must impose a supplementary independent relation on two of the model~parameters. 

\paragraph{{\it Energy balance equation.  
}} One of the basic consequences of the standard Friedmann cosmology is the conservation of the matter energy--momentum tensor. But~as one can easily observe from the Friedmann equations~(\ref{Fr1B}) and (\ref{Fr2B}), this property does not hold anymore in the Barthel--Kropina cosmology. The~matter non-conservation equation as well as the energy density balance equation can be obtained after the multiplication of  equation~(\ref{Fr1B}) with $a^3$, and~applying the time derivation operator $d/dx^0$ on the result. By~using  the second generalized Friedmann equation in the obtained  relation, the~energy non-conservation equation in the Barthel--Kropina cosmology is obtained as~\cite{Hama2}
\begin{eqnarray}\label{balB}
\frac{8 \pi G}{c^4}\left[\frac{d}{dx^0}\left(\rho c^2a^3\right)+p\frac{d}{dx^0}a^3\right]=6a^5\Big[\mathcal{H} \left(\eta '\right)^2
+\left(\eta '+\mathcal{H} \eta\right)\eta ''+\mathcal{H} ^2\eta \eta '\Big].
\end{eqnarray}
Equation~(\ref{balB}) can be written in a form similar to the standard general relativistic conservation equation as
\begin{eqnarray}
\frac{4 \pi G}{c^4}\left[\frac{d}{dx^0}\left(\rho c^2a^3\right)+p\frac{d}{dx^0}a^3\right]
=3a^5\left[\frac{8\pi G}{2c^4}\left(\frac{5}{3}\rho c^2+p\right)\frac{\mathcal{H}}{a^2}+\eta ' \eta ''\right].
\end{eqnarray}

\paragraph{{The general relativistic limit.
}} An interesting and important property of the generalized Friedmann equations of the Barthel--Kropina cosmological model, given by equations~(\ref{Fr1B}) and (\ref{Fr2B}),  is that they admit a general relativistic limit, in~which they take the form of the standard Friedmann equations of general relativity. The~general relativistic limit is given by~\cite{Hama2}
\be
\eta \rightarrow \pm \frac{1}{a}, \beta \rightarrow (1,0,0,0).
\ee
Then from 
 equations~(\ref{Fr1B}) and (\ref{Fr2B}), we immediately reobtain the Friedmann equations of standard general relativity 
\be\label{Lim}
\frac{3(a')^2}{a^2}=\frac{8 \pi G}{c^2}\rho,\;\;
2\frac{a ''}{a}+\frac{(a')^2}{a^2}=-\frac{8\pi G}{c^4}p.
\ee

From equations~(\ref{Lim}), it follows that the energy density $\rho$ is conserved, with~the conservation equation given by
$\dot{\rho}+3H\left(\rho +p/c^2\right)=0$.

\subsection{Conformal Barthel--Kropina~Cosmology}

Let us consider that an  $(\alpha,\beta)$-metric with $F=F(\alpha,\beta)$ is given.  The~conformal transformation of the metric is defined as~\cite{Hama4}
\be\label{Cftrans}
\widetilde{F}(x,y):=e^{\sigma(x)}F(x,y)=\widetilde{F}\left(\widetilde{\alpha},\widetilde{\beta} \right). 
\ee
The metric  
 (\ref{Cftrans}) is again an $\left(\widetilde{\alpha},\widetilde{\beta}\right)$ metric, with
\be\label{Cftrans1}
\widetilde{\alpha}=e^{\sigma(x)}\alpha,\ \widetilde{\beta}=e^{\sigma(x)}\beta.
\ee
The fundamental  
 tensor of $\widetilde{F}$ is calculated with the help of the Hessian~\cite{Hama4}
\be\label{Hess}
\widetilde{g}_{IJ}:=\frac{1}{2}\frac{\partial^2\widetilde{F}^2}{\partial y^I\partial y^J}.
\ee
The conformal
 transformation of the Kropina metric is obtained as 
$$
\widetilde{F}:=e^{\sigma(x)}\frac{\alpha^2}{\beta}=\frac{\widetilde{\alpha}^2}{\widetilde{\beta}},
$$
where $\widetilde{\alpha}=e^{\sigma(x)}\alpha$, $\widetilde{\beta}=e^{\sigma(x)}\beta$.
The osculating Riemannian metric is obtained as
\be\label{14}
\widehat{\widetilde{g}}_{IJ}(x)=e^{2\sigma(x)}\widehat{g}_{IJ}(x),
\ee
where $\widehat{g}_{IJ}(x)$ is given by
\begin{eqnarray}\label{eq_Kropina Hessian}
\widehat{g}_{IJ}(x,b(x))&=&\frac{2\alpha^2}{\beta^2}a_{IJ}(x)+\frac{3\alpha^4}{\beta^4}b_Ib_J-\frac{4\alpha^2}{\beta^3}(b_Ib_J+b_Jb_I)
+\frac{4}{\beta^2}b_Ib_J,
\end{eqnarray}
where ${b_I:=a_{IJ}b^J}=\widetilde{a}_{0I}(x,b(x))$ (see Figure \ref{flow}). 
For cosmological applications we choose the conformal factor as $\sigma (x)=\phi \left(x^0\right)$. Hence in the study of the dynamical evolution of the Universe,  we will restrict our investigations to conformal transformations of the Kropina metric that depend only on~time. 

\paragraph{{\it The generalized Friedmann equations. 
}} In the conformal Barthel--Kropina geometry, the generalized Friedmann equations  take the form~\cite{Hama4}
\be\label{B1}
\frac{3(\eta')^2}{\eta^2}=\frac{8\pi G}{c^2}\frac{1}{a^2\eta ^2}{\widetilde{\rho}}+3\left(\phi '\right)^2-6\frac{\eta '}{\eta }\phi ',
\ee
and
\begin{eqnarray}\label{B2}
&&\frac{2}{\eta^2}[-3(\eta')^2+2\eta\eta'\mathcal{H}+2\eta\eta'']
=\frac{16\pi G}{c^4}\frac{1}{a^2\eta ^2}{\widetilde{p}}\nonumber\\
&&-4\left[\phi ''+\frac{1}{2}\left(\phi '\right)^2\right]
+\left(\frac{\eta'}{\eta}-\mathcal{H}\right)\phi',
\end{eqnarray}
respectively. We eliminate now the term $-3\left(\eta '\right)^2/\eta ^2$ between equations~(\ref{B1}) and (\ref{B2}), and~thus we find the relation
\begin{eqnarray}
2\frac{1}{a\eta}\frac{d}{dx^0}\left(a\eta '\right)&=&\frac{4\pi G}{c^4}\frac{1}{a^2\eta ^2}\left({\widetilde{\rho}}c^2+{\widetilde{p}}\right)-\left(\phi ''-\left(\phi '\right)^2\right)
-\frac{11}{4}\frac{\eta '}{\eta} \phi'
-\frac{1}{4}\phi ' \mathcal{H}.
\end{eqnarray}

The general relativistic limit of the system (\ref{B1}) and  (\ref{B2}) is obtained by taking  $\eta \rightarrow 1/a$, and~$\left(b_I(x)\right)=(1,0,0,0)$, respectively. Consequently,  $\beta =y^0$. Thus, in~this limit, the~generalized Friedmann cosmological evolution equations of the conformal Barthel--Kropina model become~\cite{Hama4}
\be
3\mathcal{H} ^2=\frac{8\pi G}{c^4}{\widetilde{\rho}}c^2+3\left(\phi '\right)^2+6\mathcal{H} \phi',
\ee
and
\be
2\mathcal{H} '+3\mathcal{H} ^2=-\frac{8\pi G}{c^4}{\widetilde{p}}+2\left[\phi ''+\frac{1}{2}\left(\phi '\right)^2\right]+\mathcal{H} \phi',
\ee
respectively. For~$\phi=0$ we fully recover the standard Friedmann equations of general~relativity.

\subsection{Thermodynamic Interpretation of the \texorpdfstring{$(\alpha,\beta)$}{ab}  Cosmologies}

It is a general property of several Finslerian-type cosmological models that the matter energy--momentum is not conserved. For~example, Equation~(\ref{consR}) illustrates this situation for the case of the Barthel--Randers cosmological models. Hence, contrary to the general relativistic case, in~the Barthel--Randers-type cosmological model, as~well as in the Barthel--Kropina and conformally transformed Barthel--Kropina cosmologies, the~baryonic matter content of the Universe is not conserved anymore. This intriguing property of the models raises the problem of the physical interpretation of the non-conservation of the matter energy--momentum tensor, and~the problem of the cosmological significance of this~effect. 

A possible physical understanding of the energy--momentum non-conservation can be obtained by interpreting this effect by using the thermodynamics of irreversible processes, and~assuming that it describes particle creation and annihilation in a cosmological environment. In~the following, we briefly introduce first the foundations of the thermodynamics of irreversible processes, and~then we illustrate the general formalism by considering the specific case of the Barthel--Randers-type cosmological model. The~non-conservation of the energy--momentum tensor is a specific feature of several modified gravity theories, and~specifically in approaches to gravity involving the presence of geometry--matter coupling~\cite{Ha14}. Examples of such theories are the  $f\left(R,L_m\right)$ \cite{fRLm} and the $f(R,T)$ \cite{fRT} theories.

The non-conservation of the matter energy--momentum tensor, as~shown, for~example, for~the Barthel--Randers case by Equation~(\ref{consR}), can thus be interpreted as showing that due to the existence of the Finslerian geometric effects, during~cosmological evolution, matter creation processes may take place. This indicates the possibility of creating matter from geometry. Quantum field theories in curved spacetime also predict the same particle creation effect,  as~initially proposed and investigated in~\cite{Parker, Parker1, Parker1a, Zel, Parker2, Parker3}. In~quantum field theory, particle creation is due to the time variation of the gravitational field. In~an anisotropic Bianchi Type I metric, quantum particle creation was considered in~\cite{Zel}, and~for a quantum scalar field with a non-zero mass, the renormalized expression  of the energy--momentum tensor was determined. Hence, the~osculating Finsler--Barthel-type gravity theories, in~which the creation of matter is also allowed by the general formalism,  could also be interpreted as providing an effective, semiclassical description of the quantum effects in the gravitational field. It is important to note, however, that the nature of the particles is not necessarily known, unless~quantum theoretical effects are taken into~account.

\subsubsection{Irreversible Thermodynamics and Matter~Creation}

Equation~(\ref{consR}), obtained within the framework of the Barthel--Randers cosmology,  shows that the covariant divergence of the energy--momentum tensor, which is a function 
of the equilibrium quantities of the thermodynamic system represented by the baryonic matter, is different from zero.  Similar effects appear for the case of other thermodynamical quantities, like, for~example, the~particle and entropy fluxes. Hence, in~the presence of matter creation, all the balance equilibrium equations must be modified to account for this effect~\cite{P-M,Lima,Su}. We will present first the general formalism of irreversible thermodynamics in the presence of matter creation,  by~adopting a cosmological perspective. We will consider all the results in the Riemann space, with~metric $a_{AB}(x)$. Moreover, we interpret the Finslerian effects as generating a set of specific physical events in the background Riemann geometry. Hence, in~the following, all the physical and geometrical quantities will be expressed with the help of the FLRW metric (\ref{metr}). Consequently,  all considered physical and geometrical quantities are functions of the cosmological time $t$ only.

 \paragraph{{Particle balance equations. 
}} To describe particle dynamics, we introduce the particle number density $n$, and~the four-velocity $u^{I}$ of matter. From~these quantities, we construct the particle flux $N^{I} \equiv nu^I$. All these physical parameters are defined in Riemannian geometry.  The~particle balance equation in the presence of matter creation is given by
\begin{equation}
\nabla _{I}N^{I}=\dot{n}+3Hn=n\Gamma,
\end{equation}
where $\nabla _I$ is the covariant derivative defined in the Riemann space  with the help of the Levi--Civita connection associated with the FLRW metric (\ref{metr}), while $\Gamma $ denotes the matter creation rate. For~$\Gamma \ll 3H$, the~source term in the particle balance equation is negligible, and~we reobtain the standard particle conservation law of general~relativity. 

\paragraph{{The entropy flux. 
}} We now introduce the entropy density $\tilde{s}$, and~the entropy per particle  $\sigma =\tilde{s}/n$. From~these quantities we construct the entropy flux vector defined as $S^{I} \equiv \tilde{s}u^{I} = n\sigma u^{I}$.  The~divergence of the entropy flux gives the relation
\begin{equation}  \label{62b}
\nabla _{I}S^{I}=n\dot{\sigma}+n\sigma \Gamma \geq 0,
\end{equation}
where the positivity condition is a direct consequence of the second law of thermodynamics. The~case  $\sigma ={\rm constant}$ gives the relation
\begin{equation}\label{82}
\nabla _{I}S^{I}=n\sigma \Gamma =\tilde{s}\Gamma \geq 0.
\end{equation}
Equation~(\ref{82}) shows that if the entropy per particle can be taken as a constant, then the variation of the entropy is only due to the matter generated via the transfer of the energy of the gravitational field to matter. Since $\tilde{s}>0$, the~particle production rate $\Gamma$ must satisfy the basic thermodynamic condition $\Gamma  \geq 0$. From~a physical point of view, this condition can be interpreted as permitting the creation of matter from the gravitational field, but~suppressing the opposite~process. 

\paragraph{{\it The creation pressure. 
}} If particle creation takes place,  the~matter energy--momentum tensor must also be modified to take into account the presence  of the irreversible processes and the second law of thermodynamics. Generally, in~the thermodynamical description of open systems,  the~energy--momentum tensor can be represented as~\cite{Bar}
\begin{equation}\label{64}
T^{IJ}=T^{I J}_\text{eq}+\Delta T^{IJ},
\end{equation}
where $T^{IJ}_\text{eq}$  is the equilibrium thermodynamic component~\cite{Bar}, and~$\Delta T^{IJ}$ describes the supplementary terms induced by particle creation. For~a homogeneous and isotropic  spacetime geometry, $\Delta T^{IJ}$, giving the particle creation contribution to  $T^{IJ}$, can be generally represented in the form
\begin{equation}
\Delta T_{\; 0}^0=0, \quad \Delta T_{\; I}^J=-P_c\delta_{\; I}^J,
\end{equation}
where $P_c$ denotes the creation pressure, an~effective thermodynamic quantity, which in a macroscopic physical system describes phenomenologically particle creation. Moreover, in~a fully covariant representation,  the tensor $\Delta T^{IJ}$ is given by~\cite{Bar}
\begin{equation}
\Delta T^{I J}=-P_ch^{IJ}=-P_c\left(a^{I J}-u^{I}u^{J}\right),
\end{equation}
where $h^{IJ}=a^{IJ}-u^Iu^J$ is the projection operator defined in the FLRW geometry. Thus, we  straightforwardly obtain the relation  $u_{I}\nabla _{J}\Delta T^{I J}=3HP_c$.

In the presence of particle production by the gravitational field, from~the scalar component of the energy balance
equation $u_{I}\nabla _{J}T^{I J}=0$, obtained from Equation~\eqref{64}, we obtain the time variation of the energy density $\rho$ of the cosmological fluid as
\begin{equation}\label{cons1}
\dot{\rho}+3H\left[\rho+\frac{1}{c^2}\left(p+P_c\right)\right]=0.
\end{equation}

The  thermodynamic quantities describing baryonic matter must also satisfy the Gibbs law~\cite{Lima}
\begin{equation}\label{Gibbs}
n T d \left(\frac{\tilde{s}}{n}\right)=nTd\sigma=d\rho -\frac{\rho+p/c^2}{n}dn,
\end{equation}
where $T$ is the temperature of the cosmological~matter.

\subsubsection{Application: Particle Creation in Barthel--Randers~Cosmology}

As an example of the application of the irreversible thermodynamic of open systems, we consider now the interpretation of the Barthel--Randers cosmology as a cosmological theory also describing particle creation during the evolution of the~Universe.  

\paragraph{{\it Creation pressure in Barthel--Randers cosmology.
}} After some simple algebraic transformations, the~particle energy balance equation~\eqref{consR} of the Barthel--Randers cosmology can be rewritten as
\begin{equation}\label{cons3}
\dot{\rho}+3H\left[\left(\rho +\frac{p}{c^2}\right)\left(1+\frac{\dot{\phi}}{2H\phi}\right)\right] =0.
\end{equation}%

By comparing equations~(\ref{cons1}) and (\ref{cons3}), we find the expression of the  Barthel--Randers creation pressure  as
\be\label{pc}
P_{c}=\frac{\dot{\phi}}{2H\phi}\left(\rho c^2+p\right)=\frac{c^2\rho \dot{\phi}}{2H\phi}\left(1+\omega\right),
\ee
where $\omega=p/\rho c^2$. The~energy density balance Equation~\eqref{consR} can now be derived equivalently 
from the divergence  of the total energy momentum tensor $T^{IJ}$, given by
\begin{equation}
T^{I J }=\left(\rho c^2 +p+P_{c}\right) u^{I }u^{J }-\left(p+P_{c}\right) a^{I J },
\end{equation}
where all the mathematical operations must be performed in the Riemann space with the FLRW metric (\ref{metr}).
\paragraph{{\it The particle creation rate.
}} By assuming  adiabatic particle production, which requires that  $\dot{\sigma}=0$,  the~Gibbs relation (\ref{Gibbs}) gives
\begin{equation}
\dot{\rho}
=\left(\rho+\frac{p}{c^2}\right)\frac{\dot{n}}{n}
=\left(\rho+\frac{p}{c^2}\right)\left(\Gamma-3H\right).
\end{equation}
By combining 
 this relation with the energy balance equation~(\ref{cons3}), we obtain the particle creation rate as a function of the creation pressure, the~Hubble function and the equilibrium thermodynamic quantities  as
\begin{equation}
\Gamma=\frac{-3HP_c}{\rho c^2+p}.
\end{equation}

By using equation~(\ref{pc}), we find  for the Barthel--Randers cosmological particle  creation rate the expression
\begin{eqnarray}
\Gamma=-\frac{3}{2}\frac{\dot{\phi}}{\phi}=-\frac{3}{2}\frac{d}{dt}\ln \phi=-\frac{3}{2}\frac{\dot{a}\eta+a\dot{\eta}}{1+a\eta}.
\end{eqnarray}

If $1+a\eta >0$, the~condition $\Gamma \geq 0$, which is equivalent with the existence of particle creation, leads to the   constraint 
\be
\dot{\phi}=\dot{a}\eta+a\dot{\eta}=a\left(\dot{\eta}+H\eta\right)\leq 0,
\ee
on the scale factor $a$, and~the temporal component of the one-form  $\eta$. The~condition does not depend explicitly on the equation of state of the baryonic  matter, but~an indirect dependence via the Hubble function does exist. The~condition $\dot{\phi}<0$ implies the existence  of a negative creation pressure, as~it follows from Equation~(\ref{pc}).  Therefore particle production is thermodynamically allowed only if the creation pressure is negative,  $P_c<0$.

The Barthel--Randers particle balance equation can thus be reformulated to take the form
\be
\dot{n}+3Hn=-\frac{3}{2}n\frac{\dot{\phi}}{\phi}=\Gamma n,
\ee
and it can be integrated to give the expression for the particle number density
\be
n=\frac{n_0}{\phi ^{3/2}a^3},
\ee
where $n_0$ is a constant of~integration.

For a Universe consisting of pressureless dust with $p=0$, the~creation pressure is given by
\be
P_{c}=\frac{\dot{\phi}}{2H\phi}\rho c^2=\frac{\rho c^2}{2H}\frac{d}{dt}\ln \phi,
\ee
and it depends linearly on the baryonic matter~density.

The divergence of the entropy flux vector $S^I$ takes the form
\begin{equation}
\nabla _{I}S^{I}=\frac{-3 n \sigma H P_c}{\rho c^2 +p}=-\frac{3}{2}n\sigma \frac{\dot{\phi}}{\phi}=n\sigma \Gamma.
\end{equation}

\paragraph{{\it The matter temperature. 
}} The temperature $T$ of the baryonic matter represents an important  characteristic of physical systems. To~obtain the temperature evolution in the presence of particle creation,  we consider the general thermodynamic case in which the density and the pressure are functions of the particle number density, and~of the temperature. Hence $\rho$ and $p$ can be generally represented in parametric form $\rho =\rho (n, T)$ and $p=p(n,T)$, respectively. Then, we obtain
\begin{equation}
\dot{\rho}=\left(\frac{\partial \rho }{\partial n} \right)_T\dot{n}+\left(%
\frac{\partial \rho }{\partial T} \right)_n\dot{T}.
\end{equation}

By using the energy and particle balance equations, we find first the relation
\begin{equation}  \label{78a}
-3H\left(\rho c^2+p+P_c\right)=\left(\frac{\partial \rho }{\partial n}
\right)_T n\left(\Gamma-3H\right)
+\left(\frac{\partial \rho }{\partial T} \right)_n\dot{T}.
\end{equation}

As a second step we use the thermodynamic relation~\cite{Bar}
\begin{equation}
T\left(\frac{\partial p}{\partial T}\right)_n=\rho c^2+p-n\left(\frac{\partial
\rho}{\partial n}\right)_T,
\end{equation}
which together with equation~\eqref{78a} gives the temperature evolution of the matter in the Barthel--Randers cosmology in the presence of particle production as
\begin{equation}
\frac{\dot{T}}{T}=\frac{1}{c^2}\left(\frac{\partial p}{\partial \rho}\right)_n\frac{\dot{n}}{n}=\omega \frac{\dot{n}}{n}.
\end{equation}

 From the particle balance equation, we obtain the ratio $\dot{n}/n$ as
 \be
 \frac{\dot{n}}{n}=-3\left(\frac{\dot{\phi}}{\phi}+\frac{\dot{a}}{a}\right).
 \ee
 By assuming  
 that $\dot{\phi}<0$, we obtain for the temperature evolution of the particles in the Barthel--Randers Universe the relation
 \be
 T=T_0\frac{\phi ^{3\omega/2}}{a^3}.
 \ee

 Generally,  $\omega $ is an arbitrary scale factor-dependent parameter $\omega =\omega (a)$ \cite{r3s}. In~a thermodynamically consistent cosmological model, all geometrical and physical quantities must be well-defined and regular for all  $\omega (a)$.

\subsubsection{Creation of Exotic~Matter}

 In our discussion of the thermodynamics of open systems, as~presented in the previous section, we have assumed that matter is created in the form of baryonic matter,  satisfying the energy condition $\rho c^2+p\geq 0$, and~the restriction on the parameter of the matter equation of state  $\omega \geq 0$. But~we cannot a priori exclude the possibility case in which an exotic fluid with $\omega <0$ is created during the cosmological evolution of the  Barthel--Randers Universe. The~open system thermodynamic approach to particle creation considered previously is also applicable if $\omega <0$.  For~this case the creation pressure becomes negative for $\dot{\phi}>0$, with~the particle creation rate $\Gamma $ becoming positive. Therefore, matter creation processes generating exotic particles can also be included in the Barthel--Randers cosmological~model.

 A particular and important case is represented  by an exotic fluid satisfying the equation of state $\rho_{(ex)} c^2+p_{(ex)}=0$, with~$\omega =-1$, which is equivalent  to the presence of a cosmological constant. Then from equation~(\ref{cons1}), we obtain
\begin{equation}
  \dot{\rho}_{(ex)} = -3H P_c.
\end{equation}

 If the exotic particle creation process is adiabatic, with~$\dot{\sigma}=0$, from~the Gibbs relation, we find
\begin{equation}
\dot{\rho}_{(ex)} = \left(\rho _{(ex)} c^2+p_{(ex)} 
 \right)\frac{\dot{n}_{(ex)}}{n_{(ex)}} = 0.
\end{equation}
 The above 
 two equations  independently give
\begin{equation}
  \dot{\rho}_{(ex)} = P_c = 0,
\end{equation}
which describe a Universe with constant exotic matter density and~vanishing creation pressure. On~the other hand, if  $\omega =-2$, and~$\rho _{(ex)} >0$, then $P_c=-\left(c^2\dot{\phi}/2H\phi\right)\rho_{(ex)}$,  $\Gamma=(3/2)\dot{\phi}/\phi$, and~matter production can take place if the conditions $\dot{\phi}>0$ and $\phi >0$ are~satisfied.

Thus, in a Barthel--Randers Universe, the creation of exotic matter or possibly of scalar fields is allowed in a way consistent with the laws of thermodynamics. The~creation processes take place in the Riemannian geometry characterized by the FLRW metric.

\section{Cosmological Implications of Barthel--Randers and Barthel--Kropina~Models}\label{sect3}

In this section, we investigate the cosmological implications of the Barthel--Randers and Barthel--Kropina models by exploring three distinct variants of the Barthel--Randers framework alongside the Barthel--Kropina cosmological model. To~constrain the free parameters of these models, we employ a combination of observational datasets, including Type Ia Supernovae, Baryon Acoustic Oscillations, and~Hubble parameter measurements. By~extracting the posterior distributions of the model parameters through Bayesian inference, we are able to assess the observational viability of the Barthel--Randers and Barthel--Kropina models. This comparison with the standard \(\Lambda\)CDM cosmology enables us to explore possible deviations from the conventional expansion history and to evaluate whether these geometrically extended models offer a competitive or improved description of the Universe’s~evolution.

\subsection{Specific Cosmological~Models}

In this subsection, we present the normalized Hubble functions associated with both the Barthel--Randers~\cite{Hama} and Barthel--Kropina~\cite{Hama2} cosmological models. Our starting point is the family of cosmological scenarios proposed in~\cite{Hama}, where three distinct variants of the Barthel--Randers model were introduced. Each of these variants is characterized by a specific choice of the function \(\varphi(z)\). The~normalized Hubble function for all three models can be expressed in a unified form as
\begin{equation}\label{hubble_1}
h(z) = \frac{2\varphi(z)\sqrt{(1+z)^3 \sqrt{\varphi(z)}\, \Omega_{m0} + \varphi(z)^2 \Omega_{\Lambda0} + (1+z)^4 \Omega_{r0}}}{(1+z)\frac{d\varphi(z)}{dz} - 2\varphi(z)},
\end{equation}
where \(z\) is the redshift, and \(\Omega_{m0}\), \(\Omega_{\Lambda0}\), and~\(\Omega_{r0}\) are the present-day matter, dark energy, and~radiation density parameters, respectively. The~dark energy density parameter \(\Omega_{\Lambda0}\) is obtained through the constraint
\begin{equation}
\Omega_{\Lambda0} = \left(1 - \frac{1}{2}  \varphi'(0)  \right)^2 - \Omega_{m0} - \Omega_{r0},
\end{equation}
where \(\varphi'(0)\) is the derivative of the function \(\varphi(z)\) evaluated at \(z = 0\). The~form of \(\varphi(z)\) determines the behavior of each specific model and encodes the influence of the underlying Finslerian geometry. In~the following, we present the explicit forms of \(\varphi(z)\) corresponding to each of the three models proposed in~\cite{Hama}:

\begin{enumerate}
  \item \textbf{Linear model: 
 \(\varphi(z) = 1 + \beta z\)}.
  \item \textbf{Logarithmic model: 
 \(\varphi(z) = 1 + \ln(1 + \beta z)\)}.
  \item \textbf{Exponential model:  
\(\varphi(z) = e^{2\beta z}\)}.
\end{enumerate}

We can obtain the corresponding normalized Hubble function by plugging the corresponding form of \(\varphi(z)\) into equation~\eqref{hubble_1}.

In the case of the Barthel--Kropina geometry, we adopt the model proposed by~\cite{Hama2}. The~corresponding normalized Hubble function is expressed as a system of differential equations in the redshift representation as follows:
\begin{equation}\label{linear1}
\frac{df(z)}{dz} = -\frac{u(z)}{(1 + z)h(z)},
\end{equation}
\begin{equation}\label{linear2}
\begin{split}
\frac{du(z)}{dz} = \frac{1 + f(z)}{2(1 + z)h(z)} \Big[ \,
& 2h(z)\left(\frac{2}{1 + f(z)} - 3\omega(1 + f(z)) \right) u(z)\\
&+ 3\left(\omega - \frac{1}{(1 + f(z))^2} \right) u(z)^2 
 + 3\omega f(z)(2 + f(z))h(z)^2 \, \Big],
\end{split}
\end{equation}
\begin{equation}\label{linear3}
\frac{dh(z)}{dz} = \frac{1}{2(1+z)h(z)} \left[
3h(z)^2
- \frac{4h(z)u(z)}{1+f(z)}
+ \frac{3u(z)^2}{(1+f(z))^2}
+ \frac{2(1+z)h(z)}{1+f(z)} \frac{du(z)}{dz}
\right].
\end{equation}
The system 
 of equations above has to be solved with initial conditions $f(0) = f_0$, $u(0)=u_0$, and~$h(0)=1$.

\subsection{Methodology and~Datasets}
{In this subsection, we provide a detailed explanation of the methodology used to estimate the posterior distributions of the model parameters, employing the Markov Chain Monte Carlo (MCMC) approach.   This approach allows us to constrain model parameters by analyzing various observational datasets. The~MCMC method explores the parameter space of theoretical models by sampling from the likelihood function and incorporating prior information, leading to a robust estimation of the posterior probability distribution~\cite{Bernardo1979}. Through this method, the~model's parameter space is thoroughly explored.
The MCMC {algorithm} relies on Bayes’ theorem, which is given by}
\begin{equation}  
P(\theta|D) = \frac{L(D|\theta) P(\theta)}{P(D)},  
\end{equation}  
{where \(P(\theta|D) \) represents the probability of the parameters \(\theta \) given the observational data \(D \). The~term \(L(D|\theta) \) is the likelihood function, which measures how well the model fits the data. \(P(\theta) \) is the prior distribution, incorporating any previous knowledge about the parameters, while \(P(D) \) is the evidence, acting as a normalization factor~\cite{Joyce2021}.}

{One of the key advantages of the MCMC method is that it not only finds the most likely values for the parameters but also accounts for uncertainties in both the models. We define the likelihood function in such a way that it compares the theoretical predictions of the Hubble parameter (\(H(z) \)), luminosity distance (\(\mu(z) \)), transverse distance (\(D_H(z) \)), comoving angular diameter distance (\(D_M(z) \)), and~comoving volume distance (\(D_V(z) \)) with observational data, taking into account uncertainties through covariance matrices.}

{The MCMC sampling is carried out using the \texttt{emcee} library\endnote{\protect\url{https://emcee.readthedocs.io/en/stable/}   
} \cite{ForemanMackey2013,ForemanMackey2019}, which includes an affine-invariant ensemble sampler. This sampler is used to efficiently explore the parameter space. Specifically, we use 30 walkers, with~a total of 100,000 steps for the chain. To~ensure the removal of biases, the~first 10,000 steps are discarded as burn-in, and~the remaining steps are thinned by a factor of 30 to reduce autocorrelation between samples.}

{After running the MCMC chains, the~posterior distributions of the parameters are analyzed, and~the best-fit values are identified. To~visualize and analyze the posterior distributions, we utilize the \texttt{GetDist} package\endnote{\protect\url{https://getdist.readthedocs.io/en/latest/plot_gallery.html}  
} \cite{GetDist}, which provides an extensive set of tools for generating 1D and 2D posterior distributions. In~this work, we use three different observational datasets: Cosmic Chronometers, Type Ia Supernovae, and~Baryon Acoustic Oscillations.}

Below, we provide a detailed description of each dataset and how the likelihood has been formed based on~them. 
\begin{itemize}
        \item \textbf{Cosmic Chronometers:}  
 {\textls[-15]{In this study, we use the Hubble measurements extracted based on the differential age approach. This technique leverages passively evolving massive galaxies, which formed at redshifts around \(z \sim 2-3 \), enabling a direct and model-independent determination of the Hubble parameter using the relationship \(\Delta z / \Delta t \). This method significantly reduces the reliance on astrophysical assumptions~\cite{Jimenez}. For~our analysis, we use 15 Hubble measurements selected from the 31 Hubble measurements, which cover a redshift range from \(0.179 \leq z \leq 1.965 \) \cite{Moresco2012,Moresco2015,Moresco2016}. We use the likelihood function provided by Moresco in his GitLab repository\endnote{\protect\url{https://gitlab.com/mmoresco/CCcovariance}   
} which uses the full covariance matrix, accounting for both statistical and systematic uncertainties~\cite{Moresco2018,Moresco2020} .}}
        \item \textbf{Type Ia Supernova:
} We also use the Pantheon$^{+}$ dataset without the SHOES calibration, which consists of {1701 light curves from 1550} Type Ia Supernovae (SNe Ia) covering a redshift range of \(0 \leq z \leq 2.3 \) \cite{brout2022}. To~analyze this data, we adopt the likelihood function described in~\cite{astier2006}, which incorporates the total covariance matrix, \(\mathbf{C}_{\text{total}}\), which includes both statistical (\(\mathbf{C}_{\text{stat}}\)) and systematic (\(\mathbf{C}_{\text{sys}}\)) uncertainties~\cite{conley2010}. The~likelihood function is given by $\mathcal{L_{\text{SNe Ia}}} = e^{\left(-\frac{1}{2} \mathbf{r}^T \mathbf{C}^{-1}_{\text{total}} \mathbf{r} \right)},$ where \(\mathbf{r}\) represents the residual vector, defined as the difference between the observed and theoretical distance moduli: $r_i = \mu_{\text{obs}}(z_i) - \mu_{\text{th}}(z_i, \theta)$, where $\theta$. Here, \(\mathbf{C}^{-1}_{\text{total}}\) is the inverse of the total covariance matrix. The~model-predicted distance modulus is calculated as $\mu_{\text{model}}(z_i) = 5 \log_{10} \left(\frac{d_L(z)}{\text{Mpc}} \right) + \mathcal{M} + 25,$ where the luminosity distance \(d_L(z) \) in a flat FLRW Universe is given by $d_L(z) = c(1 + z) \int_0^z \frac{dz'}{H(z')}.$ Here, \(c \) is the speed of light, and~\(H(z) \) is the Hubble parameter. This formulation highlights the degeneracy between the nuisance parameter \(\mathcal{M}\) and the Hubble constant \(H_0 \). 
        \item \textbf{Baryon Acoustic Oscillation: 
}  {In our analysis, we also include the 13 recent Baryon Acoustic Oscillation (BAO) measurements from the Dark Energy Spectroscopic Instrument (DESI) Data Release 2 (DR2) \cite{karim2025}. These measurements cover a redshift range of $0.295 \leq z \leq 2.330$. They were obtained from observations of the Bright Galaxy Sample (BGS), Luminous Red Galaxies (LRG1, LRG2, LRG3), Emission Line Galaxies (ELG1 and ELG2), Quasars (QSO), and~Lyman-$\alpha$ tracers\endnote{\url{https://github.com/CobayaSampler/bao_data}  
}. The~measurements are reported using three distance indicators: the Hubble distance $D_H(z)$, the~comoving angular diameter distance $D_M(z)$, and~the volume-averaged distance $D_V(z)$. To~compare these with cosmological models, we compute the ratios $D_H(z)/r_d$, $D_M(z)/r_d$, and~$D_V(z)/r_d$, where $r_d$ represents the sound horizon at the drag epoch, occurring around redshift $z \sim 1060$. In~a flat \(\Lambda \)CDM model, \(r_d = 147.09 \pm 0.26 \) Mpc~\cite{fit4}. However, in~this study, we treat \(r_d \) as a free parameter, allowing late-time observations to constrain the corresponding model parameters~\cite{Pogosian,Jedamzik,Pogosian2,Lin,vagnozzi2023seven}. The~chi-squared statistic for the BAO measurements is given by $\chi^2_i = (\vec{O}^{\,\text{th}}_i - \vec{O}^{\,\text{obs}}_i)^{\!\!T} \, \mathbf{C}_i^{-1} \, (\vec{O}^{\,\text{th}}_i - \vec{O}^{\,\text{obs}}_i),$ where $\vec{O}^{\,\text{th}}_i$ and $\vec{O}^{\,\text{obs}}_i$ denote the vectors of theoretical predictions and observed measurements, respectively, and~$\mathbf{C}_i$ is the associated covariance matrix\endnote{\url{https://github.com/CobayaSampler/bao_data/blob/master/desi_bao_dr2/desi_gaussian_bao_ALL_GCcomb_cov.txt}   
}.} 
\end{itemize}
The posterior 
 distribution of model parameters is obtained by maximizing the likelihood function. The~total likelihood function is given by
\begin{equation}
\mathcal{L}_{\text{tot}} = \mathcal{L}_{\text{CC}} \times \mathcal{L}_{\text{SNe Ia}} \times \mathcal{L}_{\text{BAO}}.
\end{equation} 
First, we  
 discuss how to extract the posterior distribution of the model parameters in the Barthel--Randers framework. To~achieve these, we apply the standard approach outlined above. In~these models, we treat the parameters \(H_{0} \), \(\Omega_{m0} \), \(\beta \), \(\mathcal{M} \), and~\(r_d \) as free parameters with the following {priors: 
$H_0\ (\text{km}\,\text{s}^{-1}\,\text{Mpc}^{-1}) \in [50, 100], \quad
\Omega_{mo} \in [0, 1], \mathcal{M} \in [-20, -18],  \quad
r_d\ (\text{Mpc}) \in [100, 200]$, and~\(\beta \in [0, 0.1]\),} as the model is sensitive to this parameter. The~\(\Omega_{r0} \) is extracted using the following relation: $\Omega_{r0} = \left(4.183699 \times 10^{-5}\right) h^{-2},$ where \(h = \frac{H_0}{100} \). In~the case of the linear and logarithmic models, \(\Omega_{\Lambda0} \) is extracted using the  relation $\Omega_{\Lambda0} = 1 - \beta + \frac{\beta^2}{4} - \Omega_{m0} - \Omega_{r0},$ while in the case of the exponential model, \(\Omega_{\Lambda0} \) is extracted using the relation $\Omega_{\Lambda0} = 1 - 2\beta + \beta^2 - \Omega_{m0} - \Omega_{r0}$. This makes the explicit variation of \(\Omega_{r0} \) and \(\Omega_{\Lambda0} \) redundant, as~both are fully determined by the other~parameters.

To constrain the Barthel--Kropina model, we begin by numerically solving the system of differential equations given in Equations \eqref{linear1}–\eqref{linear3}. These equations are integrated using the \textit{solve\_ivp} function from \textit{SciPy}, employing the Radau method with a fifth-order implicit Runge–Kutta scheme particularly effective for handling the stiff differential equations. 
For~numerical stability and accuracy over the redshift range \(z \in [0, 3]\), we set the relative and absolute tolerances to \(10^{-3}\) and \(10^{-6}\), respectively. After~obtaining the numerical solutions, we use an MCMC approach to constrain the model parameters. For~more information on how to handle these kinds of Hubble-like functions that appear within differential equations, please consult~\cite{SSMG}. In~the case of the Barthel--Kropina model, {we take the $H_0\ (\text{km}\,\text{s}^{-1}\,\text{Mpc}^{-1}) \in [50, 100], \quad
u_{o} \in [0.0, 3.0], \mathcal{M} \in [-20, -18], \quad
r_d\ (\text{Mpc}) \in [100, 200], \text{and } f_0 \in [0, 0.1]$.} It is important to note that in both cases, we work with the normalized Hubble function. The~final Hubble parameter \(H(z) \) can be obtained by multiplying the normalized function \(h(z) \) by the Hubble constant \(H_0 \), i.e.,~\(H(z) = H_0 \cdot h(z) \).
\subsubsection*{{\texorpdfstring{Convergence~Test}{convtest}}}
\paragraph{{{\it Gelman--Rubin Statistic} 
 \texorpdfstring{$\left(\widehat{R} \right) $}{gelmanrubin}}}
{ To assess the convergence of multiple MCMC chains, we use the Gelman--Rubin criterion $\left(\widehat{R} \right)$, which compares the within-chain variance $\left(\widehat{W} \right)$ and the between-chain variance $(B)$ to estimate the total variance $\left(\widehat{V} \right)$. These quantities are related by
\begin{equation}
    \widehat{R}=\sqrt{\frac{\widehat{V}}{ \widehat{W}}}.
\end{equation}
If $\widehat{R} \approx 1$, it indicates convergence, suggesting that the chains are sampling from the same distribution. On~the other hand, if~$\widehat{R} > 1$, it suggests non-convergence, and~values significantly above 1 (e.g., $\widehat{R} > 1.1$) indicate that more iterations are needed to ensure proper convergence.}
\paragraph{{{\it Convergence of Chains and Trace Plots}}}
{In addition to the Gelman--Rubin statistic, trace plots are used to visualize whether the chains have converged. These plots show the evolution of each parameter over time, helping to verify if the chains are evolving in the convergence region or still exploring different regions of the parameter space.}


\section{{Comparing Barthel--Randers, Barthel--Kropina and \boldmath{\texorpdfstring{\(\Lambda\)}{lambda}}CDM Models}}\label{sect4}
{After obtaining the mean parameter values from the MCMC simulations, we proceed to compare the predictions of the Barthel--Randers and Barthel--Kropina cosmological models with those of the standard \(\Lambda\)CDM model. This comparison is essential to assess how well the Barthel--Randers and Barthel--Kropina frameworks are consistent with the observed cosmic expansion history.} 
\subsection{{\texorpdfstring{Evolution of the Hubble Parameter, Hubble Residual, and~BAO Distance~Scales}{hubresid}}}
{In this subsection, we compare the Hubble parameter $H(z)$ predicted by three Barthel--Randers models and the Barthel--Kropina model with those of the standard $\Lambda$CDM model. The~$\Lambda$CDM model is given by the equation
$$
H_{\Lambda \text{CDM}}(z) = H_0 \sqrt{\Omega_{m0} (1 + z)^3 + \Omega_{\Lambda0}},
$$
where $H_0 = 67.8\, \mathrm{km\,s^{-1}\,Mpc^{-1}}$, $\Omega_{m0} = 0.3092$, and~$\Omega_{\Lambda0} = 0.608$. To~quantify the deviation between the Barthel--Randers models and \(\Lambda\)CDM, we define the Hubble residual as
\[
\Delta H(z) = H_{\text{Barthel-Randers Models}}(z) - H_{\Lambda\text{CDM}}(z),
\]
where \(H_{\text{Barthel-Randers Models}}(z)\) denotes the Hubble parameter predicted by the respective Barthel--Randers model. Similarly, for~the Barthel--Kropina model, we define
\[
\Delta H(z) = H_{\text{Barthel-Kropina Model}}(z) - H_{\Lambda\text{CDM}}(z),
\]
where \(H_{\text{Barthel-Kropina Model}}(z)\) denotes the Hubble parameter predicted by the respective Barthel--Kropina model. We plot both \(H(z)\) and \(\Delta H(z)\) for the Barthel--Randers models and Barthel--Kropina model. A~small or nearly constant residual indicates close agreement with the $\Lambda$CDM model, while significant deviations suggest potential extensions to the standard cosmological model. In~addition to analyzing the Hubble parameter, we also consider the BAO distance scales, specifically the ratio between the angle-averaged distance $D_V(z) = (z D_M^2 D_H)^{1/3}$ and the ratio of transverse to line-of-sight comoving distances, $FAP(z) = D_M / D_H$, which are scaled by $z^{-2/3}$ and $z^{-1}$, respectively. By~comparing these BAO distance ratios with the predictions of the best-fit flat $\Lambda$CDM model, we gain further insight into the behavior of the Barthel--Randers and Barthel--Kropina models in relation to the standard cosmological framework. The~results of these comparisons are visualized by plotting $H(z)$, $\Delta H(z)$, and~the BAO distance scales, providing a comprehensive view of how each model behaves relative to $\Lambda$CDM. We consider the CC data~\cite{Moresco2012,Moresco2015,Moresco2016} in our analysis, and~the BAO measurements are taken from~\cite{karim2025}.}


\subsection{Cosmographic Analysis of Barthel--Randers and Barthel--Kropina~Models}\label{sect5}
Cosmography provides a model-independent framework to describe the kinematic behavior of the Universe by expanding cosmological quantities in terms of redshift~\cite{Cosmographic1,Cosmographic2,Cosmographic3,Cosmographic4}. In~this work, we apply this approach to analyze the deceleration and jerk parameters within the Barthel--Randers and Barthel--Kropina cosmological models, which modify standard cosmic dynamics through different geometrical correction~terms.
\subsubsection*{Deceleration Parameter \texorpdfstring{\(q(z) \)}{qz} and Jerk Parameter \texorpdfstring{\(j(z) \)}{jz}}

The deceleration parameter \(q(z) \) describes the acceleration or deceleration of the cosmic expansion and is defined as
\[
q(z) = -\frac{1}{H^2(z)}\frac{dH(z)}{dz} - 1.
\]

A negative value of \(q(z)\) indicates that the Universe is undergoing accelerated expansion. Two important phases derived from \(q(z)\) are the present-day deceleration parameter, \(q_0 = q(z = 0)\), which describes the current rate of acceleration, and~the transition redshift, \(z_{\text{tr}}\), defined by the condition \(q(z_{\text{tr}}) = 0\), marking the point at which the Universe transitioned from decelerated to accelerated expansion. The~jerk parameter \(j(z) \) is a third-order cosmographic quantity defined as~\cite{jerksnap}
\[
j(z) = \frac{1}{H^3(z)}\frac{d^2H(z)}{dz^2},
\]
which characterizes the rate of change of the acceleration. Importantly, in~the standard \(\Lambda\)CDM model, the~jerk parameter remains constant at $j(z) = 1.$ Any deviation from this value can signal dynamics beyond the standard cosmological model. For~the three Barthel--Randers models and the Barthel--Kropina model, we evaluate both the deceleration parameter \(q(z)\) and the jerk parameter \(j(z)\) numerically and assess their behavior across the redshift range.

\subsection{Statistical Assessment of Barthel--Randers and Barthel--Kropina~Models}\label{sect6}
In this subsection, we apply a set of model selection criteria to evaluate the goodness of fit and model complexity of three variants of the Barthel--Randers cosmological models (linear, logarithmic  and exponential) 
 and Barthel--Kropina model. It is crucial to assess how well each model aligns with the data when compared to the \(\Lambda\)CDM~model.
\subsection{Goodness of~Fit}
{The primary measure of model fit is the chi-squared statistic, \(\chi^2\), which quantifies the discrepancy between theoretical predictions and observational data. For~a suitable comparison, as~models generally have different numbers of parameters, we compute the reduced chi-square}
\begin{equation}
\chi^2_{\text{red}} = \frac{\chi^2_{\text{tot}}}{\text{DOF}},
\end{equation}
where \(\chi^2_{\text{tot}}\) is the total chi-squared value and DOF denotes the degrees of freedom, calculated as the number of data points minus the number of fitted parameters. A~value of \(\chi^2_{\text{red}} \approx 1\) indicates a good statistical fit~\cite{ASM1}. Lower values can suggest overfitting, while significantly higher values may reflect poor model~performance.

\subsubsection{Model Comparison Using AIC and~BIC}
In addition to chi-squared statistics, we apply information criteria that account for both fit quality and model complexity. These include the Akaike information criterion (AIC) and the Bayesian information criterion (BIC), defined respectively as~\cite{Liddle,AIC1,AIC2,AIC3,BIC1}
\begin{align}
\text{AIC} &= -2 \log \mathcal{L}_{tot, max} + 2\mathcal{P}, \\
\text{BIC} &= -2 \log \mathcal{L}_{tot, max} + \mathcal{P} \ln(\mathcal{N}),
\end{align}
where $\mathcal{P}$ is the number of free parameters, $\mathcal{N}$ is the total number of data points used in the analysis and $\mathcal{L}_{tot,max}$ is the total maximum likelihood. In~our analysis, the~total number of data points used is the sum of three different observational datasets: {$\mathcal{N} = 1729$.} For the model comparison, we note that the \(\Lambda\)CDM model has four free parameters, whereas the Barthel--Randers cosmological model has five free parameters, and~the Barthel--Kropina model has six free parameters. These criteria penalize overly complex models to prevent overfitting. BIC generally applies a stricter penalty than AIC, especially in large~datasets.

\subsubsection{Relative Comparison: \texorpdfstring{\(\Delta\)}{daic}AIC and \texorpdfstring{\(\Delta\)}{dbic}BIC}
To directly compare the Barthel--Randers models with the \(\Lambda\)CDM model, we compute
\begin{align}
\Delta \text{AIC} &= \text{AIC}_{\text{Barthel-Randers Models}} - \text{AIC}_{\Lambda \text{CDM Model}}, \\
\Delta \text{BIC} &= \text{BIC}_{\text{Barthel-Randers Models}} - \text{BIC}_{\Lambda \text{CDM Model}}.
\end{align}
Similarly,  
 for~the comparison between the Barthel--Kropina model and the \(\Lambda\)CDM model, we compute
\begin{align}
\Delta \text{AIC} &= \text{AIC}_{\text{Barthel-Kropina Model}} - \text{AIC}_{\Lambda \text{CDM Model}}, \\
\Delta \text{BIC} &= \text{BIC}_{\text{Barthel-Kropina Model}} - \text{BIC}_{\Lambda \text{CDM Model}}.
\end{align}
According 
 to the calibrated Jeffreys' scales~\cite{JP1}, these differences offer insight into model~preference:
\begin{itemize}
  \item \(|\Delta \text{AIC}| \leq 2\): Comparable support.
  \item \(4 \leq |\Delta \text{AIC}| < 10\): Considerably less support.
  \item \(|\Delta \text{AIC}| \geq 10\): Strongly disfavored.
  \item \(|\Delta \text{BIC}| \leq 2\): Weak evidence against the model.
  \item \(2 < |\Delta \text{BIC}| \leq 6\): Moderate evidence against the model.
  \item \(|\Delta \text{BIC}| > 6\): Strong evidence against the model.
\end{itemize}

{A negative value of \(\Delta \text{AIC}\) or \(\Delta \text{BIC}\) indicates that the Barthel--Randers and Barthel--Kropina models are preferred over \(\Lambda\)CDM, whereas positive values indicate that these models are less favored}

\subsubsection{\emph{p}-Value~Statistics}
{In our analysis, we also  computed the \emph{p}-value to assess the statistical significance of the fit for the studied cosmological} models, calculated as follows:
\[
p = 1 - \mathcal{F}_{\chi^2_{\text{min}}}(\chi \mid \nu),
\]
where \(\mathcal{F}_{\chi^2_{\text{min}}}(\chi \mid \nu)\) is the cumulative distribution function (CDF) of the chi-squared distribution with \(\nu\) degrees of freedom, and~\(\nu\) represents the number of data points minus the number of free parameters~\cite{Andrade2019}. 

{The \emph{p}-value quantifies the likelihood of observing results as extreme as those seen in the data, assuming the null hypothesis holds true. A~\emph{p}-value smaller than 0.05 (\(p < 0.05\)) indicates statistical significance, providing strong evidence against the null hypothesis and suggesting that the model is a good fit for the data. These corresponding statistical metrics (\(\chi^2_{\text{min}}\), \(\chi^2_{\text{red}}\), AIC, BIC, \(\Delta\)AIC, \(\Delta\)BIC, and~\emph{p}-value) are computed and compared with those of the \(\Lambda\)CDM model. These statistical metrics ensure that both the goodness of fit and the complexity of each model are thoroughly evaluated, enabling a comprehensive statistical assessment of the Barthel--Randers and Barthel--Kropina models in the context of cosmology.}

\section{{Summary and Discussion of the~Results}}\label{sect7}

{In the present section, we discuss the main results obtained by confronting the proposed cosmological models with observations, and~by comparing their predictions and statistical significance with those of the $\Lambda$CDM model}

\subsection{\texorpdfstring{{MCMC and Convergence~Results}}{convergence}}
{Figure~\ref{fig_1aaa} shows the corner plot of the parameters of the Barthel--Randers and Barthel--Kropina models. The~diagonal panels illustrate the one-dimensional marginalized posterior distributions for each parameter, highlighting their most probable values. The~off-diagonal panels depict the two-dimensional marginalized contour plots, showing the 68\% and 95\% confidence intervals. These contour plots also reveal correlations or degeneracies between each model parameter. Figure~\ref{fig_2bbb} shows the corresponding trace plots, which depict the convergence behavior of the Markov chains for each parameter.} {As a convergence diagnostic, the~Gelman–Rubin statistic was used for each model parameter. The~results obtained are given by
$$
\text{BR}_{\text{Linear Case}}^{\text{Gelman-Rubin statistics}} \{H_0, \Omega_{mo}, b_1,  \mathcal{M}, r_d\} = \{1.006, 1.001, 1.001, 1.006, 1.006\},
$$
$$
\text{BR}_{\text{Logarithmic Case}}^{\text{Gelman-Rubin statistics}} \{H_0, \Omega_{mo}, b_1,  \mathcal{M}, r_d\} = \{1.005, 1.002, 1.001, 1.005, 1.005\},
$$
$$
\text{BR}_{\text{Exponential Case}}^{\text{Gelman-Rubin statistics}} \{H_0, \Omega_{mo}, b_1,  \mathcal{M}, r_d\} = \{1.004, 1.002, 1.001, 1.004, 1.004\},
$$
$$
\text{BK}^{\text{Gelman-Rubin statistics}} \{H_0, f_0, u_0, \omega, \mathcal{M}, r_d\} = \{1.007, 1.001, 1.002, 1.002, 1.007 1.007\}.
$$

}

In our case, the~$\widehat{R}$ values for the Barthel--Randers (linear, logarithmic, and~exponential) and Barthel--Kropina models are all very close to 1, indicating excellent convergence of the MCMC chains. However, it is important to note that in the BR models, $\Omega_{\Lambda0}$ and $\Omega_{r0}$ are not independent parameters. Instead, they are determined algebraically once the other parameters constrained through the MCMC analysis are known. For~this reason, we do not list $\Omega_{\Lambda0}$ and $\Omega_{r0}$ among the constrained parameters, as~they are derived quantities rather than directly fitted~ones.

One can observe that the predicted values of $H_0$ from the $\Lambda$CDM and Barthel--Randers models (linear, logarithmic, and~exponential cases), as~well as from the osculating Barthel--Kropina dark energy model, are in close agreement with those obtained from {Moresco's results. Specifically, Moresco's measurements, considering both the full covariance matrix and without systematics, are as follows: Without systematics: $H_0 = 66.2^{+3.8}_{-3.9}\, \mathrm{km\,s^{-1}\,Mpc^{-1}}$. With~systematics, we get $H_0 = 66.2^{+5.5}_{-5.6}\, \mathrm{km\,s^{-1}\,Mpc^{-1}}$. While the central value of our predicted $H_0$ is close to the estimates from Planck, DESI DR1, and~DESI DR2, the~associated uncertainty is notably larger.}

\begin{figure}[H]
 
\begin{tabular}{cc}
  \begin{subfigure}{0.44\textwidth}
    \includegraphics[width=\linewidth]{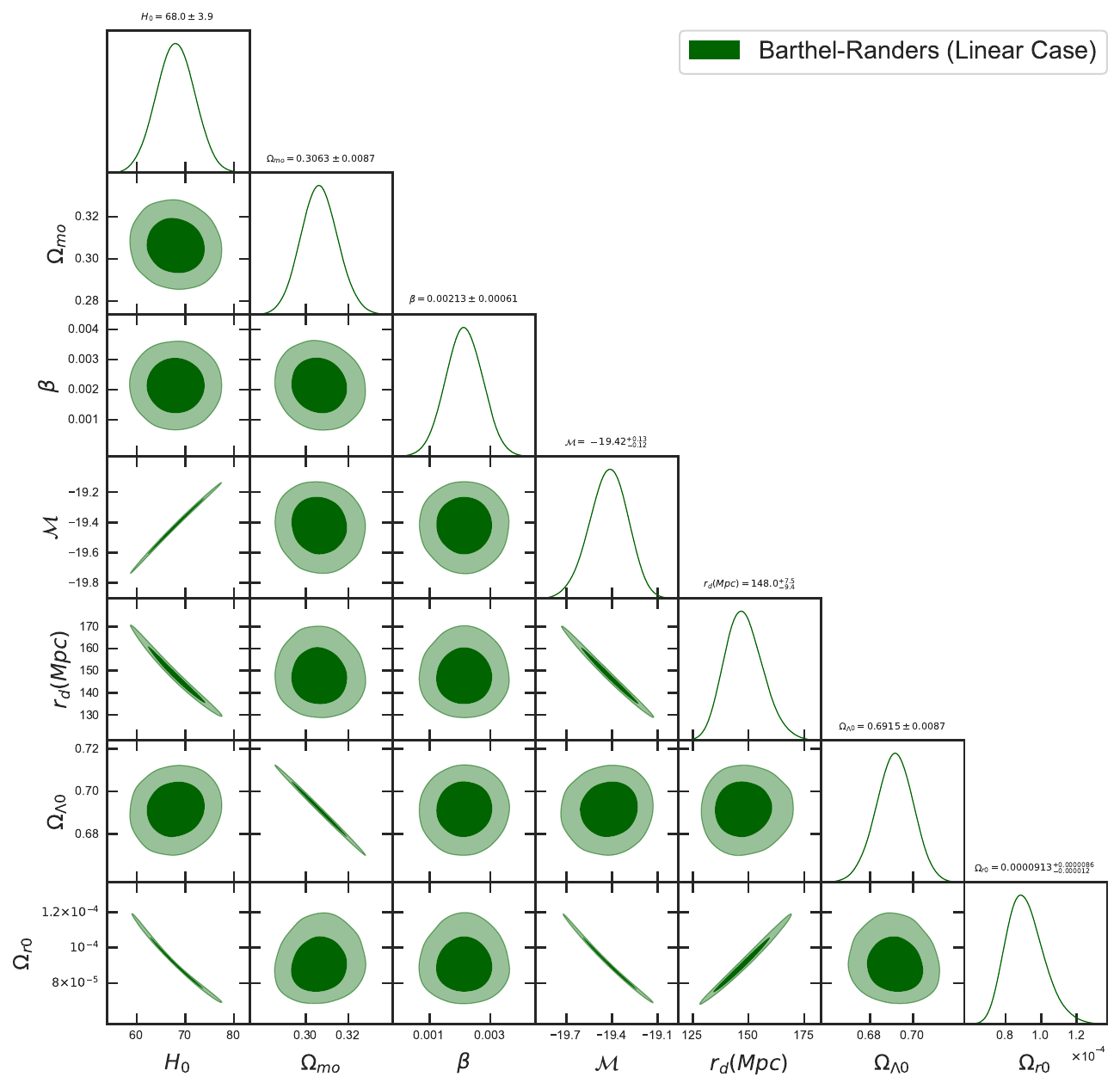}
  \end{subfigure} &
  \begin{subfigure}{0.44\textwidth}
    \includegraphics[width=\linewidth]{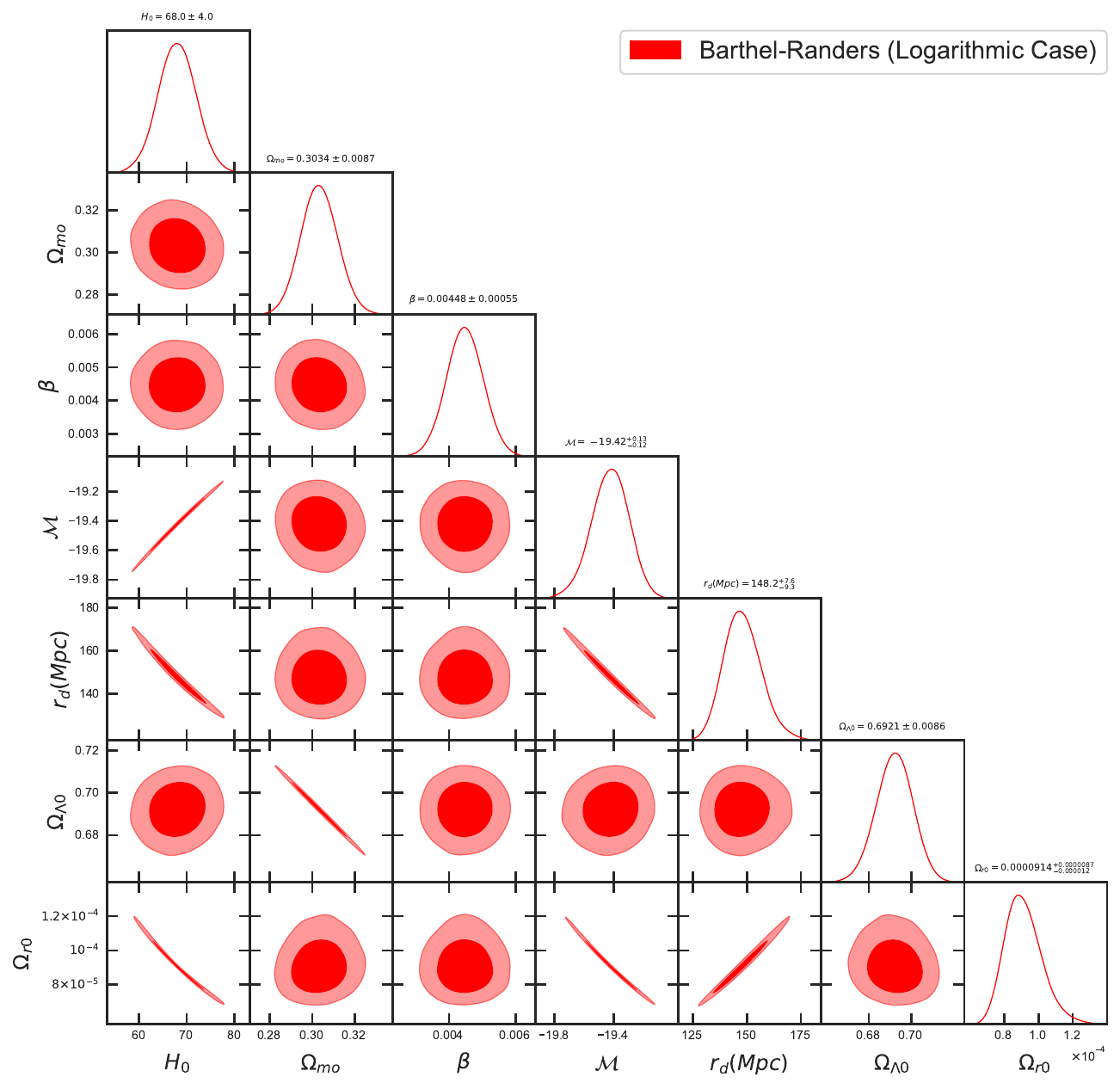}
  \end{subfigure} \\
  \begin{subfigure}{0.44\textwidth}
    \includegraphics[width=\linewidth]{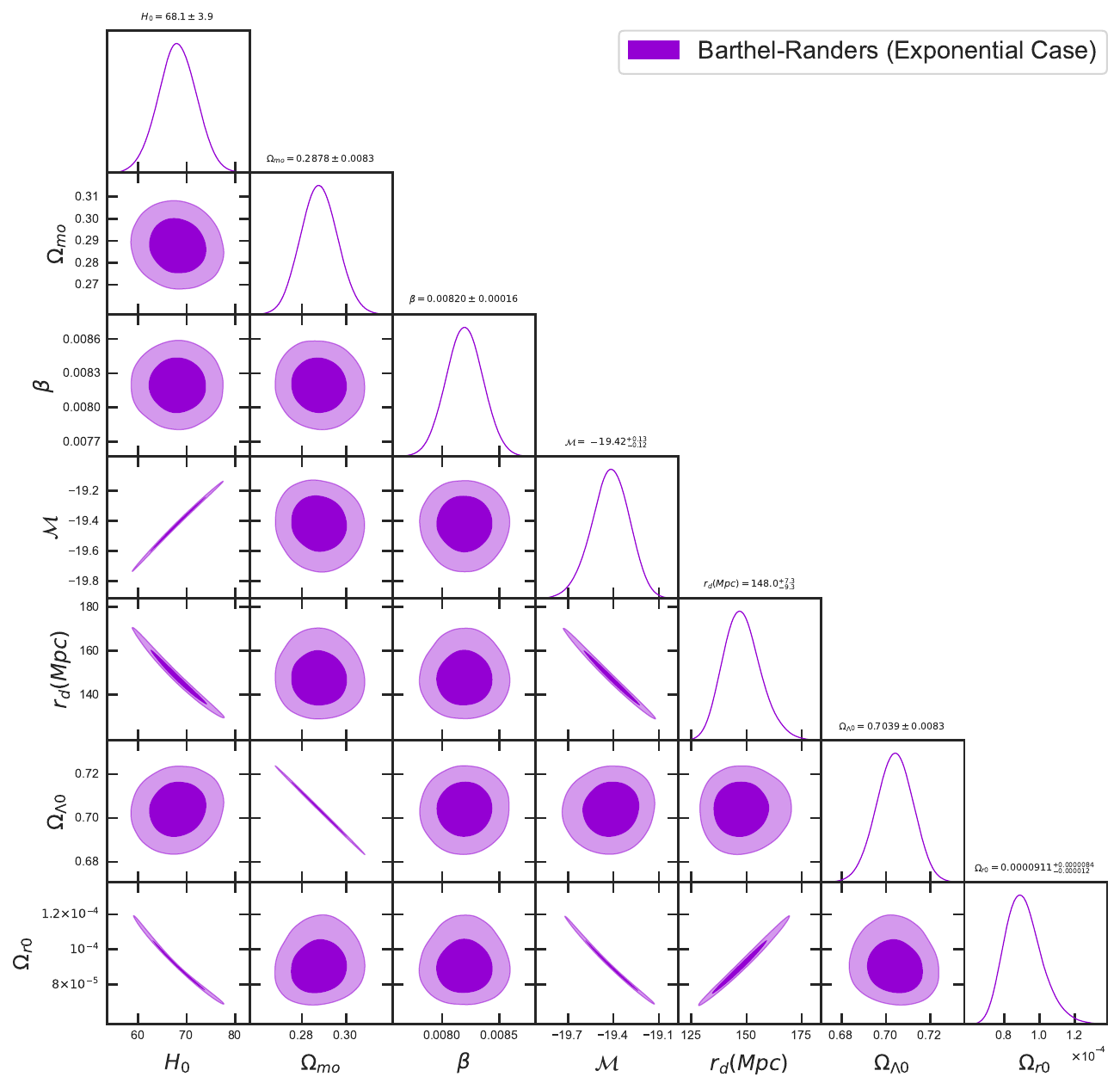}
  \end{subfigure} &
  \begin{subfigure}{0.44\textwidth}
    \includegraphics[width=\linewidth]{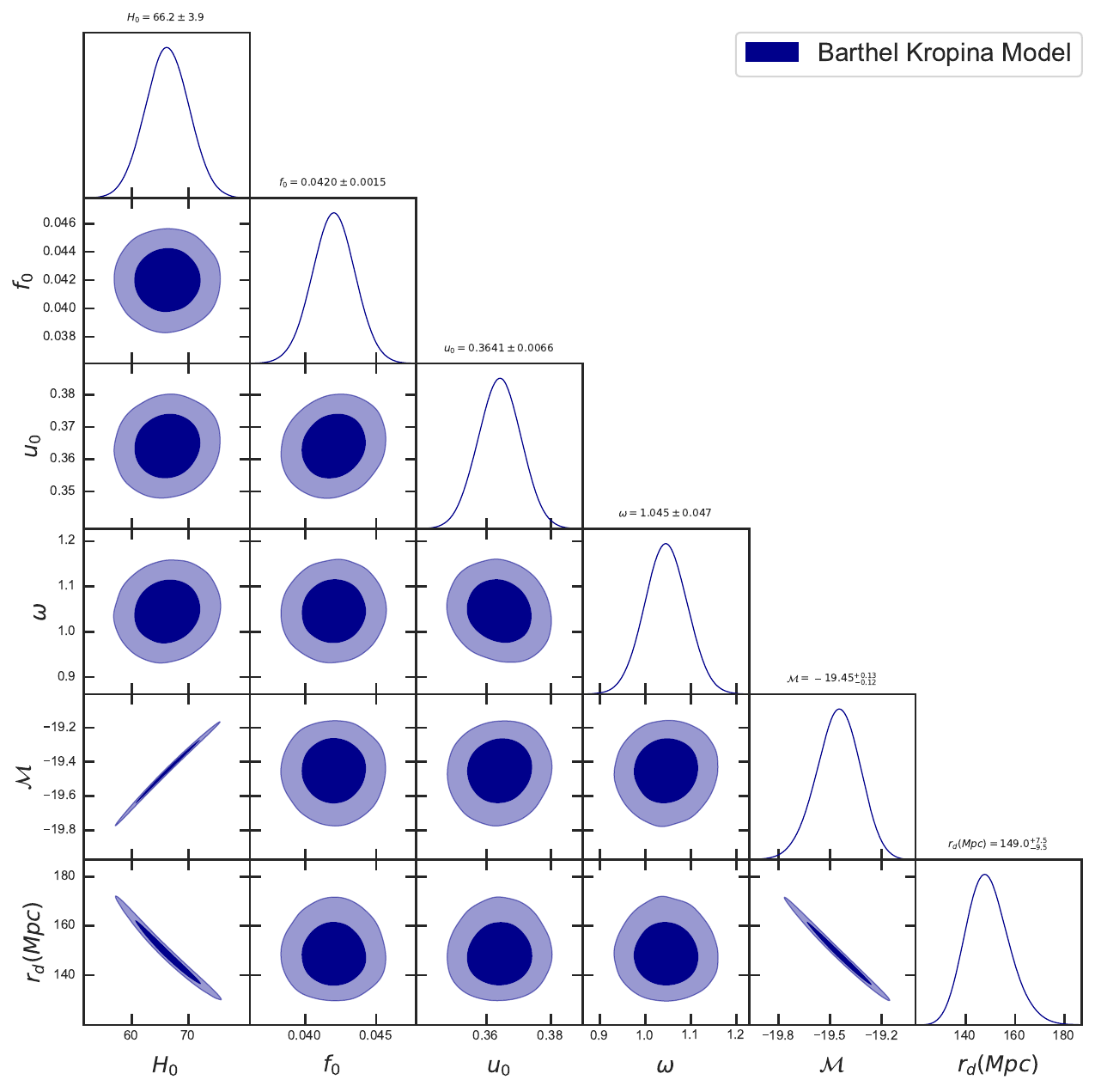}
  \end{subfigure} \\
\end{tabular}
\caption{{The figure 
 shows the parameter constraints of the Barthel--Randers and Barthel--Kropina models using the CC + SNe Ia and BAO datasets together, and displaying both $1\sigma$ and $2\sigma$ confidence intervals. The~contours show the correlations between these parameters, with~marginalized probability distributions along the diagonal.}}
\label{fig_1aaa}
\end{figure}
 

{A similar trend is observed for $r_d$, where the predicted central values across all models are close to the Planck estimation, $r_d^{\text{Planck}} = (147.09 \pm 0.26)\ \mathrm{Mpc}$, although~the associated uncertainty is also noticeably larger.  The~primary reason for the higher uncertainty is the inclusion of the full cosmic covariance matrix, which accounts for uncertainties in several factors, including the estimation of stellar metallicity, star formation history, the~adopted initial mass function (IMF), the~stellar library used, the~stellar population synthesis model, and~potential contributions from any residual young components in the galaxy spectra.}

{Table~\ref{tab_2} presents the mean values at corresponding 68\% confidence intervals for the model parameters of the Barthel--Randers models (linear, logarithmic, and exponential cases), as~well as for the osculating Barthel--Kropina dark energy model.} The predicted value of the present-day matter density parameter in the case of the $\Lambda$CDM model, as~well as the linear case of the Barthel--Randers model, is lower than the Planck estimation, $\Omega_{m0}^{\text{Planck}} = 0.315 \pm 0.007,$  
and is closer to the values predicted by DESI DR1 and DESI DR2, $\Omega_{m0}^{\text{DESI DR1}} = 0.295 \pm 0.015 \quad \text{and} \quad \Omega_{m0}^{\text{DESI DR2}} = 0.2975 \pm 0.0086.$

Similarly, the~predicted value of \(\Omega_{m0}\) in the logarithmic case of the Barthel--Randers model is also in good agreement with the DESI DR1 and DR2 estimations.  
In contrast, the~exponential case of the Barthel--Randers model predicts a matter density parameter that is lower than the values reported by both DESI DR1 and DR2. It is interesting to observe that the $\Lambda$CDM model, as~well as the linear, logarithmic, and exponential cases of the Barthel--Randers model, predict a present-day dark energy density parameter, \(\Omega_{\Lambda0}\), that is greater than the value predicted by the Planck estimation: $\Omega_{\Lambda0}^{\text{Planck}} = 0.685 \pm 0.007.$

\begin{figure}[H]
 
\begin{tabular}{cc}
  \begin{subfigure}{0.46\textwidth}
    \includegraphics[width=\linewidth]{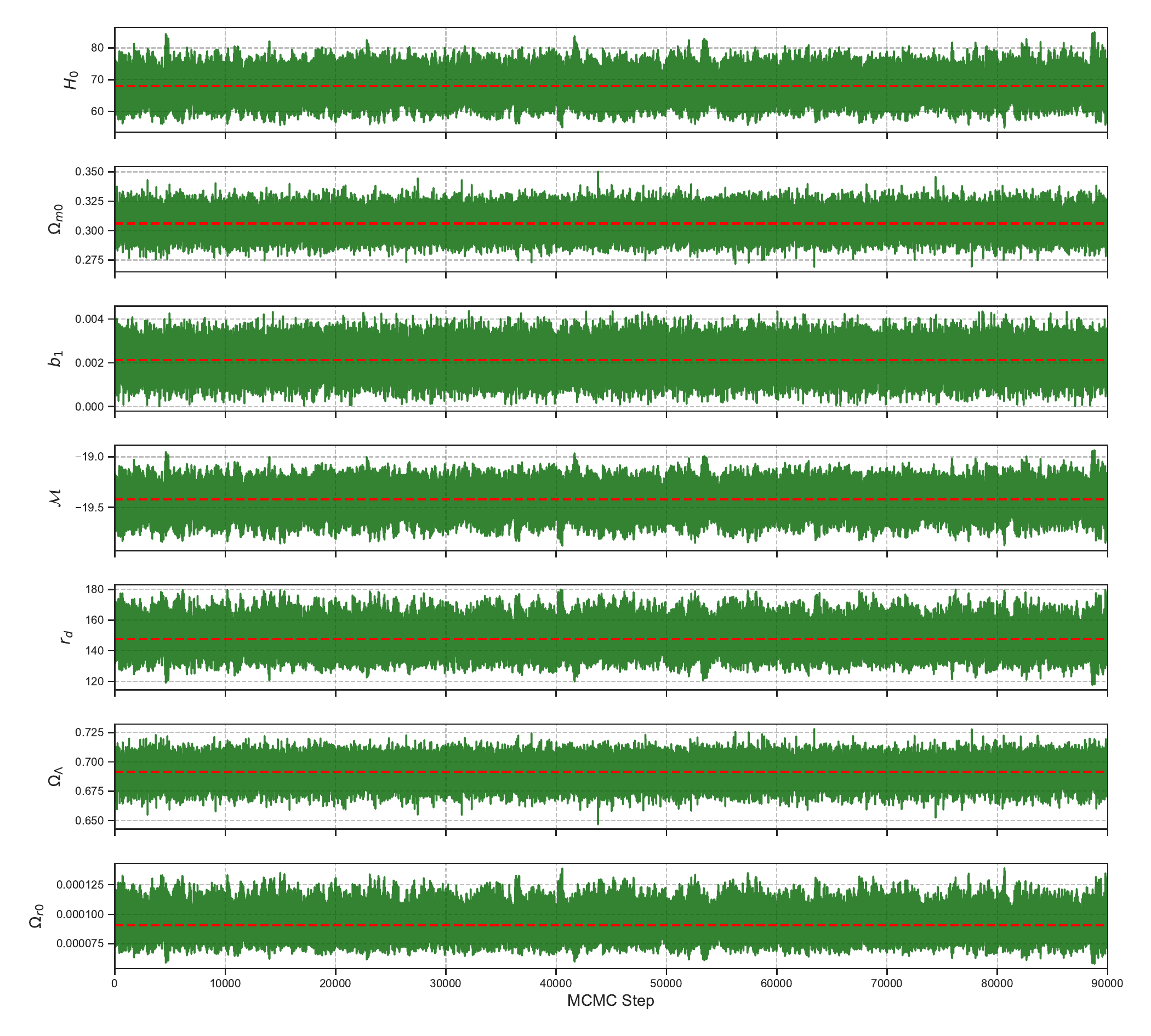}
  \end{subfigure} &
  \begin{subfigure}{0.46\textwidth}
    \includegraphics[width=\linewidth]{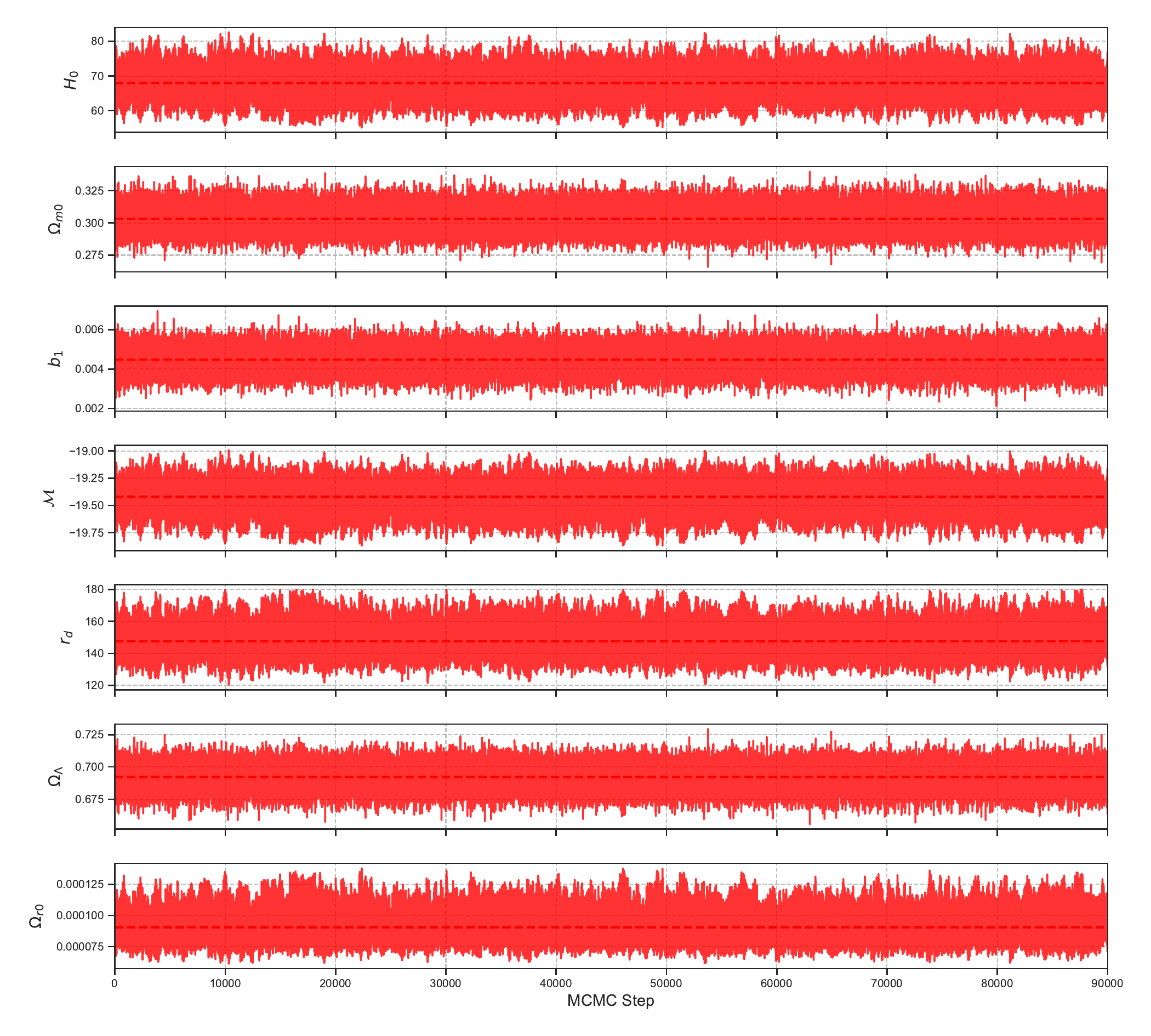}
  \end{subfigure} \\

  \begin{subfigure}{0.46\textwidth}
    \includegraphics[width=\linewidth]{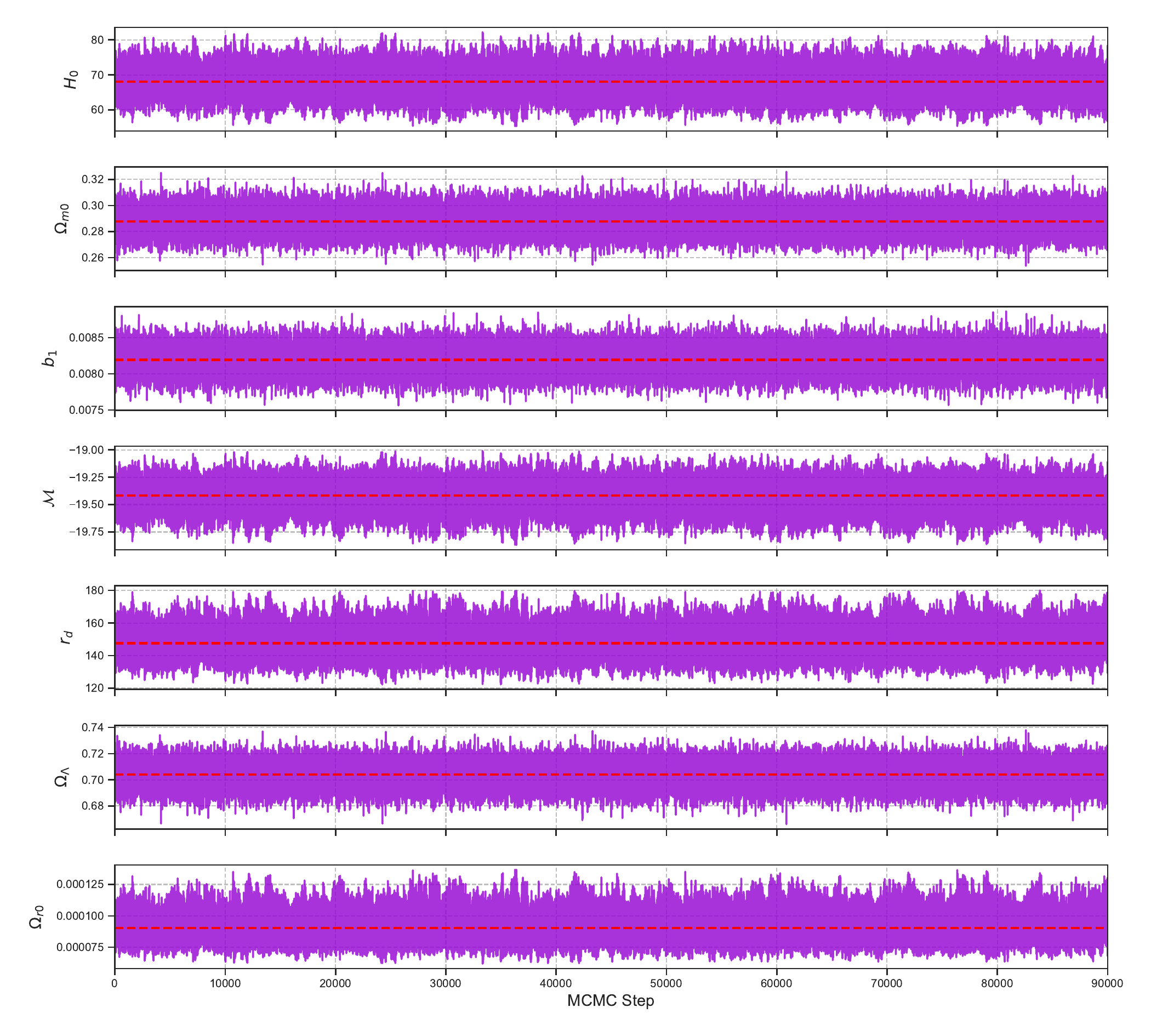}
  \end{subfigure} &
  \begin{subfigure}{0.46\textwidth}
    \includegraphics[width=\linewidth]{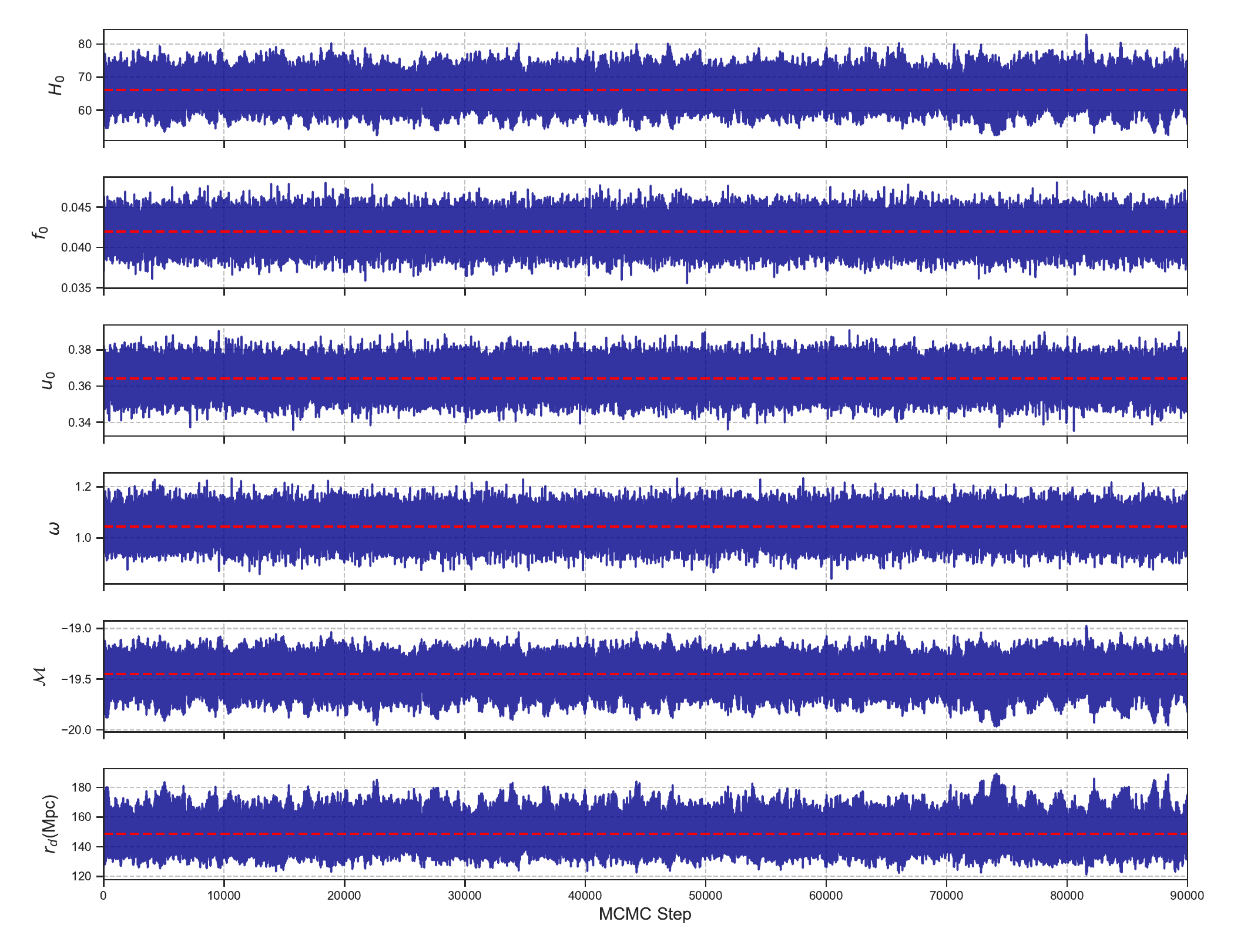}
  \end{subfigure} \\
\end{tabular}
\caption{{The figure 
 illustrates the trace plots, which show the convergence behavior of the Markov chains for each parameter in the Barthel--Randers and Barthel--Kropina cosmological models.}}
\label{fig_2bbb}
\end{figure}

\vspace{-9pt}

\begin{table}[H]
\caption{Summary of the mean values and {68\% confidence intervals (1$\sigma$)} 
	for the parameters of the $\Lambda$CDM, Barthel--Randers, and~Barthel--Kropina cosmological~models.}\label{tab_2}
\renewcommand{\arraystretch}{1.2} 
\setlength{\tabcolsep}{10pt} 
{
\begin{tabularx}{\textwidth}{cCC}
    \midrule
    \textbf{Cosmological Models} & \textbf{Parameter} & \textbf{JOINT} \\
    \midrule
    \multirow{5}{*}{$\Lambda$CDM Model} 
     & $H_0$ & $67.8 \pm 3.7$ \\
     & $\Omega_{m0}$ & $0.3092 \pm 0.0086$ \\
     & $\mathcal{M}$ & $-19.43 \pm 0.12$ \\
     & $r_d$ & $148.5 \pm 7.5$ \\
     & $\Omega_{\Lambda0}$ & $0.6908 \pm 0.0086$ \\
    \midrule
    \multirow{7}{*}{BR (Linear Case)}
    & $H_0$ & $68.1 \pm 4.0$ \\
    & $\Omega_{m0}$ & $0.3064 \pm 0.0087$ \\
    & $\beta$ & $0.00213 \pm 0.00061$ \\
    & $\mathcal{M}$ & $-19.42 \pm 0.12$ \\
    & $r_d$ & $147.9 \pm 7.6$ \\
    & $\Omega_{\Lambda0}$ & $0.6914 \pm 0.0087$ \\
    & $\Omega_{r0}$ & $(9.11 \pm 1.20) \times 10^{-5}$ \\

\bottomrule
\end{tabularx}}
\end{table}
\unskip
\begin{table}[H]\ContinuedFloat
\caption{\textit{Cont.}}
{\begin{tabularx}{\textwidth}{cCC}
    \midrule
    \textbf{Cosmological Models} & \textbf{Parameter} & \textbf{JOINT} \\
    \midrule

    \multirow{7}{*}{BR (Logarithmic Case)}
    & $H_0$ & $68.9 \pm 3.9$ \\
    & $\Omega_{m0}$ & $0.3034 \pm 0.0087$ \\
    & $\beta$ & $0.00447 \pm 0.00055$ \\
    & $\mathcal{M}$ & $-19.42 \pm 0.12$ \\
    & $r_d$ & $148.1 \pm 7.4$ \\
    & $\Omega_{\Lambda0}$ & $0.6921 \pm 0.0086$ \\
    & $\Omega_{r0}$ & $(9.13 \pm 1.20) \times 10^{-5}$ \\
    \midrule

    \multirow{7}{*}{BR (Exponential Case)}
    & $H_0$ & $68.2 \pm 4.0$ \\
    & $\Omega_{m0}$ & $0.2878 \pm 0.0083$ \\
    & $\beta$ & $0.00820 \pm 0.00016$ \\
    & $\mathcal{M}$ & $-19.42 \pm 0.12$ \\
    & $r_d$ & $147.9 \pm 7.4$ \\
    & $\Omega_{\Lambda0}$ & $0.7040 \pm 0.0083$ \\
    & $\Omega_{r0}$ & $(9.00 \pm 1.20) \times 10^{-5}$ \\
    \midrule
    \multirow{6}{*}{Barthel--Kropina Model}
    & $H_0$ & $66.2 \pm 3.9$ \\
    & $f_0$ & $0.0420 \pm 0.0015$ \\
    & $u_0$ & $0.3641 \pm 0.0066$ \\
    & $\omega$ & $1.045 \pm 0.047$ \\
    & $\mathcal{M}$ & $-19.45 \pm 0.12$ \\
    & $r_d$ & $149.0 \pm 7.5$ \\
    \midrule
\end{tabularx}
}
\end{table}
\unskip

\subsection{\texorpdfstring{{Hubble Parameter, Hubble Residual, and~BAO Distance Scale~Results}}{Hubbleparam}}
Figure~\ref{fig_3} shows the evolution of the Hubble function and the Hubble residuals for the linear, logarithmic, and exponential cases of the Barthel--Randers model, as~well as the Barthel--Kropina model. These are compared with the predictions from the $\Lambda$CDM model and CC measurements. The~left panel shows that at high redshift ($z > 0.7$), all the considered models deviate noticeably from the $\Lambda$CDM model. However, these deviations are not significant. At~low redshift ($z < 0.7$), each model aligns closely with the $\Lambda$CDM~predictions.

\begin{figure}[H]
\begin{subfigure}{.48\textwidth}
\includegraphics[width=\linewidth]{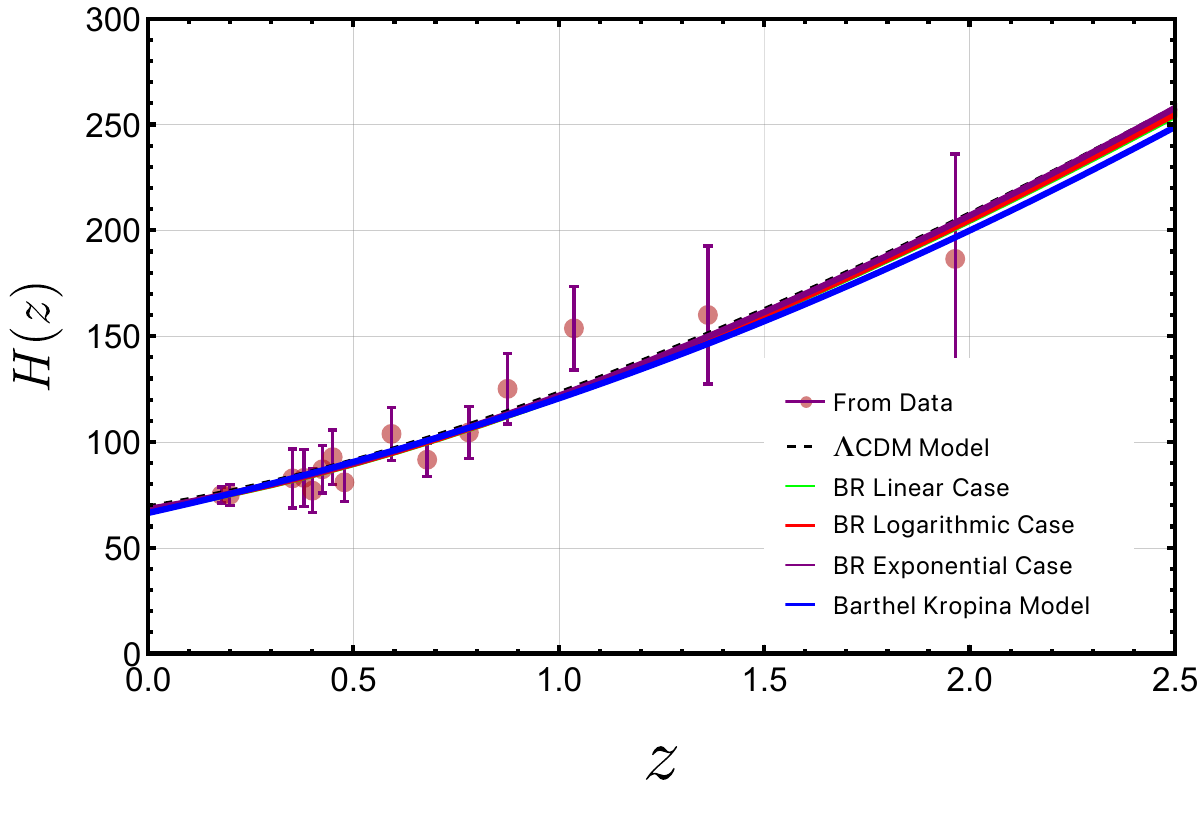}
\end{subfigure}
\hfil
\begin{subfigure}{.48\textwidth}
\includegraphics[width=\linewidth]{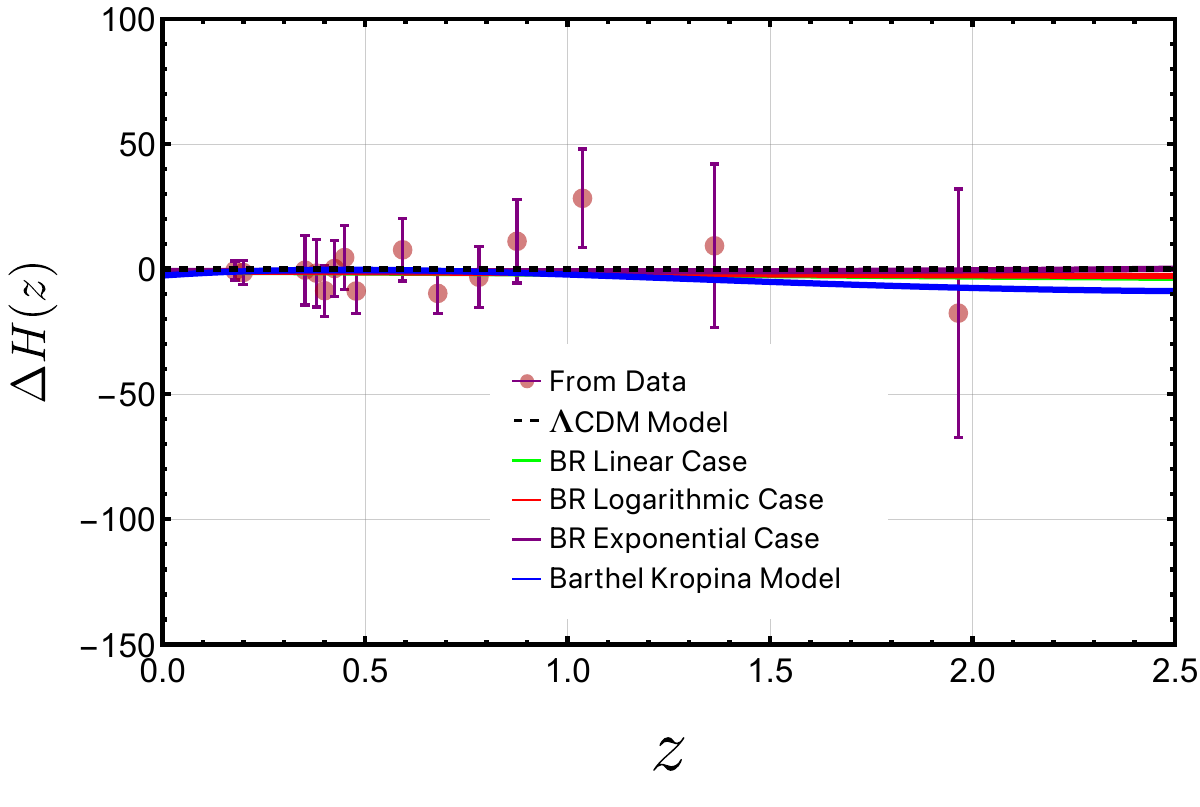}
\end{subfigure}
\caption{{Comparison of the Barthel--Randers (linear, logarithmic, and exponential) and~Barthel--Kropina models, with~the \(\Lambda\)CDM model {and CC measurements}. The~left panel shows the evolution of the Hubble parameter \(H(z)\) using mean MCMC values. The~right panel shows the Hubble residuals \(\Delta H(z)\), indicating deviations from the standard model.}}\label{fig_3}
\end{figure}

On the right plane, we can also observe similar behavior in the Hubble residual plots, where each model shows a slight deviation from the $\Lambda$CDM model at high redshift. However, at~low redshift, all models exhibit behavior that closely matches the $\Lambda$CDM predictions. All considered models (Barthel--Randers variants and Barthel--Kropina) closely follow $\Lambda$CDM at low redshifts, indicating consistency with current observations. Their mild deviations at higher redshifts ($z > 0.7$) suggest a potentially different early Universe expansion history, offering alternative insights that may help address cosmological tensions such as the Hubble~tension.

{Figure~\ref{fig_4} shows the evolution of different BAO distance scales. All three Barthel--Randers variants exhibit minimal deviation in both the volume-averaged distance $D_V / (r_d z^{2/3})$ and the transverse distance measure $D_M / (z D_H)$, indicating strong alignment with cosmological observations. The~logarithmic and exponential BR models closely match $\Lambda$CDM at low to intermediate redshifts ($z < 1.5$), though~some discrepancies appear beyond $z \approx 1.5$, particularly with the exponential model diverging in the $D_V$ observable range 
	and both models showing reduced accuracy for high-redshift QSO and Ly-$\alpha$ data. The~Barthel--Kropina model displays a distinct profile, with~the $D_V / (r_d z^{2/3})$ curve generally below the $\Lambda$CDM prediction, especially in the mid-redshift range ($z \sim 0.5-1.2$). Despite this, it remains within observational error for $z < 1$. In~terms of $D_M / (z D_H)$, the~Barthel--Kropina model performs similarly to $\Lambda$CDM but shows more pronounced deviations in residuals, particularly around $z \sim 1$, indicating less precise modeling in this region.}

\begin{figure}[H]
\centering
\includegraphics[width=\linewidth]{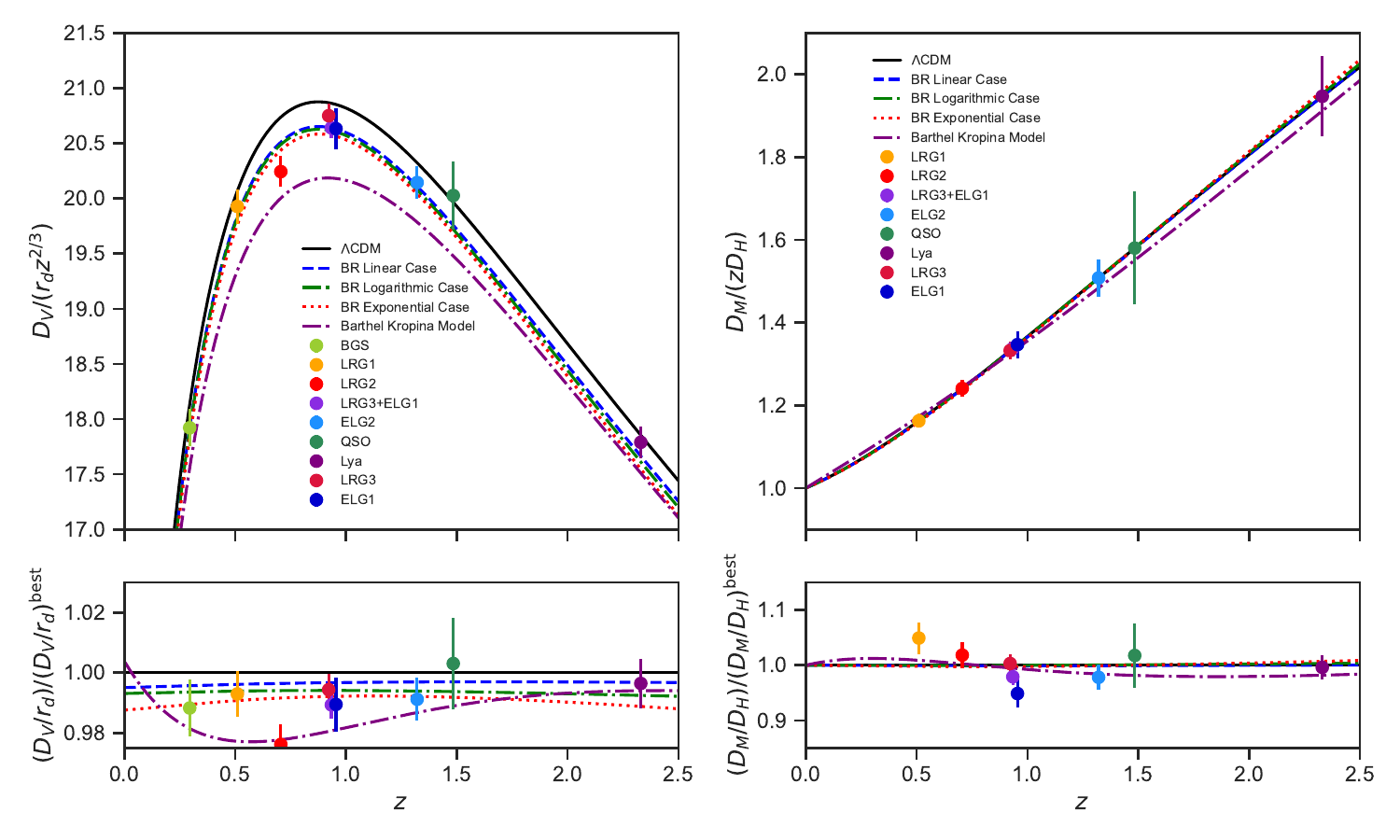} 
\caption{{Comparison of the Barthel--Randers (linear, logarithmic, and exponential) and~Barthel--Kropina models, with~the $\Lambda$CDM model across various distance measures, with~data from tracers such as LRG, QSO, and~Lya. Measurement uncertainties are represented by error bars. In~the left panel, the~evolution of the ratio between the angle-averaged distance $D_V \equiv (z D_M^2 D_H)^{1/3}$ and the sound horizon at the baryon drag epoch, $r_d$, is shown. The~right panel illustrates the ratio of transverse to line-of-sight comoving distances $FAP \equiv D_M / D_H$. To~enhance clarity and compress the dynamic range, an~arbitrary scaling of $z^{-2/3}$ has been applied to the left panel, and~$z^{-1}$ to the right. The~bottom row presents the same models and data points as the top row, but~now scaled as the ratio relative to the best-fit flat $\Lambda$CDM model predictions.}}\label{fig_4}
\end{figure}
\unskip
\subsection{Cosmographic~Results}
Figure~\ref{fig_5} shows the evolution of the cosmography parameters for the linear, logarithmic, and exponential cases of the Barthel--Randers model, along with the Barthel--Kropina model, compared to the $\Lambda$CDM model. The~left panel shows the evolution of the deceleration parameter, $q(z)$.

\begin{figure}[H]
\begin{subfigure}{.48\textwidth}
\includegraphics[width=\linewidth]{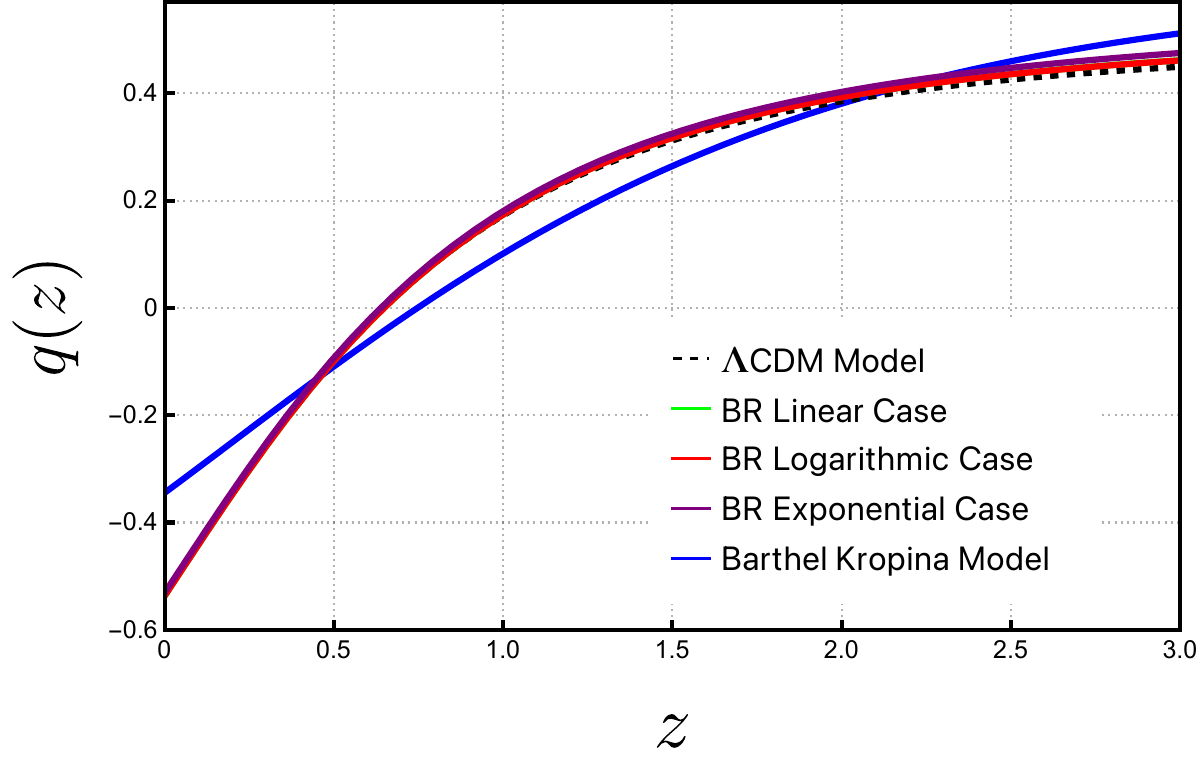}
\end{subfigure}
\hfil
\begin{subfigure}{.48\textwidth}
\includegraphics[width=\linewidth]{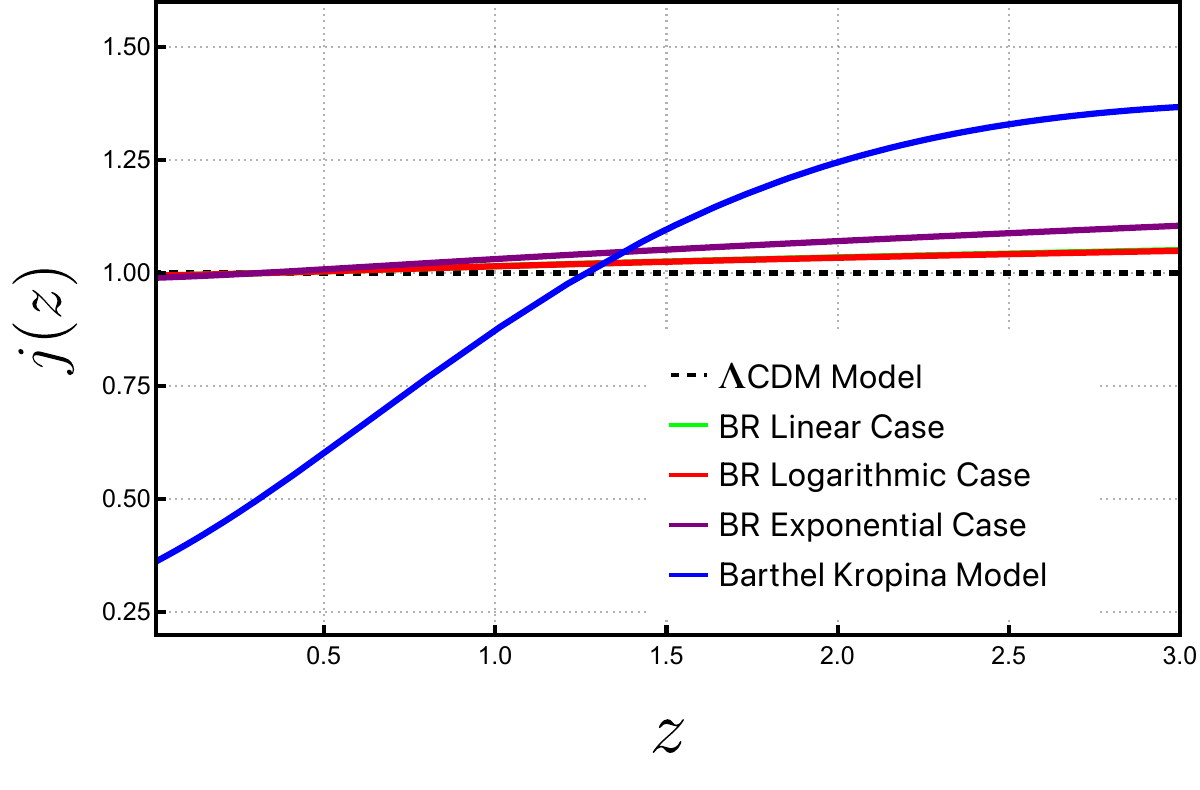}
\end{subfigure}
\caption{Evolution of the deceleration parameter \(q(z)\) and jerk parameter \(j(z)\) for the Barthel--Randers (linear, logarithmic, and exponential) and~Barthel--Kropina models, and~of the  \(\Lambda\)CDM model. The~left panel shows the evolution of \(q(z)\) as a function of redshift, while the right panel presents \(j(z)\), showing deviations from the \(\Lambda\)CDM value \(j = 1\).}\label{fig_5}
\end{figure}

{Notably, the~Barthel--Kropina model exhibits a distinct evolution compared to both the standard $\Lambda$CDM model and the Barthel--Randers variants. At~the present time ($z = 0$), all Barthel--Randers variants and $\Lambda$CDM predict a deceleration parameter of approximately $q_0 \approx -0.511$, while the Barthel--Kropina model predicts a less negative value of $q_0 \approx -0.366$. We also analyze an important cosmological feature: the transition redshift \(z_{\text{tr}} \). 

For the $\Lambda$CDM model, the~transition from decelerated to accelerated expansion occurs at $z_{\text{tr}}^{\Lambda \text{CDM}} \approx 0.623.$  
The Barthel--Randers variants exhibit slightly later transitions: $z_{\text{tr}}^{\text{BR Linear}} \approx 0.610,\quad z_{\text{tr}}^{\text{BR Logarithmic}} \approx 0.607,\quad \text{and} \quad z_{\text{tr}}^{\text{BR Exponential}} \approx 0.606.$ }

{In contrast, the~Barthel--Kropina model shows a significantly earlier transition at $z_{\text{tr}}^{\text{Barthel-Kropina}} \approx 0.723.$ While at high redshift, all models show deviations from the $\Lambda$CDM model in their evolution, these deviations are not significant.} The right panel shows the evolution of the jerk parameter for each model. At~the present time ($z = 0$), all variants of the Barthel--Randers models exhibit good agreement with the $\Lambda$CDM model. Although~each variant predicts a slightly different value for the jerk parameter, these deviations are not significant. {Moreover, the~Barthel--Kropina model shows a substantial deviation from the other models, predicting a present-day jerk value of $j_0^{\text{Barthel-Kropina}} \approx 0.448.$}

At high redshift, all variants of the Barthel--Randers models, as~well as the Barthel--Kropina model, deviate from the predictions of the $\Lambda$CDM~model.

\subsection{Statistical~Results}
{Table~\ref{tab_3} presents a comparative analysis of the linear case, logarithmic case, and~exponential case of the Barthel--Randers model, as~well as the Barthel--Kropina model, against~the $\Lambda$CDM model, using several statistical metrics. The~$\chi^2_{\text{tot}, \min}$ value for~the $\Lambda$CDM model is 1780.94. The~Barthel--Randers models (linear, logarithmic, and exponential cases) show similar total $\chi^2_{\text{tot}, \min}$ values ranging from 1781.22 to 1783.38, which are slightly higher than those of the $\Lambda$CDM model. The~Barthel--Kropina model shows the lowest $\chi^2_{\text{tot}, \min}$ value of 1762.37. This indicates that, in~terms of overall goodness of fit, the~Barthel--Kropina model provides the best fit, while the Barthel--Randers models perform slightly worse than the $\Lambda$CDM~model.

The $\chi^2_{\text{red}}$ value provides insight into the goodness of fit adjusted for the number of data points and parameters in the model. The~$\Lambda$CDM model has a $\chi^2_{\text{red}}$ of 1.032, which is very close to the Barthel--Randers models, all of which have values slightly above 1.033. The~lowest $\chi^2_{\text{red}}$ value is from the Barthel--Kropina model (1.022), indicating that it is marginally the best in terms of adjusting for the complexity of the model and fitting the data~well.

\begin{table}[H]
\caption{{Summary of $\chi^2_{\text{tot}}, \chi^2_{\text{red}},$ AIC, $\Delta$AIC, BIC, $\Delta$BIC, and~$p$-value for the $\Lambda$CDM, Barthel--Randers, and~Barthel--Kropina models.}}\label{tab_3}
\centering
\resizebox{\textwidth}{!}{%
{
\begin{tabular}{lcccccccc}
    \toprule
    \textbf{Models} & \boldmath{$\chi^2_{\text{tot}}$} & \textbf{DoF} & \boldmath{$\chi^2_{\text{red}}$} & \textbf{AIC} & \boldmath{$\Delta$}\textbf{AIC }& \textbf{BIC} & \boldmath{$\Delta$}\textbf{BIC} & \textbf{\emph{p}-Value} \\
    \midrule
    $\Lambda$CDM & 1780.94 & 1725 & 1.032 & 1788.94 & 0 & 1810.76 & 0.170 & 0.167 \\
    BR (Linear Case) & 1781.22 & 1724 & 1.033 & 1791.22 & 2.27 & 1818.49 & 7.74 & 0.164 \\
    BR (Logarithmic Case) & 1781.60 & 1724 & 1.033 & 1791.60 & 2.65 & 1818.87 & 8.11 & 0.163 \\
    BR (Exponential Case) & 1783.38 & 1724 & 1.034 & 1793.38 & 4.44 & 1820.65 & 9.89 & 0.155 \\
    Barthel--Kropina & 1762.37 & 1723 & 1.022 & 1774.37 & -14.57 & 1807.10 & -3.65 & 0.249 \\
    \bottomrule
\end{tabular}
}
}
\end{table}

The AIC is a measure of the relative quality of a model, accounting for both the goodness of fit and the number of parameters. The~$\Lambda$CDM model has an AIC of 1788.94. The~Barthel--Randers models (linear, logarithmic, and exponential) have AIC values around 1791.22, 1791.60, and~1793.38, which are higher by 2.27, 2.65, and~4.44, respectively, compared to the $\Lambda$CDM model, indicating a slightly worse balance between fit and model complexity. The~Barthel--Kropina model has the lowest AIC value of 1774.37, which is 14.57 points lower than the $\Lambda$CDM model. 
According to the AIC, the~Barthel--Kropina model provides the best balance between fit and complexity, while the Barthel--Randers models perform slightly worse than the $\Lambda$CDM~model.

The BIC is another metric for model selection, similar to AIC but with a stronger penalty for the number of parameters. The~$\Lambda$CDM model has a BIC of 1810.76, which is lower than the Barthel--Randers models but slightly higher than the Barthel--Kropina model. The~Barthel--Randers models show BIC values between 1818.49 and 1820.65, with~an increase of approximately 7.74 to 9.89 compared to the $\Lambda$CDM model. The~Barthel--Kropina model has the lowest BIC of 1807.10, a~decrease of 3.65 compared to the $\Lambda$CDM model. This suggests that the Barthel--Kropina model is the best according to the BIC criterion, while the Barthel--Randers models perform worse than the $\Lambda$CDM~model.

The \emph{p}-value indicates the statistical significance of the model in terms of explaining the data. The~$\Lambda$CDM model has a \emph{p}-value of 0.167, indicating that the model is statistically significant at a conventional level (i.e., it cannot be rejected at the 5\% significance level). All the other models (Barthel--Randers linear, logarithmic, exponential, and~Barthel--Kropina) have \emph{p}-values ranging from 0.155 to 0.249, which are slightly higher or comparable, suggesting that none of the models can be rejected at the 5\% significance level. The~\emph{p}-values imply that all models, including the base $\Lambda$CDM model, are statistically significant in explaining the~data.

The Barthel--Kropina model outperforms the $\Lambda$CDM model in terms of goodness of fit, AIC, and~BIC, offering a better balance between fit and model complexity. The~Barthel--Randers models (linear, logarithmic, exponential), however, perform slightly worse than the $\Lambda$CDM model in this updated~comparison.

Despite these improvements, all models, including the $\Lambda$CDM model, are statistically significant, with~\emph{p}-values indicating that none of them should be rejected. In~terms of simplicity, however, the~$\Lambda$CDM model remains the most straightforward and easiest to interpret. Therefore, while the Barthel--Kropina model appears to be the most favorable overall, the~differences between the models are relatively small, and~all offer comparable statistical significance.}
\section{Discussion and Final~Remarks}\label{sect8}

One of the fundamental assumptions, and~results, of~present day physics is that the gravitational interaction can be successfully described only in geometric terms. However, which geometry can best describe gravity is still a matter of debate. The~initial Riemannian framework of general relativity was extended to include geometries with non-metricity, torsion, or~both. {While non-metricity naturally arises in the Finslerian setting, the~torsional equivalent is not as immediate. Hence, formulating teleparallel theories in the Finslerian setting could be an interesting line of research. A~mathematical study of teleparallel connections in Finsler geometry is presented in~\cite{Vargas}, and~it has been shown that general teleparallel gravity could be obtained from Finsler geometry~\cite{Tong2023}. This opens up new possibilities of research: the exploration of $f(T)$ theories in the Finslerian setting, which, to~the best of our knowledge, has not yet been achieved. While the discussion above is entirely classical, it is widely believed that quantum effects should become significant at the Planck scale. A~central focus of quantum gravity phenomenology—aimed at identifying observable signatures of quantum gravity—is the study of modified dispersion relations~\cite{AmelinoCamelia2013, Caroff2025}. The~most well-known examples arise from frameworks such as doubly special relativity (DSR) \cite{AmelinoCamelia2002, Letizia2017} and the  $\kappa$-Poincaré algebra~\cite{MajidRuegg1994}. Interestingly, the~symmetries of doubly special relativity can be realized within the framework of Finsler geometry~\cite{AmelinoCamelia2014}. Therefore, further investigation into the connection between quantum gravity and Finsler geometry represents a promising direction for research.}

In the present work, we have briefly reviewed some recent advances in the application of another geometric framework that could provide some insight into the understanding of the gravitational force, namely Finsler geometry, and~its various particular cases. In~Finsler geometry, instead of the point $x$ of Riemann geometry, the~metric depends also on an internal degree of freedom $y$, which is often physically interpreted as velocity. Hence, the~$y$ dependence is the essential characteristic of  Finsler spaces, and~the presence of this internal degrees of freedom opens some new perspective on the physical interpretation of the Finsler geometric theories.  Thus, Finsler geometries are generally anisotropic and nonlocal, and~these properties significantly enlarge the possibilities of the physical~applications.   

On the other hand, from~a physical point of view it is natural to assume that in some specific physical processes, the internal vector $y$ becomes a function of $x$. Such specific situations may occur  if we interpret $y$  as describing spacetime fluctuations, and~we perform an averaging over it, according to the rule $\left<\phi (x)\right>=\int{\phi (x,y)f(y)d^4y}$, where $f(y)$ is a specific distribution function of $y$ \cite{Ikeda}.  Hence,  Finsler geometry with $\widehat{g}(x)=\left.g(x,y)\right |_{y=Y(x)}$ has a deep physical origin, which points towards interesting physical consequences. The~averaging or~the osculating  process leads to a reduction of a Finsler geometry $F$ to a corresponding Riemann geometry. The~interpretation of the vector $y$ as related to some quantum effects, like the vacuum fluctuation, is also supported by the fact that generally, in $(\alpha,\beta)$-type gravitational theories, the matter energy--momentum tensor is not conserved. The~most natural physical interpretation of this result is related to the possibility of particle production from the Finslerian vacuum in an expanding Universe, an~effect that is also specific to quantum field theories in curved spacetimes~\cite{Parker, Parker1, Zel, Parker1a, Parker2}. Hence, Finslerian-type gravitational theories can offer a glimpse of the structure and classical limit of quantum field theories in Riemann~geometries.    

In our presentation, we have concentrated on a specific class of osculating geometries, the~osculating $(\alpha, \beta)$ geometries, which have the remarkable property that their Barthel connection is the Levi--Civita connection. This property allows a significant simplification of the mathematical formalism, which becomes easily tractable by using the standard methods of Riemannian geometry.  The~osculating $(\alpha, \beta)$-type theories can be interpreted as two metric theories. First, we have the Finsler metric function $F(\alpha, \beta)$, which generates its own metric, while $\alpha$ is a Riemannian metric given a priori. For~cosmological applications, the FLRW metric can be adopted as the Riemann metric $\alpha$, and~we assume that this is the physical metric in which the real, observable  gravitational processes take place. However, the~Finsler metric $\widehat{g}(x)=g(x,Y(x))$ also has an important imprint on gravitation, since its presence induces new terms in the gravitational field equations. We have investigated the effects of these new terms from a cosmological perspective only, by~interpreting them as describing an effective dark energy that is responsible for the late acceleration of the~Universe.   

 An important test of a cosmological model is represented by its consistency with the observations. We have used three observational datasets (Cosmic Chronometers, Type Ia Supernovae, Baryon Acoustic Oscillations) to perform a detailed analysis, and~comparison of the Barthel--Randers, Barthel--Kropina and $\Lambda$CDM models. Generally, one can appreciate that all the considered models give a good description of the observational data, but~important differences still appear between the various Finslerian approaches and~the $\Lambda$CDM model. The~differences appear, for~example, in~the prediction of the critical redshift $z_{crit}$ at which the transition from deceleration to acceleration occurs, with~the Barthel--Randers models predicting a value of $z_{crit}$ of the order of $z_{crit}\approx 0.60$, while that in the Barthel--Kropina cosmology is $z_{crit}\approx 0.72$. Significant differences appear in the present values of the deceleration parameter $q(0)$, which in the Barthel--Kropina cosmology takes the value $q(0)\approx-0.37$, while in $\Lambda$CDM and Barthel--Randers models, it has the value $q(0)\approx -0.50$. Moreover, at~redshifts $z>3$, the~predictions of all considered models show significant differences with respect to each other. Hence, cosmological observations at high redshift will give the possibility of discriminating between the various cosmological models based on modified gravity theory constructed with the help of Finsler (and other) geometries.
 
 The main goal of the present review is to draw attention to the possibilities offered by Finsler geometry in the explanation of the cosmological phenomena, and~to give a brief introduction to the existing results in a very specific and~simple  class of~this very complex geometry. We have presented in some detail the generalized Friedmann cosmological equations, and~the general theoretical framework in which cosmological models can be constructed. We would like to point out that from the point of view of calculations, the~derivations of the field equations and of the evolution equations in these particular geometries are no more complicated than those in standard general relativity. This is related especially to the important fact that the connection in osculating-point Finsler geometries is, 
 for~the $(\alpha, \beta)$ metrics, the~Barthel connection, which has the same mathematical form as the Levi--Civita connection. In~our review, we have presented some basic theoretical tools that can be used for the further exploration of the applications of Finsler geometry in the description of gravitational~phenomena.

\vspace{6pt}

\authorcontributions{ Conceptualization, S. V. S., R. H., T. H. and S. S. ; methodology L. C. and S. V. S.; software A.B and H. C.; review and editing T. H., L. C., A. B.,  H. C., and S.S.. All authors have read and agreed to the published version of the manuscript.}

\funding{The work of L.Cs. was supported by a grant of the Ministry of Research, Innovation and Digitization, CNCS/CCCDI-UEFISCDI, project number PN-IV-P8-8.1-PRE-HE-ORG-2023-0118, within~PNCDI IV and Collegium Talentum~Hungary. T. H. was supported by a grant from Collegium Talentum~Hungary.}



\dataavailability{The data underlying this article is already given with references during the analysis of this work.}

\acknowledgments{We would like to thank the anonymous referees for comments and suggestions that helped to improve our~manuscript. }

\conflictsofinterest{The authors declare no conflicts of interest. The~funders had no role in the design of the study; in the collection, analyses, or~interpretation of data; in the writing of the manuscript; or in the decision to publish the~results}

\newpage

\begin{adjustwidth}{-\extralength}{0cm}

\printendnotes[custom]
\reftitle{References}

\PublishersNote{}
\end{adjustwidth}

\end{document}